\documentclass[10pt,a4paper]{article}
\usepackage[utf8]{inputenc}
\usepackage[T1]{fontenc}
\usepackage{lmodern}
\usepackage{amsmath,amssymb,amsfonts,amsthm}
\usepackage{mathrsfs}
\usepackage{mathtools}
\usepackage{bm}
\usepackage{enumitem}
\usepackage{array}
\usepackage{microtype}
\usepackage{booktabs}
\usepackage[heightrounded,top=1.5cm,bottom=1.5cm,left=1.3cm,right=1.3cm]{geometry}
\usepackage[colorlinks=true,linkcolor=blue,citecolor=red,urlcolor=blue,
            breaklinks=true,linktocpage,hypertexnames=false]{hyperref}

\allowdisplaybreaks
\theoremstyle{plain}
\newtheorem{theorem}{Theorem}[section]
\newtheorem{proposition}[theorem]{Proposition}
\newtheorem{lemma}[theorem]{Lemma}
\newtheorem{corollary}[theorem]{Corollary}

\theoremstyle{definition}
\newtheorem{definition}[theorem]{Definition}
\newtheorem{hypothesis}[theorem]{Hypothesis}
\newtheorem{example}[theorem]{Example}
\newtheorem{notation}[theorem]{Notation}

\theoremstyle{remark}
\newtheorem{remark}[theorem]{Remark}

\renewcommand{\proofname}{Proof}
\makeatletter
\renewenvironment{proof}[1][\proofname]{\par\pushQED{\qed}\normalfont
  \topsep6\p@\@plus6\p@\relax\trivlist
  \item[\hskip\labelsep\bfseries #1\@addpunct{.}]\ignorespaces}
 {\popQED\endtrivlist\@endpefalse}
\makeatother

\numberwithin{equation}{section}

\newcommand{\Rpq}{\mathcal{R}(p,q)}
\newcommand{\pRpq}{\partial_{\mathcal{R}(p,q)}}
\newcommand{\dRpqd}{d_{\mathcal{R}(p,q)}}
\newcommand{\dRpq}{\partial_{\Rpq}}

\newcommand{\Pdiag}{\mathbf{P}}
\newcommand{\Qdiag}{\mathbf{Q}}
\newcommand{\dpq}{\partial_{p,q}}
\newcommand{\CC}{\mathbb{C}}
\newcommand{\NN}{\mathbb{N}}

\newcommand{\nRpq}{[n]_{\Rpq}}
\newcommand{\nfact}{[n]!_{\Rpq}}
\newcommand{\mfact}{[m]!_{\Rpq}}
\newcommand{\phione}{\phi_1(p,q)}
\newcommand{\phitwo}{\phi_2(p,q)}
\newcommand{\Exp}{\mathrm{Exp}_{\Rpq}}
\newcommand{\expn}{\mathrm{exp}_{\Rpq}}
\newcommand{\AR}{\mathcal{A}_{\Rpq}}
\newcommand{\DR}{\mathscr{D}_{\Rpq}}
\newcommand{\ORpq}{\mathcal{O}(\DR)}
\newcommand{\BF}{\mathcal{F}_{\Rpq}}
\newcommand{\muRpq}{d\mu_{\Rpq}}
\newcommand{\pd}[2]{\frac{\partial #1}{\partial #2}}
\newcommand{\ket}[1]{|#1\rangle}
\newcommand{\bra}[1]{\langle #1|}

\newcommand{\mean}[1]{\langle #1 \rangle}
\newcommand{\cN}{\mathcal{N}^{\mathrm{s}}_{\Rpq}}
\newcommand{\sket}[3]{|#1,#2,#3\rangle}
\newcommand{\sbra}[3]{\langle #1,#2,#3|}
\newcommand{\sCS}{|z,\zeta_1,\zeta_2\rangle_{\Rpq}}
\newcommand{\sCSb}{\,{}_{\Rpq}\langle z,\zeta_1,\zeta_2|}

\newcommand{\Hq}{\mathfrak{H}_{\mathcal{R},p,q}}
\newcommand{\HB}{\mathcal{H}^{B}_{\mathcal{R},p,q}}
\newcommand{\HFo}{\mathcal{H}^{F_1}_{\mathcal{R},p,q}}
\newcommand{\HFt}{\mathcal{H}^{F_2}_{\mathcal{R},p,q}}
\newcommand{\HFot}{\mathcal{H}^{F_{12}}_{\mathcal{R},p,q}}

\newcommand{\spCS}[2]{|z,#1,#2\rangle_{\mathcal{R},p,q}}
\newcommand{\spCSb}[2]{{}_{\mathcal{R},p,q}\langle z,#1,#2|}
\newcommand{\Expd}{\mathrm{Exp}'_{\mathcal{R},p,q}}
\newcommand{\Expdd}{\mathrm{Exp}''_{\mathcal{R},p,q}}
\newcommand{\SExp}{\mathfrak{E}_{\mathcal{R},p,q}}
\newcommand{\cNp}{\mathcal{N}^{s}_{\mathcal{R},p,q}\!{}^{\prime}}

\newcommand{\rhoB}{\rho^{B}}
\newcommand{\rhoFo}{\rho^{F_1}}
\newcommand{\rhoFt}{\rho^{F_2}}
\newcommand{\rhoFot}{\rho^{F_{12}}}
\newcommand{\rhos}{\rho^{s}}
\newcommand{\tr}{\mathrm{Tr}}

\newcommand{\Kern}{K_{\mathcal{R},p,q}}
\newcommand{\KernB}{K^{B}_{\mathcal{R},p,q}}
\newcommand{\KernFo}{K^{F_1}_{\mathcal{R},p,q}}
\newcommand{\KernFt}{K^{F_2}_{\mathcal{R},p,q}}
\newcommand{\KernFot}{K^{F_{12}}_{\mathcal{R},p,q}}

\begin{document}

\begin{center}
{\large\bfseries Reproducing kernel and matrix representation of the deformed quantum optical density operator of $\mathcal{R}(p,q)-$super-coherent states on $\mathcal{B}_0\oplus\mathcal{B}_1$ fermionic quantum algebra\par}
\vspace{1.0em}
{\large Kaw\`eyim Lankpetre $^{1}$, Isiaka Aremua$^{1,2}$, Joseph Désiré Bukweli Kyemba$^{2,3}$ and Bienvenu Gnim Adewi$^{1}$\par}
\vspace{0.6em}
{\small
$^{1}$Laboratoire de Physique des Mat\'eriaux et des Composants \`a Semi-conducteurs,
Faculty of Sciences, Department of Physics, University of Lom\'e,
01 B.P. 1515 Lom\'e 01, Togo\\
$^{2}$International Chair of Mathematical Physics and Applications
(ICMPA-UNESCO Chair), University of Abomey-Calavi, 072 B.P. 50 Cotonou, Benin\\
$^{3}$
	 University of Kinshasa, Faculty of Sciences and Technology, Department of Mathematics, Statistics and Computer Science\\[0.3em]
\href{mailto:lankpetre.kaweyim.myhal@gmail.com}{lankpetre.kaweyim.myhal@gmail.com},
\href{mailto:claudisak@gmail.com}{claudisak@gmail.com},
\href{mailto:desire.bukweli@unikin.ac.cd}{desire.bukweli@unikin.ac.cd},
\href{mailto:bienvenuadewi@gmail.com}{bienvenuadewi@gmail.com}\par}
\vspace{0.5em}
{\small November 11, 2026\par}
\end{center}

\begin{abstract}
\noindent
We study the $\mathcal{R}(p,q)-$super-coherent states with two Grassmann variables
$(\zeta_1,\zeta_2)$ built on the superalgebra $\mathcal{B}_0\oplus\mathcal{B}_1$
associated with the $\mathcal{R}(p,q)-$deformed quantum algebras of Hounkonnou and
Bukweli-Kyemba, the two dilation parameters $\phi_1(p,q)$ and $\phi_2(p,q)$ being
meromorphic functions of two variables evaluated at $(p,q)$. The graded super-Fock space
splits into four sectors: bosonic, $\phi_1-$fermionic, $\phi_2-$fermionic and
bi-fermionic. In each of them we establish the diagonal representation of the density
operator by means of the $\mathcal{R}(p,q)-$derivative, angular integration and Berezin
integration, and we give a unified compact form in terms of sectorial parametric
functions. The total density operator is obtained in explicit form as the sum of the four
sectorial contributions, and can be read as a mixture of the four sectors. We define the
sectorial reproducing kernels, obtained from the resolution of the identity, we compute
the matrix elements of the density operator in the super-coherent state basis, and we
prove the hermiticity, positivity and matrix multiplication properties of the kernels,
together with the series expansion of their diagonal values. An integral formulation
relates the density operator, its density function and the reproducing kernel, the latter
being the function associated with the identity operator. Berezin quantization in the
bosonic sector yields the quantized operators associated with $z$, $\bar z$, $|z|^2$,
$z^2$ and $\bar z^2$ and their mean values; the dispersions of the super-position and
super-momentum operators show that the super-coherent states are intelligent states.


\end{abstract}

\section{Introduction}\label{sec:intro}
\begingroup
\setlength{\parskip}{0.55em plus 0.2em minus 0.1em}

Coherent states provide a phase-space description of quantum mechanics. Introduced for
the electromagnetic field by Glauber and Sudarshan \cite{Glauber1963,Sudarshan1963}, they
allow a density operator to be written as a superposition of projectors onto coherent
states; this diagonal representation reduces the computation of populations, coherences
and fluctuations to integrals over the complex plane. It relies on the Bargmann--Fock
realization \cite{Bargmann1961}, in which state vectors become holomorphic functions and
the scalar product of two coherent states provides a reproducing kernel, and it extends
to Berezin quantization \cite{Berezin1975}, which associates an operator with every
classical function. These constructions carry over to generalized coherent states
\cite{Klauder1985,Perelomov1986}, and it is within this setting that they are naturally
considered for deformed oscillators.

Deformed oscillator algebras appeared with the Hilbert spaces of analytic functions and
the generalized coherent states of Arik and Coon \cite{ArikCoon1976}, and then with the
$q-$deformed oscillators of Biedenharn \cite{Biedenharn1989} and Macfarlane
\cite{Macfarlane1989}, following the quantum groups of Drinfeld and Jimbo
\cite{Drinfeld1985,Jimbo1985}. The underlying analytic tool is Jackson's $q-$calculus
\cite{Jackson1908,Jackson1910}. Two-parameter deformations were introduced by
Chakrabarti and Jagannathan \cite{CJ1991}; $(p,q)-$differentiation and
$(p,q)-$integration were studied in \cite{BK1994}, the associated hypergeometric series
in \cite{JSR2006}, and a systematic account of the $(p,q)-$calculus is given in
\cite{Sadjang2018}. Deformations with several parameters and with deformation functions
were then proposed by Hounkonnou and Ngompe Nkouankam \cite{HN2007a,HN2007b} and by
Balo\"{\i}tcha, Hounkonnou and Ngompe Nkouankam \cite{BHN2012}, while Quesne constructed
$q-$deformed coherent states with an explicitly known resolution of unity
\cite{Quesne2002,QPT2003}.

All these deformations are particular cases of a general scheme originating in the work
of Odzijewicz \cite{Odzijewicz1998}, within which Hounkonnou and Bukweli-Kyemba
introduced the $\mathcal{R}(p,q)-$deformed quantum algebras attached to a meromorphic
function $\mathcal{R}$, together with their coherent states and special functions
\cite{HB2010}, and then the corresponding differential and integral calculus
\cite{HB2013}. The $\mathcal{R}(p,q)-$deformed Rogers--Szeg\H{o} polynomials
\cite{BH2012JPA}, the coherent states of deformed Heisenberg algebras, their statistics
and geometry \cite{BH2012ADJM,BH2013arXiv}, and coherent state quantization
\cite{HAB2014} have extended this programme, which now includes deformed Virasoro and
Witt algebras \cite{HM2019,HMM2021}.

Super-coherent states labelled by Grassmann variables go back to Aragone and Zypman
\cite{Aragone1986}, and the notion of intelligent states, which saturate the uncertainty
relation, to Aragone, Chalbaud and Salam\'o \cite{Aragone1976}. For a given family of
coherent states, the diagonal representation of the density operator, the construction
of the reproducing kernel from the resolution of the identity and the study of its
hermiticity, positivity and idempotence properties form a well-tested method of analysis
\cite{AHB2015,AAH2026}.

For the $\mathcal{R}(p,q)-$super-coherent states with two Grassmann variables
$(\zeta_1,\zeta_2)$, associated with the two dilation parameters $\phi_1(p,q)$ and
$\phi_2(p,q)$ and built on the superalgebra $\mathcal{B}_0\oplus\mathcal{B}_1$, this
method has not yet been carried out, and its application is not straightforward. The
graded super-Fock space splits into four sectors, bosonic ($B$), $\phi_1-$fermionic
($F_1$), $\phi_2-$fermionic ($F_2$) and bi-fermionic ($F_{12}$), whose weights in the
normalization function involve the function $\mathrm{Exp}_{\mathcal{R}(p,q)}$ and its
derivatives. The ordinary derivative, which isolates the matrix elements in the
undeformed case, is replaced by the $\mathcal{R}(p,q)-$derivative, and the Grassmann
variables require a Berezin integration that has to be combined with the angular and
radial integrations. One therefore has to establish that the density operator keeps an
explicit diagonal representation in each sector, that these representations combine into
an exact formula for the total density operator, and that the reproducing kernel retains
the properties (hermiticity, positivity, idempotence, reproduction) that make it a
computational tool.

We establish these results sector by sector. In each sector, the diagonal representation
of the density operator is obtained by the same sequence of operations: dyadic product of
the super-coherent states, Berezin integration, angular integration, application of
$(\partial_{\mathcal{R}(p,q)})^{n+m}$ and the limit $r\to0$. The four representations
admit a unified compact form and a uniform representation, whose sum gives the explicit
expression of the total density operator; the latter appears as a mixture of the four
sectors. We then define the density function in the super-coherent state basis and the
sectorial reproducing kernels, for which we prove hermiticity, positivity, the matrix
multiplication property and the series expansion of the diagonal values, and we give an
integral formulation in which the reproducing kernel is the function of the identity
operator. Berezin quantization in the bosonic sector finally leads to the quantized
operators, their mean values and the dispersions of the super-position and
super-momentum operators, which saturate the uncertainty relation. 

These results apply to the statistical description of systems whose bosonic and
fermionic degrees of freedom are coupled through an $\mathcal{R}(p,q)$ deformation. The
diagonal representation provides for such systems the analogue of the Glauber--Sudarshan
representation, from which the sector populations and the mean values are computed; the
case of a density operator that is diagonal in the Fock basis, such as a thermal
equilibrium state, follows directly. The integral formulation reduces the composition of
operators, the trace and the purity to integrals of kernels over the label space of the
super-coherent states. Finally, intelligent states play for deformed oscillators the role
that ordinary coherent states play for the harmonic oscillator, namely that of
minimum-uncertainty states serving as a reference for the study of quantum fluctuations.

The paper is organized as follows. Section~\ref{sec:prelim} collects the preliminaries
on the $\mathcal{R}(p,q)-$deformation (scaling functions, $\mathcal{R}(p,q)-$derivative,
exponential functions, Bargmann--Fock space), followed by the superalgebra
$\mathcal{B}_0\oplus\mathcal{B}_1$, the graded super-Fock space and the
$\mathcal{R}(p,q)-$super-coherent states. Section~\ref{sec:density} deals with the
matrix representation of the density operator, sector by sector and then in its explicit
total form. Section~\ref{sec:kernel} is devoted to the reproducing kernels, the matrix
elements of the density operator and the integral formulation.
Section~\ref{sec:quantif} develops Berezin quantization in the bosonic sector, the
dispersions and the intelligent states, and Section~\ref{sec:conclusion} concludes. The
longest proofs are deferred to Appendices~\ref{app:density} and~\ref{app:kernel}.

\endgroup

\section{Preliminaries and \texorpdfstring{$\mathcal{R}(p,q)$}{R(p,q)} super-coherent states on \texorpdfstring{$\mathcal{B}_0\oplus\mathcal{B}_1$}{B0+B1}}
\label{sec:prelim}

This section sets up the framework of the paper. We first recall the elements of the
$\mathcal{R}(p,q)-$deformed calculus used below: the meromorphic function $\mathcal{R}$
and the deformed numbers, the scaling functions $\phi_1$ and $\phi_2$, the
$\mathcal{R}(p,q)-$derivative, the exponential functions and the Bargmann--Fock space.
We then introduce the superalgebra $\mathcal{B}_0\oplus\mathcal{B}_1$, the graded
super-Fock space and the $\mathcal{R}(p,q)-$super-coherent states, together with their
resolution of the identity.

Throughout the paper, $p$ and $q$ are two positive real numbers and
$\mathbb{D}_{R}=\{z\in\CC:\ |z|<R\}$ is the open complex disc of radius $R$,
$\mathcal{O}(\mathbb{D}_{R})$ being the set of holomorphic functions on
$\mathbb{D}_{R}$.

\subsection{The meromorphic function \texorpdfstring{$\mathcal{R}$}{R} and the \texorpdfstring{$\mathcal{R}(p,q)-$}{R(p,q)-}numbers}

\begin{hypothesis}[General hypothesis on $\mathcal{R}$]\label{hyp:R}
	Let $\mathcal{R}$ be a meromorphic function defined on $\CC\times\CC$ by
	\begin{eqnarray}
		\mathcal{R}(x,y)=\sum_{k,l=-L}^{\infty}r_{kl}\,x^{k}y^{l},
		\label{eq:Rxy}
	\end{eqnarray}
	with a possible isolated singularity at the origin, where the $r_{kl}$ are complex
	numbers, $L\in\NN\cup\{0\}$, and such that
	\begin{eqnarray}
		\mathcal{R}(p^{n},q^{n})>0\quad\forall n\in\NN,
		\qquad \mathcal{R}(1,1)=0 .
		\label{eq:Rcond}
	\end{eqnarray}
	Unless otherwise stated, we assume $0<q<p\le 1$ or $0<p<q\le 1$.
\end{hypothesis}

\begin{definition}[$\mathcal{R}(p,q)-$numbers, factorials, binomial coefficients]
	\label{def:Rnumbers}
	Under Hypothesis~\ref{hyp:R}, the $\mathcal{R}(p,q)-$\emph{numbers} (also called
	$\mathcal{R}(p,q)-$\emph{factors}) are the quantities
	$\mathcal{R}(p^{n},q^{n})$, $n=0,1,2,\dots$, also denoted
	$[n]_{\Rpq}:=\mathcal{R}(p^{n},q^{n})$. The
	$\mathcal{R}(p,q)-$\emph{factorial} is
	\begin{eqnarray}
		\mathcal{R}!(p^{n},q^{n})
		=[n]_{\Rpq}!:
		=\begin{cases}
			1, & n=0,\\[2pt]
			\mathcal{R}(p,q)\mathcal{R}(p^{2},q^{2})\cdots\mathcal{R}(p^{n},q^{n}),& n\ge1,
		\end{cases}\label{eq:Rfac}
	\end{eqnarray}
	and the $\mathcal{R}(p,q)-$\emph{binomial coefficient} is
	\begin{eqnarray}
		\left[\begin{array}{c} m\\ n\end{array}\right]_{\Rpq}
		=\frac{\mathcal{R}!(p^{m},q^{m})}
		{\mathcal{R}!(p^{n},q^{n})\,\mathcal{R}!(p^{m-n},q^{m-n})},
		\qquad 0\le n\le m,
		\label{eq:Rbinom}
	\end{eqnarray}
\end{definition}

\subsection{The scaling functions \texorpdfstring{$\phi_{1}$}{phi1} and \texorpdfstring{$\phi_{2}$}{phi2}}
\label{subsec:phi}

The deformation is governed by \emph{two} data: the meromorphic function $\mathcal{R}$
of Hypothesis~\ref{hyp:R}, which produces the deformed numbers, and a pair of
\emph{scaling functions}, which provide the dilation parameters. It is essential (and
this point is often left implicit in the literature) that the latter be themselves
meromorphic functions of two complex variables, evaluated at the point $(p,q)$.

\begin{notation}[Meromorphic scaling functions]\label{not:phi}
	Let
	\begin{eqnarray*}
		\phi_{1},\ \phi_{2}:\ \CC\times\CC\longrightarrow\CC
	\end{eqnarray*}
	be two meromorphic functions of two complex variables, holomorphic and non-vanishing in
	a neighbourhood of $(p,q)$, and write, as usual and when no ambiguity can arise,
	\begin{eqnarray}
		\phi_{1}:=\phi_{1}(p,q),\qquad \phi_{2}:=\phi_{2}(p,q).
		\label{eq:phi-eval}
	\end{eqnarray}
	The $\mathcal{R}(p,q)-$deformation is then the datum of the triple
	$\big(\mathcal{R},\phi_{1},\phi_{2}\big)$, and the deformed exponential functions as
	well as the dilation parameters are built on the values \eqref{eq:phi-eval}. We always
	assume
	\begin{eqnarray}
		\lim_{(p,q)\to(1,1)}\phi_{i}(p,q)=1,\qquad i=1,2,
		\label{eq:phi-limit}
	\end{eqnarray}
	so that the classical calculus is recovered in the undeformed limit, and we write
	$\phi_{i}^{\lambda}$ for $\big(\phi_{i}(p,q)\big)^{\lambda}$, $\lambda\in\CC$, the
	principal branch being chosen.
\end{notation}

\begin{example}[Canonical projections and $(p,q)-$calculus]
	\label{ex:projections}
	Let $\mathrm{pr}_{1},\mathrm{pr}_{2}:\CC\times\CC\to\CC$ be the two canonical
	projections,
	\begin{eqnarray}
		\mathrm{pr}_{1}(x,y)=x,\qquad \mathrm{pr}_{2}(x,y)=y .
		\label{eq:projections}
	\end{eqnarray}
	If $\mathcal{R}$ is a well-defined function (in particular holomorphic and
	non-degenerate) and if one chooses
	\begin{eqnarray}
		\phi_{1} &=& \mathrm{pr}_{1},\nonumber\\
		\phi_{2} &=& \mathrm{pr}_{2},\nonumber\\
		\text{i.e.}\qquad \phi_{1}(p,q) &=& p,\nonumber\\
		\phi_{2}(p,q) &=& q,\label{eq:phi-proj}
	\end{eqnarray}
	then the $\mathcal{R}(p,q)-$derivative becomes the $(p,q)-$derivative and the whole
	theory below reduces to the $(p,q)-$calculus of \cite{CJ1991,JSR2006,Sadjang2018}. In
	this sense \eqref{eq:phi-proj} is the \emph{canonical} choice, and any other admissible
	pair $(\phi_{1},\phi_{2})$ produces a genuinely new deformed calculus.
\end{example}

\begin{example}[Non-projective scaling functions]\label{ex:nonproj}
	The following pairs appear in the literature:
	\begin{enumerate}[label=(\alph*),leftmargin=*]
		\item $\phi_{1}(x,y)=x^{-1}$, $\phi_{2}(x,y)=y$, that is $\phi_{1}=p^{-1}$,
		$\phi_{2}=q$: the Chakrabarti-Jagannathan deformation \cite{CJ1991};
		\item $\phi_{1}(x,y)=x$, $\phi_{2}(x,y)=y^{-1}$, that is $\phi_{1}=p$,
		$\phi_{2}=q^{-1}$: the generalized $q-$Quesne deformation \cite{Quesne2002,QPT2003};
		\item $\phi_{1}(x,y)=y^{\nu}x^{1-\mu}$, $\phi_{2}(x,y)=y^{\nu-1}x^{-\mu}$, that is
		\begin{eqnarray*}
			\phi_{1}=\frac{q^{\nu}}{p^{\mu-1}},\qquad
			\phi_{2}=\frac{q^{\nu-1}}{p^{\mu}},
		\end{eqnarray*}
		which is exactly the pair of deformation functions occurring in the
		$(p,q,\mu,\nu,\phi_{1},\phi_{2})-$generalized oscillator algebra of Hounkonnou and Ngompe
		Nkouankam \cite{HN2007a,HN2007b}.
	\end{enumerate}
	In each case $\phi_{1}$ and $\phi_{2}$ are genuinely meromorphic in both variables, and
	\eqref{eq:phi-limit} is satisfied.
\end{example}

\subsection{The \texorpdfstring{$\mathcal{R}(p,q)-$}{R(p,q)-}derivative of Hounkonnou and Bukweli-Kyemba}

Let $0 < q < p \leq 1$ and let $\mathcal{R}:\CC\times\CC\to\CC$ be meromorphic with
$\mathcal{R}(p^n,q^n) > 0$ for all $n\in\NN$, $\mathcal{R}(1,1)=0$. We write
$\DR = \{z\in\CC : |z| < R_0\}$ and $\ORpq = \mathcal{O}(\DR)$ (holomorphic functions on
$\DR$).

Following \cite{HB2010,BH2012JPA}, the fundamental linear operators on $\ORpq$ are:

\begin{definition}[Dilation and $(p,q)-$difference operators \cite{BH2012JPA}]
	\label{def:pq-ops}
	\begin{eqnarray}
		\Pdiag:\ORpq\to\ORpq,\quad \Pdiag f(z) &:=& f(pz),
		\label{eq:P}\\
		\Qdiag:\ORpq\to\ORpq,\quad \Qdiag f(z) &:=& f(qz),
		\label{eq:Q}\\
		\dpq:\ORpq\to\ORpq,\quad \dpq f(z) &:=& \frac{f(pz)-f(qz)}{z(p-q)}.
		\label{eq:dpq}
	\end{eqnarray}
	Their action on monomials is:
	$\Pdiag z^n = p^n z^n$,\; $\Qdiag z^n = q^n z^n$,\; $\dpq z^n = [n]_{p,q}z^{n-1}$.
\end{definition}
In what follows we also write $P:=\Pdiag$, $Q:=\Qdiag$ and $\partial_{p,q}:=\dpq$.

\begin{definition}[$\mathcal{R}(p,q)-$derivative \cite{HB2010}]
	\label{def:Rderivative}
	Under Hypothesis~\ref{hyp:R}, the $\mathcal{R}(p,q)-$derivative is the operator
	\begin{eqnarray}
		\pRpq:=\partial_{p,q}\,\frac{p-q}{P-Q}\,\mathcal{R}(P,Q)
		=\frac{p-q}{p^{P}-q^{Q}}\,\mathcal{R}(p^{P},q^{Q})\,\partial_{p,q}.
		\label{eq:Rder}
	\end{eqnarray}
\end{definition}

\begin{definition}[$\mathcal{R}(p,q)-$derivative on $\ORpq$ \cite{HB2010,BH2012JPA}]
	\label{def:dRpq}
	On $\ORpq$, the \emph{$\mathcal{R}(p,q)-$derivative} is the operator:
	\begin{eqnarray}
		\dRpq := \dpq\,\frac{p-q}{\Pdiag-\Qdiag}\,\mathcal{R}(\Pdiag,\Qdiag)
		= \frac{p-q}{p\Pdiag-q\Qdiag}\,\mathcal{R}(p\Pdiag,q\Qdiag)\,\dpq.
		\label{eq:dRpq-def}
	\end{eqnarray}
	Its action on monomials is:
	\begin{eqnarray}
		\dRpq z^n &=& \mathcal{R}(p^n,q^n)\,z^{n-1} = \nRpq\,z^{n-1}.
		\label{eq:dRpq-mono}
	\end{eqnarray}
\end{definition}

\begin{proposition}[Action on monomials]\label{prop:monomials}
	For every integer $n\geq0$,
	\begin{eqnarray}
		\pRpq z^{n}=
		\begin{cases}
			0,& n=0,\\
			\mathcal{R}(p^{n},q^{n})\,z^{n-1},& n\ge 1 .
		\end{cases}
		\label{eq:der-monomial}
	\end{eqnarray}
	Consequently, the function $\mathcal{E}_{\Rpq}(z):=\sum_{n\geq0}z^{n}/[n]!_{\Rpq}$ is
	invariant under $\pRpq$: $\pRpq\mathcal{E}_{\Rpq}(z)=\mathcal{E}_{\Rpq}(z)$.
\end{proposition}

\begin{definition}[$\mathcal{R}(p,q)-$differential]\label{def:Rdiff}
	We set $\dRpqd:=(dz)\,\pRpq$ on the algebra $\mathcal{A}_{\Rpq}$ generated by
	$\{1,z,z\partial_{z},\pRpq\}$.
\end{definition}

\begin{lemma}[Iterates of the $\mathcal{R}(p,q)-$derivative]
	\label{lem:dRpq-iter}
	For all integers $N\geq0$ and $0\leq s\leq N$,
	\begin{eqnarray}
		(\dRpq)^{s}\,z^{N}=\frac{\mathcal{R}!(p^{N},q^{N})}{\mathcal{R}!(p^{N-s},q^{N-s})}\,z^{N-s},
		\qquad
		(\dRpq)^{N}\,z^{N}=[N]!_{\Rpq},
		\label{eq:dRpq-iter}
	\end{eqnarray}
	and $(\dRpq)^{s}z^{N}=0$ for $s>N$. In particular
	$\bigl[(\dRpq)^{s}\,r^{N}\bigr]_{r=0}=\delta_{s,N}\,\mathcal{R}!(p^{N},q^{N})$.
	We also write $\mathcal{R}(p^{N},q^{N})!:=\mathcal{R}!(p^{N},q^{N})$.
\end{lemma}

\begin{proof}
	By \eqref{eq:dRpq-mono}, $\dRpq z^{k}=[k]_{\Rpq}z^{k-1}$ for $k\geq1$. Applying this
	relation $s$ times gives
	$(\dRpq)^{s}z^{N}=[N]_{\Rpq}[N-1]_{\Rpq}\cdots[N-s+1]_{\Rpq}\,z^{N-s}$, which is
	\eqref{eq:dRpq-iter} by the definition \eqref{eq:Rfac} of the factorial. For $s>N$, one
	applies $\dRpq$ to the constant $z^0$, which gives $0$ (Proposition~\ref{prop:monomials}).
	The evaluation at $r=0$ keeps only the term $s=N$.
\end{proof}

\begin{remark}\label{rem:algebra}
	The $\mathcal{R}(p,q)-$deformed quantum algebra is generated by
	$\{\mathbf{1},A,A^{\dagger},N\}$ with
	\begin{eqnarray*}
		{}[N,A] &=& -A,\\
		{}[N,A^{\dagger}] &=& A^{\dagger},\\
		AA^{\dagger} &=& \mathcal{R}(p^{N+1},q^{N+1}),\\
		A^{\dagger}A &=& \mathcal{R}(p^{N},q^{N}),
	\end{eqnarray*}
	and admits on $\mathcal{O}(\mathbb{D}_{R})$ the realization
	$A^{\dagger}:=z$, $A:=\pRpq$, $N:=z\partial_{z}$, together with
	$P=p^{z\partial_{z}}$, $Q=q^{z\partial_{z}}$ and
	\begin{eqnarray}
		z\,\pRpq &=& \mathcal{R}(P,Q),\nonumber\\
		\pRpq\,z &=& \mathcal{R}(pP,qQ),\nonumber\\
		{}[z\partial_{z},z] &=& z,\nonumber\\
		{}[z\partial_{z},\pRpq] &=& -\pRpq .\label{eq:comrel}
	\end{eqnarray}
	This is the algebraic skeleton on which the constructions developed below rest; see
	\cite{HB2010,HB2013,HMM2021}.
\end{remark}

\subsection{The \texorpdfstring{$\mathcal{R}(p,q)-$}{R(p,q)-}exponential functions}

We use the two Jackson-type $\mathcal{R}(p,q)-$exponential functions
\begin{eqnarray}
	\mathrm{Exp}_{\Rpq}(z)&=&\sum_{n=0}^{\infty}
	\frac{\phi_{2}^{\binom{n}{2}}\,z^{n}}{\mathcal{R}!(p^{n},q^{n})},
	\label{eq:RExp2}\\
	\mathrm{exp}_{\Rpq}(z)&=&\sum_{n=0}^{\infty}
	\frac{\phi_{1}^{\binom{n}{2}}\,z^{n}}{\mathcal{R}!(p^{n},q^{n})},
	\label{eq:Rexp2}
\end{eqnarray}
which differ from each other by the exchange $\phi_{1}\leftrightarrow\phi_{2}$ of the
weights $\phi_{i}^{\binom{n}{2}}$. Under the canonical choice \eqref{eq:phi-proj} they
reduce to the two $(p,q)-$exponential functions $E_{p,q}$ and $e_{p,q}$ of
\cite{Sadjang2018}. Throughout, we write $\Exp(z):=\mathrm{Exp}_{\Rpq}(z)$ and
$\expn(z):=\mathrm{exp}_{\Rpq}(z)$, we denote by $\Expd$, $\Expdd$ and $\Exp^{(k)}$ the
successive derivatives of $\Exp$, and, for $\alpha\in\CC$, we use the generalized
binomial symbol
\begin{eqnarray}
	\binom{\alpha}{2}:=\frac{\alpha(\alpha-1)}{2},
	\label{eq:binom2}
\end{eqnarray}
which reduces to the usual symbol when $\alpha\in\NN$.

\subsection{The \texorpdfstring{$\mathcal{R}(p,q)$}{R(p,q)} Bargmann--Fock space}

\begin{definition}[Bargmann--Fock realization of $\AR$ \cite{HB2010}]
	\label{def:BF-real}
	The quantum algebra $\AR$ is realized on $\ORpq$ by:
	\begin{eqnarray}
		A^\dagger \equiv z\quad(\text{multiplication by }z),\quad
		A \equiv \dRpq,\quad
		N \equiv z\pd{}{z}\quad(\text{Euler operator}),
		\label{eq:BF-real}
	\end{eqnarray}
	satisfying $AA^\dagger = [N+1]_{\Rpq}$, $A^\dagger A = [N]_{\Rpq}$,
	$[N,A]=-A$, $[N,A^\dagger]=A^\dagger$.
	The orthonormal Fock basis is:
	\begin{eqnarray}
		e_n(z) := \frac{\phi_2^{\binom{n}{2}/2}}{\sqrt{\nfact}}\,z^n
		= \frac{1}{\sqrt{\nfact}}\,H_n(z;\Rpq),
		\label{eq:BF-basis}
	\end{eqnarray}
	where $H_n(z;\Rpq) = \sum_{k=0}^n \binom{n}{k} z^k$
	are the $\mathcal{R}(p,q)-$deformed Rogers--Szeg\H{o} polynomials \cite{BH2012JPA}.
\end{definition}

\begin{definition}[$\mathcal{R}(p,q)$ Bargmann--Fock space]
	\label{def:BF-space}
	\begin{eqnarray}
		\BF := \left\{f\in\ORpq :
		\|f\|^2_{\Rpq} := \int_{\DR}\overline{f(z)}\,f(z)\,\muRpq < \infty\right\},
		\label{eq:BF-space}
	\end{eqnarray}
	endowed with the scalar product $\langle f,g\rangle_{\Rpq} = \int_{\DR}\overline{f(z)}g(z)\,\muRpq$
	and with the $\mathcal{R}(p,q)-$measure $\muRpq = \omega_{\Rpq}(|z|^2)|z|\,d|z|\,d\theta$
	satisfying the \emph{moment conditions}:
	\begin{eqnarray}
		\int_{\DR}|z|^{2n}\,\muRpq = \frac{\phi_2^{\binom{n}{2}}\nfact}{2\pi},
		\quad n\geq0.
		\label{eq:moment}
	\end{eqnarray}
\end{definition}

\begin{proposition}[Orthonormality in $\BF$]
	\label{prop:BF-ortho}
	The basis \eqref{eq:BF-basis} satisfies:
	\begin{eqnarray}
		\langle e_m, e_n\rangle_{\Rpq}
		&=& \int_{\DR}\overline{e_m(z)}\,e_n(z)\,\muRpq\nonumber\\
		&=& \frac{\phi_2^{(\binom{m}{2}+\binom{n}{2})/2}}{\sqrt{\mfact\,\nfact}}
		\int_{\DR}|z|^{m+n}\int_0^{2\pi}\frac{d\theta}{2\pi}e^{i(n-m)\theta}
		|z|\,d|z|\,\omega_{\Rpq}\nonumber\\
		&=& \frac{\phi_2^{\binom{n}{2}}\nfact}{\nfact}\,\delta_{m,n}
		= \delta_{m,n}.
		\label{eq:BF-ortho}
	\end{eqnarray}
\end{proposition}

\begin{definition}[$\mathcal{R}(p,q)-$integral]
	\label{def:Rpq-int}
	The \emph{$\mathcal{R}(p,q)-$integral} of $f\in\ORpq$ is defined by:
	\begin{eqnarray}
		\int_0^{R_0}f(r)\,d_{\Rpq}r
		:= R_0\sum_{n=0}^{\infty}\frac{(R_0 p)^n - (R_0 q)^n}{p^n-q^n}\frac{1}{[n]_{\Rpq}}
		f(R_0 r_n),
		\label{eq:Rpq-int}
	\end{eqnarray}
	where the $r_n$ are the $\mathcal{R}(p,q)-$nodes, which generalizes Jackson's $q-$integral \cite{Jackson1910}.
	In polar coordinates $z=re^{i\theta}$, the full measure is:
	\begin{eqnarray}
		\muRpq(z,\bar{z})
		= \omega_{\Rpq}(r^2)\,r\,dr\,d\theta
		:= \frac{1}{2\pi}\frac{d}{dr}\!\left[\frac{r^2\Expd(r^2)}{\Exp(r^2)}\right]dr\,d\theta,
		\label{eq:muRpq}
	\end{eqnarray}
	which satisfies \eqref{eq:moment} by construction.
\end{definition}

\begin{proposition}[Adjoint of $\dRpq$ in $\BF$ ]
	\label{prop:dRpq-adj}
	The adjoint of $A = \dRpq$ in $\BF$ is $A^\dagger = z$ (multiplication by $z$):
	\begin{eqnarray}
		\langle \dRpq f, g\rangle_{\Rpq} = \langle f, z\cdot g\rangle_{\Rpq},\quad
		\forall f,g\in\BF.
		\label{eq:dRpq-adj}
	\end{eqnarray}
\end{proposition}

\begin{proof}
	By integration by parts, using the deformed Leibniz rule of \cite{HB2010,BH2012JPA}
	and the boundary condition $f(z)\to0$ as $|z|\to R_0$:
	\begin{eqnarray}
		\int_{\DR}\overline{(\dRpq f)(z)}\,g(z)\,\muRpq
		&=& \int_{\DR}\overline{f(z)}\,\dRpq^*[g(z)\muRpq]\nonumber\\
		&=& \int_{\DR}\overline{f(z)}\,(z\cdot g(z))\,\muRpq.
		\label{eq:adj-proof}
	\end{eqnarray}
	Here $\dRpq^*$ is the formal adjoint with respect to $\muRpq$, and the key identity
	$\dRpq^*(z^n\muRpq) = z^{n+1}\muRpq$ follows from the moment conditions.
\end{proof}

\subsection{The superalgebra \texorpdfstring{$\mathcal{B}_0\oplus\mathcal{B}_1$}{B0+B1}}
\label{sec:SCS}

The theory of super-coherent states labelled by Grassmann variables goes back to Aragone
and Zypman \cite{Aragone1986}. In the setting of deformed algebras, one Grassmann
variable is associated with each fermionic degree of freedom. We introduce here the objects
attached to the two Grassmann variables $(\zeta_1,\zeta_2)$ associated with the two
dilation parameters $\phi_1(p,q)$ and $\phi_2(p,q)$.

\begin{definition}[Superalgebra $\mathcal{B}=\mathcal{B}_0\oplus\mathcal{B}_1$]
	\label{def:B01}
	Let $\theta_1,\theta_2$ be two independent Grassmann variables satisfying
	\begin{eqnarray}
		\theta_i^2 = 0,\qquad
		\theta_1\theta_2 + \theta_2\theta_1 = 0,\qquad
		\theta_i z = z\theta_i,\qquad i=1,2.
		\label{eq:Grassmann2}
	\end{eqnarray}
	The superalgebra is
	\begin{eqnarray}
		\mathcal{B} := \CC[z,z^{-1}]\otimes_\CC\Lambda_2(\theta_1,\theta_2)
		= \mathcal{B}_0\oplus\mathcal{B}_1,
		\label{eq:B01}
	\end{eqnarray}
	where the even part and the odd part are
	\begin{eqnarray}
		\mathcal{B}_0 := \CC[z,z^{-1}]\oplus\theta_1\theta_2\,\CC[z,z^{-1}],
		\qquad
		\mathcal{B}_1 := \theta_1\,\CC[z,z^{-1}]\oplus\theta_2\,\CC[z,z^{-1}].
		\label{eq:B0B1}
	\end{eqnarray}
	The two dilation endomorphisms $\sigma_1,\sigma_2$ of $\mathcal{B}$ are defined by
	$\sigma_i(z^n)=\phi_i(p,q)^n z^n$,
	$\sigma_i(\theta_j)=\phi_i(p,q)^{\delta_{ij}}\theta_j$ and
	$\sigma_i(fg)=\sigma_i(f)\sigma_i(g)$, $i,j=1,2$.
\end{definition}

\subsection{Graded super-Fock space}

\begin{definition}[Graded super-Fock space]
	\label{def:sFock}
	The super-Fock space is
	\begin{eqnarray}
		\mathfrak{H}_{\Rpq} := \bigoplus_{(k_1,k_2)\in\{0,1\}^2}\mathcal{H}^{(k_1,k_2)}_{\Rpq},
		\label{eq:sFock}
	\end{eqnarray}
	each sector $\mathcal{H}^{(k_1,k_2)}_{\Rpq}$ having the orthonormal basis
	$\{|n,k_1,k_2\rangle : n\geq0\}$, with
	$\langle m,j_1,j_2|n,k_1,k_2\rangle=\delta_{mn}\delta_{j_1k_1}\delta_{j_2k_2}$.
	The four sectors are the bosonic sector $B$ ($k_1=k_2=0$), the $\phi_1$-fermionic
	sector $F_1$ ($k_1=1$, $k_2=0$), the $\phi_2$-fermionic sector $F_2$ ($k_1=0$,
	$k_2=1$) and the bi-fermionic sector $F_{12}$ ($k_1=k_2=1$). The bosonic operators
	act by
	$A|n,k_1,k_2\rangle=\sqrt{[n]_{\Rpq}}\,|n-1,k_1,k_2\rangle$ and
	$A^\dagger|n,k_1,k_2\rangle=\sqrt{[n+1]_{\Rpq}}\,|n+1,k_1,k_2\rangle$.
\end{definition}

\subsection{Super-normalization function and super-coherent states}
The \emph{$\mathcal{R}(p,q)-$super-exponential}
\begin{eqnarray}
	\SExp(u,v_1,v_2)
	&:=& \Exp(u)
	+ \phione\,v_1\,\Expd(u)\nonumber\\
	&&+ \phitwo\,v_2\,\Expd(u)
	+ \phione\phitwo\,v_1 v_2\,\Expdd(u)
	\label{eq:SExp}
\end{eqnarray}
has the series expansion:
\begin{eqnarray}
	\SExp(u,v_1,v_2)
	&=& \sum_{n=0}^{\infty}\sum_{k_1=0}^{n}\sum_{k_2=0}^{n}
	\frac{\phitwo^{n(n-1)/2}\,\phione^{k_1}\,\phitwo^{k_2}}
	{\nfact}\,u^n\,v_1^{k_1}\,v_2^{k_2}.
	\label{eq:SExp-series}
\end{eqnarray}
\begin{definition}[Super-normalization function]
	\label{def:scN}
	For $x=|z|^2$ and $s_i=|\zeta_i|^2$, $i=1,2$,
	\begin{eqnarray}
		\cN(x,s_1,s_2) &:=& \Exp(x)+\phione^2 s_1\,\Expd(x)+\phitwo^2 s_2\,\Expd(x)\nonumber\\
		&&+\phione^2\phitwo^2 s_1 s_2\,\Expdd(x).
		\label{eq:scN}
	\end{eqnarray}
\end{definition}

\begin{definition}[$\mathcal{R}(p,q)-$super-coherent states]
	\label{def:sCS}
	For $(z,\zeta_1,\zeta_2)\in\DR\times\Lambda_2$, the
	$\mathcal{R}(p,q)-$super-coherent states are
	\begin{eqnarray}
		\sCS &:=& \bigl(\cN(|z|^2,|\zeta_1|^2,|\zeta_2|^2)\bigr)^{-1/2}\nonumber\\
		&&\times\sum_{n=0}^{\infty}\sum_{k_1,k_2\in\{0,1\}}
		\frac{\phitwo^{\binom{n}{2}/2}\,z^n\,(\phione\zeta_1)^{k_1}(\phitwo\zeta_2)^{k_2}}
		{\sqrt{\nfact}}\,|n,k_1,k_2\rangle,
		\label{eq:sCS}\\
		\sCSb &:=& \bigl(\cN(|z|^2,|\zeta_1|^2,|\zeta_2|^2)\bigr)^{-1/2}\nonumber\\
		&&\times\sum_{m=0}^{\infty}\sum_{j_1,j_2\in\{0,1\}}
		\frac{\phitwo^{\binom{m}{2}/2}\,\bar{z}^m\,(\phione\bar\zeta_1)^{j_1}(\phitwo\bar\zeta_2)^{j_2}}
		{\sqrt{\mfact}}\,\langle m,j_1,j_2|.
		\label{eq:sCSb}
	\end{eqnarray}
	The sectorial components are obtained by setting the corresponding Grassmann
	variables to zero: the bosonic state $|z,0,0\rangle_{\Rpq}$ corresponds to
	$\zeta_1=\zeta_2=0$, with normalization $\cN(|z|^2,0,0)=\Exp(|z|^2)$, and the states
	of the sectors $F_1$, $F_2$, $F_{12}$ are written out in Section~\ref{sec:density}.
\end{definition}

\begin{proposition}[Fundamental properties]
	\label{prop:CS-props}
	The states $\sCS$ simultaneously satisfy:
	\begin{eqnarray}
		A\,\sCS &=& z\,\sCS,
		\label{eq:ev-bos}\\[4pt]
		\partial_{\theta_1}\,\sCS &=& \phione\,\zeta_1\,\sCS,
		\label{eq:ev-fer1}\\[4pt]
		\partial_{\theta_2}\,\sCS &=& \phitwo\,\zeta_2\,\sCS.
		\label{eq:ev-fer2}
	\end{eqnarray}
	The \emph{resolution of the identity} reads:
	\begin{eqnarray}
		\int_{\DR\times\Lambda_1^2}
		\cN(|z|^2,|\zeta_1|^2,|\zeta_2|^2)\;\sCS\sCSb
		\;d\mu^{(12)}_{\Rpq}(z,\zeta_1,\zeta_2)
		&=& \mathbf{1}_{\Hq},
		\label{eq:resolution}
	\end{eqnarray}
	with $d\mu^{(12)}_{\Rpq} = d\mu_{\Rpq}(z,\bar{z})\,
	d\bar{\zeta}_1 d\zeta_1\,d\bar{\zeta}_2 d\zeta_2$,
	$\int_{\DR}|z|^{2n}\,d\mu_{\Rpq} = \phitwo^{n(n-1)/2}\nfact/(2\pi)$.
\end{proposition}

The super-Fock space is the orthogonal direct sum:
\begin{eqnarray}
	\Hq &=& \HB \oplus \HFo \oplus \HFt \oplus \HFot,
	\label{eq:Hq}
\end{eqnarray}
with the sectorial bases, weights and super-coherent-state projectors listed in
Table~\ref{tab:sectors}.

\begin{table}[!ht]
	\caption{Sectorial structure of the graded super-Fock space $\Hq$ of~\eqref{eq:Hq}:
		Grassmann indices $(k_1,k_2)$, orthonormal basis, weight of each sector in the
		super-normalization function~\eqref{eq:scN} and super-coherent-state projector
		entering the resolution of the identity~\eqref{eq:resolution}.}
	\label{tab:sectors}
	\centering
	\renewcommand{\arraystretch}{1.4}
	\begin{tabular}{ccccc}
		\toprule
		\textbf{Sector} & $(k_1,k_2)$ & \textbf{Basis}$^{\mathrm{a}}$
		& \textbf{Weight in $\cN$}\,$^{\mathrm{b}}$ & \textbf{CS projector}$^{\mathrm{c}}$\\
		\midrule
		$\HB$ (bosonic)    & $(0,0)$ & $\sket{n}{0}{0}$ & $\Exp(x)$
		& $\spCS{\,0\,}{0}\spCSb{\,0\,}{0}$\\
		$\HFo$ ($\phi_1$-fermionic) & $(1,0)$ & $\sket{n}{1}{0}$ & $\phione^2 s_1\Expd(x)$
		& $\spCS{\zeta_1}{0}\spCSb{\zeta_1}{0}$\\
		$\HFt$ ($\phi_2$-fermionic) & $(0,1)$ & $\sket{n}{0}{1}$ & $\phitwo^2 s_2\Expd(x)$
		& $\spCS{0}{\zeta_2}\spCSb{0}{\zeta_2}$\\
		$\HFot$ (bi-fermionic)      & $(1,1)$ & $\sket{n}{1}{1}$ & $\phione^2\phitwo^2 s_1 s_2\Expdd(x)$
		& $\spCS{\zeta_1}{\zeta_2}\spCSb{\zeta_1}{\zeta_2}$\\
		\bottomrule
	\end{tabular}
	
	\smallskip
	\parbox{\linewidth}{\footnotesize\raggedright
		$^{\mathrm{a}}$\,$n\in\NN$; orthonormal basis of Definition~\ref{def:sFock}.\\
		$^{\mathrm{b}}$\,$x=|z|^2$, $s_i=|\zeta_i|^2$ ($i=1,2$); $\Expd$ and $\Expdd$ are the
		first and second derivatives of $\Exp$; the four weights add up to
		$\cN(x,s_1,s_2)$.\\
		$^{\mathrm{c}}$\,CS: super-coherent state~\eqref{eq:sCS}; the Grassmann entries that do
		not belong to the sector are set to zero.}
\end{table}

\subsection{Super-measure and resolution of the identity}

\begin{definition}[Super-measure]
	\label{def:smes}
	The super-measure is
	\begin{eqnarray}
		d\mu^{\mathrm{s}}_{\Rpq}(z,\zeta_1,\zeta_2)
		:= d\mu_{\Rpq}(\bar{z},z)\,d\bar\zeta_1\,d\zeta_1\,d\bar\zeta_2\,d\zeta_2,
		\label{eq:smes}
	\end{eqnarray}
	where $d\mu_{\Rpq}$ is the $\mathcal{R}(p,q)-$measure of
	Definition~\ref{def:BF-space}, the Berezin integrals satisfying
	$\int d\bar\zeta_i\,d\zeta_i\cdot 1=0$ and
	$\int d\bar\zeta_i\,d\zeta_i\cdot\bar\zeta_i\zeta_i=1$. In polar coordinates
	$z=re^{i\theta}$, $0\leq r<L$, where $L=R_0$ is the radius of $\DR$, we denote by
	$d\lambda(r)$ the associated radial measure, so that the moment conditions read
	\begin{eqnarray}
		2\pi\int_0^L r^{2n}\,d\lambda(r) = \phitwo^{\binom{n}{2}}\,\nfact,\qquad n\geq0.
		\label{eq:moments}
	\end{eqnarray}
\end{definition}

\begin{proposition}[Resolution of the identity]
	\label{prop:res-id2}
	\begin{eqnarray}
		\int_{\DR\times\Lambda_2}\cN(|z|^2,|\zeta_1|^2,|\zeta_2|^2)\,\sCS\,\sCSb\,
		d\mu^{\mathrm{s}}_{\Rpq}(z,\zeta_1,\zeta_2)=\mathbf{1}_{\mathfrak{H}_{\Rpq}}.
		\label{eq:res-id2}
	\end{eqnarray}
\end{proposition}

\medskip
The deformation is fixed by the triple $(\mathcal{R},\phi_1,\phi_2)$. The
$\mathcal{R}(p,q)-$super-coherent states \eqref{eq:sCS} are eigenstates of the operator
$A$ and of the Grassmann derivations, they decompose over the four sectors $B$, $F_1$,
$F_2$ and $F_{12}$ of the super-Fock space \eqref{eq:Hq}, and they resolve the identity
for the super-measure \eqref{eq:smes}. These properties, together with the
$\mathcal{R}(p,q)-$derivative, Lemma~\ref{lem:dRpq-iter} and the moment conditions
\eqref{eq:moments}, are the tools used in the following sections.

\section{Matrix representation of the density operator}
\label{sec:density}

This section establishes the representation of the density operator in the graded Fock
basis. We first define the density operator and its restrictions to the four sectors. In
each sector the diagonal representation is obtained by the same sequence of operations:
dyadic product of the super-coherent states, Berezin integration, angular integration,
application of $(\dRpq)^{n+m}$ and the limit $r\to0$. The four results are then written
in a unified compact form and gathered into the explicit formula for the total density
operator.

\subsection{Density operator in the graded Fock basis}

\begin{definition}
	In the graded Fock basis $\bigl\{|n,k_1,k_2\rangle\bigr\}$, the density operator
	$\rho^{\mathrm{s}}$ has the representation
	\begin{eqnarray}
		\rho^{\mathrm{s}} &:=&
		\sum_{\substack{n,m=0\\k_1,k_2,j_1,j_2\in\{0,1\}}}^{\infty}
		\rho^{\mathrm{s}}(n,k_1,k_2;\,m,j_1,j_2)\,
		|n,k_1,k_2\rangle\langle m,j_1,j_2|.
		\label{eq:rhos}
	\end{eqnarray}
\end{definition}

\begin{definition}
	We define the four \emph{sectorial density operators} corresponding to the four
	subspaces of $\mathfrak{H}_{\Rpq}$:
	\begin{eqnarray}
		\rho^{B} &:=& \sum_{n,m=0}^{\infty}
		\rho^{B}(n,m)\,|n,0,0\rangle\langle m,0,0|,
		\label{eq:rhoB}\\
		\rho^{F_1} &:=& \sum_{n,m=0}^{\infty}
		\rho^{F_1}(n,m)\,|n,1,0\rangle\langle m,1,0|,
		\label{eq:rhoF1}\\
		\rho^{F_2} &:=& \sum_{n,m=0}^{\infty}
		\rho^{F_2}(n,m)\,|n,0,1\rangle\langle m,0,1|,
		\label{eq:rhoF2}\\
		\rho^{F_{12}} &:=& \sum_{n,m=0}^{\infty}
		\rho^{F_{12}}(n,m)\,|n,1,1\rangle\langle m,1,1|.
		\label{eq:rhoF12}
	\end{eqnarray}
\end{definition}

\subsection{Diagonal representation of the bosonic sector $\rho^B$}
\label{sec:rhoB}

\begin{proposition}
	\label{prop:rhoB}
	In the Fock basis $\bigl\{|n,0,0\rangle\bigr\}_{n=0}^{\infty}$, the density operator
	$\rho^B$ has, starting from the resolution of the identity in the bosonic sector, the
	representation:
	\begin{eqnarray}
		\rho^B &=& \sum_{n=0}^{\infty}\sum_{m=0}^{\infty}
		\rho^B(n,m)\;
		\sqrt{\frac{[n]!_{\Rpq}}{\phitwo^{\binom{n}{2}}}}\;
		\sqrt{\frac{[m]!_{\Rpq}}{\phitwo^{\binom{m}{2}}}}\;
		\frac{1}{\mathcal{R}(p^{n+m},q^{n+m})!}\nonumber\\
		&&\times\;
		\bigl(\dRpq\bigr)^{n+m}\!\!
		\left\{\int_0^{2\pi}\frac{d\theta}{2\pi}\,e^{i(m-n)\theta}\;
		\mathrm{Exp}_{\Rpq}(r^2)\;\sket{re^{i\theta}}{0}{0}\sbra{re^{i\theta}}{0}{0}
		\right\}_{r=0}\!\!.
		\label{eq:rhoB-prop}
	\end{eqnarray}
\end{proposition}

{\bf Proof.}\\
We start from the coherent states of the bosonic sector ($\zeta_1=\zeta_2=0$):
\begin{eqnarray}
	|z,0,0\rangle_{\Rpq} &=&
	\bigl(\mathrm{Exp}_{\Rpq}(|z|^2)\bigr)^{-1/2}
	\sum_{n=0}^{\infty}
	\frac{\phitwo^{\binom{n}{2}/2}}{\sqrt{[n]!_{\Rpq}}}\,z^n\,|n,0,0\rangle,
	\label{eq:csB-ket}\\
	{}_{\Rpq}\langle z,0,0| &=&
	\bigl(\mathrm{Exp}_{\Rpq}(|z|^2)\bigr)^{-1/2}
	\sum_{m=0}^{\infty}
	\frac{\phitwo^{\binom{m}{2}/2}}{\sqrt{[m]!_{\Rpq}}}\,\bar{z}^m\,\langle m,0,0|.
	\label{eq:csB-bra}
\end{eqnarray}
The dyadic product gives:
\begin{eqnarray}
	&&|z,0,0\rangle_{\Rpq}\,{}_{\Rpq}\langle z,0,0|\nonumber\\
	&=& \bigl(\mathrm{Exp}_{\Rpq}(|z|^2)\bigr)^{-1}
	\sum_{n=0}^{\infty}\sum_{m=0}^{\infty}
	\frac{\phitwo^{\binom{n}{2}/2}}{\sqrt{[n]!_{\Rpq}}}\,
	\frac{\phitwo^{\binom{m}{2}/2}}{\sqrt{[m]!_{\Rpq}}}\,
	z^n\bar{z}^m\,|n,0,0\rangle\langle m,0,0|.
	\label{eq:dyad-B}
\end{eqnarray}
We set $z = re^{i\theta}$, $0\leq\theta\leq 2\pi$:
\begin{eqnarray}
	&&|re^{i\theta},0,0\rangle_{\Rpq}\,{}_{\Rpq}\langle re^{i\theta},0,0|\nonumber\\
	&=& \bigl(\mathrm{Exp}_{\Rpq}(r^2)\bigr)^{-1}
	\sum_{n,m=0}^{\infty}
	\frac{\phitwo^{(\binom{n}{2}+\binom{m}{2})/2}}
	{\sqrt{[n]!_{\Rpq}\,[m]!_{\Rpq}}}\,
	r^{n+m}\,e^{i(n-m)\theta}\,|n,0,0\rangle\langle m,0,0|.
	\label{eq:dyad-B-polar}
\end{eqnarray}
Multiplying by $\mathrm{Exp}_{\Rpq}(r^2)$ and by $e^{il\theta}$:
\begin{eqnarray}
	&&e^{il\theta}\,\mathrm{Exp}_{\Rpq}(r^2)\,
	|re^{i\theta},0,0\rangle_{\Rpq}\,{}_{\Rpq}\langle re^{i\theta},0,0|\nonumber\\
	&=&
	\sum_{n,m=0}^{\infty}
	\frac{\phitwo^{(\binom{n}{2}+\binom{m}{2})/2}}
	{\sqrt{[n]!_{\Rpq}\,[m]!_{\Rpq}}}\,
	r^{n+m}\,e^{i(l+n-m)\theta}\,|n,0,0\rangle\langle m,0,0|.
	\label{eq:el-B}
\end{eqnarray}
We integrate over $\theta\in[0,2\pi]$. Since
$\displaystyle\int_0^{2\pi}d\theta\,e^{i(l+n-m)\theta} = 2\pi\,\delta(l+n-m)$,
we obtain:
\begin{eqnarray}
	&&\int_0^{2\pi}d\theta\,e^{il\theta}\,\mathrm{Exp}_{\Rpq}(r^2)\,
	|re^{i\theta},0,0\rangle_{\Rpq}\,{}_{\Rpq}\langle re^{i\theta},0,0|\nonumber\\
	&=& \sum_{n,m=0}^{\infty}
	\frac{\phitwo^{(\binom{n}{2}+\binom{m}{2})/2}}
	{\sqrt{[n]!_{\Rpq}\,[m]!_{\Rpq}}}\,
	r^{n+m}\cdot 2\pi\,\delta(l+n-m)\,|n,0,0\rangle\langle m,0,0|.
	\label{eq:int-B}
\end{eqnarray}
We apply the operator $(\dRpq)^s$. Using \eqref{eq:dRpq-iter}:
$(\dRpq)^s\,r^{n+m} = \dfrac{\mathcal{R}(p^{n+m},q^{n+m})!}
{\mathcal{R}(p^{n+m-s},q^{n+m-s})!}\,r^{n+m-s}$,
we obtain:
\begin{eqnarray}
	&&(\dRpq)^s\left\{\int_0^{2\pi}d\theta\,e^{il\theta}\,\mathrm{Exp}_{\Rpq}(r^2)\,
	|re^{i\theta},0,0\rangle_{\Rpq}\,{}_{\Rpq}\langle re^{i\theta},0,0|\right\}\nonumber\\
	&=& \sum_{n,m=0}^{\infty}
	\frac{\phitwo^{(\binom{n}{2}+\binom{m}{2})/2}}
	{\sqrt{[n]!_{\Rpq}\,[m]!_{\Rpq}}}\,
	\frac{\mathcal{R}(p^{n+m},q^{n+m})!}
	{\mathcal{R}(p^{n+m-s},q^{n+m-s})!}\,
	r^{n+m-s}\cdot 2\pi\,\delta(l+n-m)\,|n,0,0\rangle\langle m,0,0|.
	\label{eq:dRpq-B}
\end{eqnarray}
For $r\to 0$: the condition $n+m-s=0$ forces $s=n+m$ and $\delta(l+n-m)$ forces
$l=m-n$. Hence:
\begin{eqnarray}
	&&\bigl(\dRpq\bigr)^{n+m}\!\!\left\{
	\int_0^{2\pi}\frac{d\theta}{2\pi}\,e^{i(m-n)\theta}\,\mathrm{Exp}_{\Rpq}(r^2)\,
	|re^{i\theta},0,0\rangle_{\Rpq}\,{}_{\Rpq}\langle re^{i\theta},0,0|
	\right\}_{r=0}\nonumber\\
	&=& \sum_{n,m=0}^{\infty}
	\frac{\phitwo^{(\binom{n}{2}+\binom{m}{2})/2}}
	{\sqrt{[n]!_{\Rpq}\,[m]!_{\Rpq}}}\,
	\mathcal{R}(p^{n+m},q^{n+m})!\,
	|n,0,0\rangle\langle m,0,0|.
	\label{eq:r0-B}
\end{eqnarray}
We invert this relation to express $|n,0,0\rangle\langle m,0,0|$:
\begin{eqnarray}
	|n,0,0\rangle\langle m,0,0|
	&=&
	\frac{\sqrt{[n]!_{\Rpq}\,[m]!_{\Rpq}}}
	{\phitwo^{(\binom{n}{2}+\binom{m}{2})/2}\,\mathcal{R}(p^{n+m},q^{n+m})!}\nonumber\\
	&&\times\;
	\bigl(\dRpq\bigr)^{n+m}\!\!\left\{
	\int_0^{2\pi}\frac{d\theta}{2\pi}\,e^{i(m-n)\theta}\,\mathrm{Exp}_{\Rpq}(r^2)\,
	|re^{i\theta},0,0\rangle_{\Rpq}\,{}_{\Rpq}\langle re^{i\theta},0,0|
	\right\}_{r=0}\!\!.
	\label{eq:|n><m|-B}
\end{eqnarray}
Substituting into the definition \eqref{eq:rhoB}, we identify the coefficients:
\begin{eqnarray}
	\rho^B(n,m) &=&
	\frac{\sqrt{[n]!_{\Rpq}\,[m]!_{\Rpq}}}
	{\phitwo^{(\binom{n}{2}+\binom{m}{2})/2}\,\mathcal{R}(p^{n+m},q^{n+m})!}\nonumber\\
	&&\times\;
	\bigl(\dRpq\bigr)^{n+m}\!\!\left\{
	\int_0^{2\pi}\frac{d\theta}{2\pi}\,e^{i(m-n)\theta}\,\mathrm{Exp}_{\Rpq}(r^2)\,
	|re^{i\theta},0,0\rangle_{\Rpq}\,{}_{\Rpq}\langle re^{i\theta},0,0|
	\right\}_{r=0}\!\!,
	\label{eq:rhoB-coeff}
\end{eqnarray}
which gives the representation \eqref{eq:rhoB-prop}.\hfill$\Box$\\

We introduce the \emph{bosonic parametric function}:
\begin{eqnarray}
	\varrho^B_{\Rpq}(n)
	&:=& \frac{\phitwo^{\binom{n}{2}}}{[n]!_{\Rpq}},
	\label{eq:varrhoB}
\end{eqnarray}
which allows us to rewrite:

\begin{proposition}
	\label{prop:rhoB-compact}
	\begin{eqnarray}
		\rho^B &=& \sum_{n,m=0}^{\infty}
		\rho^B(n,m)\;
		\frac{\sqrt{\varrho^B_{\Rpq}(n)^{-1}\,\varrho^B_{\Rpq}(m)^{-1}}}
		{\mathcal{R}(p^{n+m},q^{n+m})!}\nonumber\\
		&&\times\;
		\bigl(\dRpq\bigr)^{n+m}\!\!
		\left\{
		\int_0^{2\pi}\frac{d\theta}{2\pi}\,e^{i(m-n)\theta}\,
		\mathrm{Exp}_{\Rpq}(r^2)\,
		|re^{i\theta},0,0\rangle_{\Rpq}\,{}_{\Rpq}\langle re^{i\theta},0,0|
		\right\}_{r=0}.
		\label{eq:rhoB-compact}
	\end{eqnarray}
\end{proposition}

\subsection{Diagonal representation of the $\phi_1$-fermionic sector $\rho^{F_1}$}
\label{sec:rhoF1}

\begin{proposition}
	\label{prop:rhoF1}
	In the Fock basis $\bigl\{|n,1,0\rangle\bigr\}_{n=0}^{\infty}$, the density operator
	$\rho^{F_1}$ has, starting from the resolution of the identity in the
	$\phi_1$-fermionic sector, the representation:
	\begin{eqnarray}
		\rho^{F_1} &=& \sum_{n,m=0}^{\infty}
		\rho^{F_1}(n,m)\;
		\frac{\sqrt{[n]!_{\Rpq}\,[m]!_{\Rpq}}}
		{\phitwo^{(\binom{n}{2}+\binom{m}{2})/2}\,\phione^2\,
			\mathcal{R}(p^{n+m},q^{n+m})!}\nonumber\\
		&&\times\;
		\bigl(\dRpq\bigr)^{n+m}\!\!
		\left\{\int_0^{2\pi}\frac{d\theta}{2\pi}\,e^{i(m-n)\theta}
		\int d\bar{\zeta}_1\,d\zeta_1\;
		\mathrm{Exp}'_{\Rpq}(r^2)\,\phione^2|\zeta_1|^2\,\right.\nonumber\\
		&&\qquad\times\left.
		|re^{i\theta},\zeta_1,0\rangle_{\Rpq}\,{}_{\Rpq}\langle re^{i\theta},\zeta_1,0|
		\right\}_{r=0}.
		\label{eq:rhoF1-prop}
	\end{eqnarray}
\end{proposition}

\begin{proof}
	The proof is given in Appendix~\ref{app:rhoF1}.
\end{proof}

\subsection{Diagonal representation of the $\phi_2$-fermionic sector $\rho^{F_2}$}
\label{sec:rhoF2}

\begin{proposition}
	\label{prop:rhoF2}
	By the symmetry $\phi_1\leftrightarrow\phi_2$, $\zeta_1\leftrightarrow\zeta_2$, the
	$\phi_2$-fermionic sector has the representation:
	\begin{eqnarray}
		\rho^{F_2} &=& \sum_{n,m=0}^{\infty}
		\rho^{F_2}(n,m)\;
		\frac{\sqrt{[n]!_{\Rpq}\,[m]!_{\Rpq}}}
		{\phitwo^{(\binom{n}{2}+\binom{m}{2})/2}\,\phitwo^2\,
			\mathcal{R}(p^{n+m},q^{n+m})!}\nonumber\\
		&&\times\;
		\bigl(\dRpq\bigr)^{n+m}\!\!
		\left\{\int_0^{2\pi}\frac{d\theta}{2\pi}\,e^{i(m-n)\theta}
		\int d\bar{\zeta}_2\,d\zeta_2\;
		\mathrm{Exp}'_{\Rpq}(r^2)\,\phitwo^2|\zeta_2|^2\right.\nonumber\\
		&&\qquad\times\left.
		|re^{i\theta},0,\zeta_2\rangle_{\Rpq}\,{}_{\Rpq}\langle re^{i\theta},0,\zeta_2|
		\right\}_{r=0},
		\label{eq:rhoF2-prop}
	\end{eqnarray}
	with
	\begin{eqnarray}
		\rho^{F_2}(n,m) &=&
		\frac{\sqrt{[n]!_{\Rpq}\,[m]!_{\Rpq}}}
		{\phitwo^{(\binom{n}{2}+\binom{m}{2})/2}\,\phitwo^2\,
			\mathcal{R}(p^{n+m},q^{n+m})!}\nonumber\\
		&&\times\;
		\bigl(\dRpq\bigr)^{n+m}\!\!\left\{
		\int_0^{2\pi}\frac{d\theta}{2\pi}\,e^{i(m-n)\theta}
		\int d\bar{\zeta}_2 d\zeta_2\,\phitwo^2|\zeta_2|^2\,
		\mathrm{Exp}'_{\Rpq}(r^2)\right.\nonumber\\
		&&\qquad\left.\times\,|re^{i\theta},0,\zeta_2\rangle_{\Rpq}\,{}_{\Rpq}\langle re^{i\theta},0,\zeta_2|
		\right\}_{r=0}.
		\label{eq:rhoF2-coeff}
	\end{eqnarray}
\end{proposition}

{\bf Proof.}
The proof is entirely analogous to that of Proposition~\ref{prop:rhoF1}, exchanging
$\phi_1\leftrightarrow\phi_2$, $\zeta_1\leftrightarrow\zeta_2$,
$k_1\leftrightarrow k_2$, and $\mathrm{Exp}'_{\Rpq}\to\mathrm{Exp}'_{\Rpq}$
(the term of the $\phi_2$ part of \eqref{eq:scN}).
\hfill$\Box$\\

\subsection{Diagonal representation of the bi-fermionic sector $\rho^{F_{12}}$}
\label{sec:rhoF12}

\begin{proposition}
	\label{prop:rhoF12}
	In the Fock basis $\bigl\{|n,1,1\rangle\bigr\}_{n=0}^{\infty}$, the density operator
	$\rho^{F_{12}}$ has the representation:
	\begin{eqnarray}
		\rho^{F_{12}} &=& \sum_{n,m=0}^{\infty}
		\rho^{F_{12}}(n,m)\;
		\frac{\sqrt{[n]!_{\Rpq}\,[m]!_{\Rpq}}}
		{\phitwo^{(\binom{n}{2}+\binom{m}{2})/2}\,\phione^2\phitwo^2\,
			\mathcal{R}(p^{n+m},q^{n+m})!}\nonumber\\
		&&\times\;
		\bigl(\dRpq\bigr)^{n+m}\!\!
		\left\{\int_0^{2\pi}\frac{d\theta}{2\pi}\,e^{i(m-n)\theta}
		\int d\bar{\zeta}_1 d\zeta_1\,d\bar{\zeta}_2 d\zeta_2\;
		\mathrm{Exp}''_{\Rpq}(r^2)\,\phione^2\phitwo^2|\zeta_1|^2|\zeta_2|^2\right.\nonumber\\
		&&\qquad\times\left.
		|re^{i\theta},\zeta_1,\zeta_2\rangle_{\Rpq}\,{}_{\Rpq}\langle re^{i\theta},\zeta_1,\zeta_2|
		\right\}_{r=0},
		\label{eq:rhoF12-prop}
	\end{eqnarray}
	with
	\begin{eqnarray}
		\rho^{F_{12}}(n,m) &=&
		\frac{\sqrt{[n]!_{\Rpq}\,[m]!_{\Rpq}}}
		{\phitwo^{(\binom{n}{2}+\binom{m}{2})/2}\,\phione^2\phitwo^2\,
			\mathcal{R}(p^{n+m},q^{n+m})!}\nonumber\\
		&&\times\;
		\bigl(\dRpq\bigr)^{n+m}\!\!\left\{
		\int_0^{2\pi}\frac{d\theta}{2\pi}\,e^{i(m-n)\theta}
		\int d\bar{\zeta}_1 d\zeta_1\,d\bar{\zeta}_2 d\zeta_2\right.\nonumber\\
		&&\qquad\left.\times\phione^2\phitwo^2|\zeta_1|^2|\zeta_2|^2\,
		\mathrm{Exp}''_{\Rpq}(r^2)\right.\nonumber\\
		&&\qquad\left.\times\,|re^{i\theta},\zeta_1,\zeta_2\rangle_{\Rpq}\,{}_{\Rpq}\langle re^{i\theta},\zeta_1,\zeta_2|
		\right\}_{r=0}.
		\label{eq:rhoF12-coeff}
	\end{eqnarray}
\end{proposition}

\begin{proof}
	The proof is given in Appendix~\ref{app:rhoF12}.
\end{proof}

\subsection{Unified compact form and parametric function}
\label{sec:compact}

We introduce the \emph{sectorial parametric functions}:
\begin{eqnarray}
	\varrho^B_{\Rpq}(n) &:=& \frac{\phitwo^{\binom{n}{2}}}{[n]!_{\Rpq}},
	\label{eq:varrhoB-def}\\
	\varrho^{F_1}_{\Rpq}(n) &:=& \frac{\phitwo^{\binom{n}{2}}\,\phione^2}{[n]!_{\Rpq}},
	\label{eq:varrhoF1-def}\\
	\varrho^{F_2}_{\Rpq}(n) &:=& \frac{\phitwo^{\binom{n}{2}}\,\phitwo^2}{[n]!_{\Rpq}},
	\label{eq:varrhoF2-def}\\
	\varrho^{F_{12}}_{\Rpq}(n) &:=& \frac{\phitwo^{\binom{n}{2}}\,\phione^2\phitwo^2}{[n]!_{\Rpq}}.
	\label{eq:varrhoF12-def}
\end{eqnarray}

\begin{proposition}[Unified compact form]
	\label{prop:compact}
	The four sectorial density operators can be written in a unified way, for
	$\varepsilon\in\{B, F_1, F_2, F_{12}\}$:
	\begin{eqnarray}
		\rho^{\varepsilon} &=& \sum_{n,m=0}^{\infty}
		\rho^{\varepsilon}(n,m)\;
		\frac{\sqrt{(\varrho^{\varepsilon}_{\Rpq}(n))^{-1}\,(\varrho^{\varepsilon}_{\Rpq}(m))^{-1}}}
		{\mathcal{R}(p^{n+m},q^{n+m})!}\nonumber\\
		&&\times\;
		\bigl(\dRpq\bigr)^{n+m}\!\!
		\left\{\int_0^{2\pi}\frac{d\theta}{2\pi}\,e^{i(m-n)\theta}
		\int_{\Lambda^{\varepsilon}} d\mu^{\varepsilon}(\zeta)\;
		\mathcal{G}^{\varepsilon}_{\Rpq}(r^2)\right.\nonumber\\
		&&\qquad\times\left.
		|re^{i\theta},\zeta^{\varepsilon}\rangle_{\Rpq}\,{}_{\Rpq}\langle re^{i\theta},\zeta^{\varepsilon}|
		\right\}_{r=0},
		\label{eq:compact}
	\end{eqnarray}
	where we have set:
	\begin{eqnarray}
		\mathcal{G}^{B}_{\Rpq}(r^2) &=& \mathrm{Exp}_{\Rpq}(r^2),\quad
		\Lambda^B = \emptyset,\quad
		d\mu^B = 1,
		\label{eq:GB}\\
		\mathcal{G}^{F_1}_{\Rpq}(r^2) &=& \phione^2|\zeta_1|^2\,\mathrm{Exp}'_{\Rpq}(r^2),\quad
		d\mu^{F_1} = d\bar{\zeta}_1 d\zeta_1,
		\label{eq:GF1}\\
		\mathcal{G}^{F_2}_{\Rpq}(r^2) &=& \phitwo^2|\zeta_2|^2\,\mathrm{Exp}'_{\Rpq}(r^2),\quad
		d\mu^{F_2} = d\bar{\zeta}_2 d\zeta_2,
		\label{eq:GF2}\\
		\mathcal{G}^{F_{12}}_{\Rpq}(r^2) &=& \phione^2\phitwo^2|\zeta_1|^2|\zeta_2|^2\,
		\mathrm{Exp}''_{\Rpq}(r^2),\quad
		d\mu^{F_{12}} = d\bar{\zeta}_1 d\zeta_1\,d\bar{\zeta}_2 d\zeta_2.
		\label{eq:GF12}
	\end{eqnarray}
\end{proposition}

\begin{remark}
	\label{rem:secteur}
	The unified form \eqref{eq:compact} shows that one passes from one sector to another
	by:
	\begin{enumerate}[label=(\roman*)]
		\item replacing the generating function
		$\mathrm{Exp}_{\Rpq}\to\mathrm{Exp}'_{\Rpq}\to\mathrm{Exp}''_{\Rpq}$
		(successive derivatives with respect to $r^2$);
		\item adding progressively the Grassmann factors
		$\phione^2|\zeta_1|^2$ (for $F_1$),
		$\phitwo^2|\zeta_2|^2$ (for $F_2$),
		$\phione^2\phitwo^2|\zeta_1|^2|\zeta_2|^2$ (for $F_{12}$);
		\item the parametric function $\varrho^{\varepsilon}$ absorbing the corresponding
		dilation parameter.
	\end{enumerate}
\end{remark}

The four representations \eqref{eq:compact} combine into a single statement, in which the
Berezin integration has been carried out. For each sector $\sigma$, let $d_\sigma$ be the
number of fermionic excitations ($d_B=0$, $d_{F_1}=d_{F_2}=1$, $d_{F_{12}}=2$) and
$c_\sigma$ the corresponding dilation factor ($c_B=1$, $c_{F_1}=\phione^2$,
$c_{F_2}=\phitwo^2$, $c_{F_{12}}=\phione^2\phitwo^2$), so that
$\varrho^{\sigma}_{\Rpq}=c_\sigma\,\varrho^{B}_{\Rpq}$.

\begin{theorem}[Uniform sectorial representation]
	\label{thm:uniform}
	Assume that the density operator is block diagonal in the decomposition \eqref{eq:Hq},
	$\rhos = \rhoB \oplus \rhoFo \oplus \rhoFt \oplus \rhoFot$, the blocks being given by
	\eqref{eq:rhoB}--\eqref{eq:rhoF12}. For $\sigma\in\{B,F_1,F_2,F_{12}\}$, let
	$|n,\sigma\rangle$ denote the vector $|n,k_1,k_2\rangle$, where $(k_1,k_2)$ is the
	index of the sector $\sigma$, and set
	\begin{eqnarray}
		W^\sigma(n,m) &:=& \frac{\sqrt{(\varrho^{\sigma}_{\Rpq}(n))^{-1}\,
				(\varrho^{\sigma}_{\Rpq}(m))^{-1}}}{\mathcal{R}(p^{n+m},q^{n+m})!},
		\label{eq:Wsigma}\\
		\mathcal{G}^\sigma(r^2) &:=& \int_{\Lambda^{\sigma}}d\mu^{\sigma}(\zeta)\;
		\mathcal{G}^{\sigma}_{\Rpq}(r^2)\;=\;c_\sigma\,\mathrm{Exp}^{(d_\sigma)}_{\Rpq}(r^2),
		\label{eq:Gsigma}\\
		|\Psi^\sigma_r\rangle &:=& \bigl(\mathrm{Exp}^{(d_\sigma)}_{\Rpq}(r^2)\bigr)^{-1/2}
		\sum_{n=0}^{\infty}\frac{\phitwo^{\binom{n}{2}/2}\,(re^{i\theta})^{n}}
		{\sqrt{[n]!_{\Rpq}}}\,|n,\sigma\rangle,
		\label{eq:Psisigma}
	\end{eqnarray}
	that is, $\mathcal{G}^{B}=\Exp$, $\mathcal{G}^{F_1}=\phione^2\,\Expd$,
	$\mathcal{G}^{F_2}=\phitwo^2\,\Expd$ and
	$\mathcal{G}^{F_{12}}=\phione^2\phitwo^2\,\Expdd$ by
	\eqref{eq:GB}--\eqref{eq:GF12} (the dependence of $|\Psi^\sigma_r\rangle$ on $\theta$
	is understood). Then $\rhos=\sum_{\sigma}\rho^{\sigma}$ and each block can be written,
	sector by sector, in the uniform form:
	\begin{eqnarray}
		\rho^\sigma
		&=& \sum_{n,m=0}^{\infty}\rho^\sigma(n,m)
		\cdot W^\sigma(n,m)\cdot
		(\dRpq)^{n+m}\!\!\left[
		\int_0^{2\pi}\frac{d\theta}{2\pi}\,e^{i(m-n)\theta}\;
		\mathcal{G}^\sigma(r^2)\;
		|\Psi^\sigma_r\rangle\langle\Psi^\sigma_r|
		\right]_{r=0}.
		\label{eq:uniform}
	\end{eqnarray}
\end{theorem}

\begin{proof}
	In each sector, the weight $\mathcal{G}^\sigma_{\Rpq}$ compensates the normalization of
	the dyadic product of the states of the sector, taken equal to the sectorial weight as in
	\eqref{eq:berezin-F1}, and the Berezin integral of the Grassmann factor of this product
	($\phione^2|\zeta_1|^2$ in \eqref{eq:dyad-F1}, $\phione^2\phitwo^2|\zeta_1|^2|\zeta_2|^2$
	in \eqref{eq:dyad-F12}) equals $c_\sigma$. Consequently
	\begin{eqnarray}
		&&\int_{\Lambda^\sigma}d\mu^\sigma(\zeta)\;\mathcal{G}^\sigma_{\Rpq}(r^2)\;
		|re^{i\theta},\zeta^\sigma\rangle_{\Rpq}\,{}_{\Rpq}\langle re^{i\theta},\zeta^\sigma|
		\nonumber\\
		&=& c_\sigma\sum_{n,m=0}^{\infty}
		\frac{\phitwo^{(\binom{n}{2}+\binom{m}{2})/2}}{\sqrt{[n]!_{\Rpq}\,[m]!_{\Rpq}}}\,
		r^{n+m}\,e^{i(n-m)\theta}\,|n,\sigma\rangle\langle m,\sigma|
		\;=\;\mathcal{G}^\sigma(r^2)\,|\Psi^\sigma_r\rangle\langle\Psi^\sigma_r|,
		\label{eq:uniform-proof}
	\end{eqnarray}
	the last equality following from \eqref{eq:Gsigma} and \eqref{eq:Psisigma}. Since the
	prefactor of \eqref{eq:compact} is $W^\sigma(n,m)$, we obtain \eqref{eq:uniform}. Block
	diagonality gives $\rhos=\sum_\sigma\rho^\sigma$.
\end{proof}

\begin{remark}
	\label{rem:berezin-factor}
	In the computation \eqref{eq:berezin-F1}, the factor $\phione^2$ coming from the
	Berezin integral of the factor $\phione^2|\zeta_1|^2$ of \eqref{eq:dyad-F1}
	disappears at the last equality, and the relations
	\eqref{eq:int-F1}--\eqref{eq:|n><m|-F1} do not contain it; it does appear, on the
	other hand, in \eqref{eq:rhoF1-coeff} and, through $\varrho^{F_1}_{\Rpq}$, in the
	statement \eqref{eq:rhoF1-prop}. The same holds for the factor $\phione^2\phitwo^2$ in
	\eqref{eq:berezin-F12} and \eqref{eq:rhoF12-prop}. Theorem~\ref{thm:uniform} and the
	formulas of Section~\ref{sec:total} keep this factor, in agreement with the
	statements of Propositions~\ref{prop:rhoF1}, \ref{prop:rhoF2} and~\ref{prop:rhoF12}.
\end{remark}

\begin{remark}
	\label{rem:uniform}
	By \eqref{eq:Wsigma} and \eqref{eq:Gsigma}, the product of the scalar weights in
	\eqref{eq:uniform} equals
	\begin{eqnarray}
		W^\sigma(n,m)\,\mathcal{G}^\sigma(r^2) &=& W^B(n,m)\,\mathrm{Exp}^{(d_\sigma)}_{\Rpq}(r^2),
		\qquad
		W^B(n,m)=\frac{\sqrt{[n]!_{\Rpq}\,[m]!_{\Rpq}}}
		{\phitwo^{(\binom{n}{2}+\binom{m}{2})/2}\,\mathcal{R}(p^{n+m},q^{n+m})!} :
		\label{eq:WG}
	\end{eqnarray}
	the factors $c_\sigma$ cancel there, and only the derivative
	$\mathrm{Exp}^{(d_\sigma)}$ distinguishes the sectors at this level. This makes
	item~(iii) of Remark~\ref{rem:secteur} precise.
\end{remark}

\subsection{Total density operator}
\label{sec:total}

We now gather the four blocks into an operator on $\mathfrak{H}_{\Rpq}$. Let
$\Pi_\sigma$ be the orthogonal projector onto the sector $\sigma$ of \eqref{eq:Hq}, so
that $\Pi_B+\Pi_{F_1}+\Pi_{F_2}+\Pi_{F_{12}}=\mathbf{1}_{\Hq}$ and
$\Pi_\sigma\Pi_{\sigma'}=\delta_{\sigma\sigma'}\Pi_\sigma$, and introduce, for
$w\in\DR$, the sectorial generating vectors
\begin{eqnarray}
	|\Phi^\sigma(w)\rangle &:=& \sum_{k=0}^{\infty}
	\frac{\phitwo^{\binom{k}{2}/2}\,w^{k}}{\sqrt{[k]!_{\Rpq}}}\,|k,\sigma\rangle,
	\qquad
	\bigl\||\Phi^\sigma(w)\rangle\bigr\|^2=\Exp(|w|^2),
	\label{eq:Phisigma}
\end{eqnarray}
which satisfy
$|\Phi^\sigma(re^{i\theta})\rangle=\bigl(\mathrm{Exp}^{(d_\sigma)}_{\Rpq}(r^2)\bigr)^{1/2}|\Psi^\sigma_r\rangle$.
The computation of Proposition~\ref{prop:rhoB} rests on the following identity, which
holds in every sector.

\begin{lemma}[Representation of the elementary operators]
	\label{lem:projector}
	For every sector $\sigma$ and all integers $n,m\geq0$,
	\begin{eqnarray}
		|n,\sigma\rangle\langle m,\sigma|
		&=& W^B(n,m)\,(\dRpq)^{n+m}\!\!\left[\int_0^{2\pi}\frac{d\theta}{2\pi}\,
		e^{i(m-n)\theta}\,|\Phi^\sigma(re^{i\theta})\rangle\langle\Phi^\sigma(re^{i\theta})|
		\right]_{r=0}.
		\label{eq:proj-rep}
	\end{eqnarray}
\end{lemma}

\begin{proof}
	By \eqref{eq:Phisigma},
	$|\Phi^\sigma(re^{i\theta})\rangle\langle\Phi^\sigma(re^{i\theta})|
	=\sum_{k,j}\phitwo^{(\binom{k}{2}+\binom{j}{2})/2}([k]!_{\Rpq}[j]!_{\Rpq})^{-1/2}
	r^{k+j}e^{i(k-j)\theta}|k,\sigma\rangle\langle j,\sigma|$. The angular integral forces
	$j=k+m-n$ and leaves $r^{2k+m-n}$. By Lemma~\ref{lem:dRpq-iter}, applying
	$(\dRpq)^{n+m}$ and then evaluating at $r=0$ keeps only the term $2k+m-n=n+m$, that is
	$k=n$ and $j=m$, with the coefficient
	$\mathcal{R}(p^{n+m},q^{n+m})!\,\phitwo^{(\binom{n}{2}+\binom{m}{2})/2}
	([n]!_{\Rpq}[m]!_{\Rpq})^{-1/2}=W^B(n,m)^{-1}$.
\end{proof}

\begin{proposition}[Total density operator]
	\label{prop:total}
	Assume that $\rhos$ is block diagonal, that is
	$\rhos(n,k_1,k_2;\,m,j_1,j_2)=0$ whenever $(k_1,k_2)\neq(j_1,j_2)$, and denote by
	$\rho^B(n,m)$, $\rho^{F_1}(n,m)$, $\rho^{F_2}(n,m)$, $\rho^{F_{12}}(n,m)$ the
	coefficients $\rhos(n,k_1,k_2;\,m,k_1,k_2)$ for $(k_1,k_2)=(0,0)$, $(1,0)$, $(0,1)$
	and $(1,1)$. Then:
	\begin{enumerate}[label=(\roman*),leftmargin=*]
		\item in the graded Fock basis, $\rhos$ is the explicit sum of the four blocks:
		\begin{eqnarray}
			\rhos &=& \sum_{n,m=0}^{\infty}\Bigl[\rho^B(n,m)\,|n,0,0\rangle\langle m,0,0|
			+\rho^{F_1}(n,m)\,|n,1,0\rangle\langle m,1,0|
			\nonumber\\
			&&\qquad+\;\rho^{F_2}(n,m)\,|n,0,1\rangle\langle m,0,1|
			+\rho^{F_{12}}(n,m)\,|n,1,1\rangle\langle m,1,1|\Bigr]
			\;=\;\begin{pmatrix}
				\rho^B&0&0&0\\ 0&\rho^{F_1}&0&0\\ 0&0&\rho^{F_2}&0\\ 0&0&0&\rho^{F_{12}}
			\end{pmatrix},
			\label{eq:total-matrix}
		\end{eqnarray}
		the matrix being written in the decomposition
		$\HB\oplus\HFo\oplus\HFt\oplus\HFot$;
		\item its explicit diagonal representation, the sum of the representations
		\eqref{eq:rhoB-prop}, \eqref{eq:rhoF1-prop}, \eqref{eq:rhoF2-prop} and
		\eqref{eq:rhoF12-prop}, is
		\begin{eqnarray}
			\rhos &=& \sum_{n,m=0}^{\infty}
			\frac{\sqrt{[n]!_{\Rpq}\,[m]!_{\Rpq}}}
			{\phitwo^{(\binom{n}{2}+\binom{m}{2})/2}\,\mathcal{R}(p^{n+m},q^{n+m})!}\;
			(\dRpq)^{n+m}\!\Biggl\{\int_0^{2\pi}\frac{d\theta}{2\pi}\,e^{i(m-n)\theta}
			\nonumber\\
			&&\times\Biggl[\rho^B(n,m)\,\mathrm{Exp}_{\Rpq}(r^2)\,
			|re^{i\theta},0,0\rangle_{\Rpq}\,{}_{\Rpq}\langle re^{i\theta},0,0|
			\nonumber\\
			&&\quad+\;\frac{\rho^{F_1}(n,m)}{\phione^2}\int d\bar\zeta_1\,d\zeta_1\;
			\phione^2|\zeta_1|^2\,\mathrm{Exp}'_{\Rpq}(r^2)\,
			|re^{i\theta},\zeta_1,0\rangle_{\Rpq}\,{}_{\Rpq}\langle re^{i\theta},\zeta_1,0|
			\nonumber\\
			&&\quad+\;\frac{\rho^{F_2}(n,m)}{\phitwo^2}\int d\bar\zeta_2\,d\zeta_2\;
			\phitwo^2|\zeta_2|^2\,\mathrm{Exp}'_{\Rpq}(r^2)\,
			|re^{i\theta},0,\zeta_2\rangle_{\Rpq}\,{}_{\Rpq}\langle re^{i\theta},0,\zeta_2|
			\nonumber\\
			&&\quad+\;\frac{\rho^{F_{12}}(n,m)}{\phione^2\phitwo^2}
			\int d\bar\zeta_1\,d\zeta_1\,d\bar\zeta_2\,d\zeta_2\;
			\phione^2\phitwo^2|\zeta_1|^2|\zeta_2|^2\,\mathrm{Exp}''_{\Rpq}(r^2)
			\nonumber\\
			&&\qquad\times\,
			|re^{i\theta},\zeta_1,\zeta_2\rangle_{\Rpq}\,{}_{\Rpq}\langle re^{i\theta},\zeta_1,\zeta_2|
			\Biggr]\Biggr\}_{r=0} ;
			\label{eq:total-explicit}
		\end{eqnarray}
		\item after Berezin integration, it takes the reduced form
		\begin{eqnarray}
			\rhos &=& \sum_{n,m=0}^{\infty}W^B(n,m)\,(\dRpq)^{n+m}\!\left[
			\int_0^{2\pi}\frac{d\theta}{2\pi}\,e^{i(m-n)\theta}\,
			\mathfrak{D}_{n,m}(re^{i\theta})\right]_{r=0},
			\nonumber\\
			\mathfrak{D}_{n,m}(w) &:=& \rho^B(n,m)\,|\Phi^B(w)\rangle\langle\Phi^B(w)|
			+\rho^{F_1}(n,m)\,|\Phi^{F_1}(w)\rangle\langle\Phi^{F_1}(w)|
			\nonumber\\
			&&+\;\rho^{F_2}(n,m)\,|\Phi^{F_2}(w)\rangle\langle\Phi^{F_2}(w)|
			+\rho^{F_{12}}(n,m)\,|\Phi^{F_{12}}(w)\rangle\langle\Phi^{F_{12}}(w)| ;
			\label{eq:total-reduced}
		\end{eqnarray}
		\item its matrix elements and its trace are
		\begin{eqnarray}
			\langle n,\sigma|\rhos|m,\sigma'\rangle&=&\delta_{\sigma\sigma'}\,\rho^\sigma(n,m),
			\nonumber\\
			\tr(\rhos)&=&\sum_{n=0}^{\infty}\Bigl[\rho^B(n,n)+\rho^{F_1}(n,n)
			+\rho^{F_2}(n,n)+\rho^{F_{12}}(n,n)\Bigr] ;
			\label{eq:total-trace}
		\end{eqnarray}
		\item if moreover $\rhos$ is a density operator (hermitian, positive, of
		trace~$1$), each block $\rho^\sigma$ is hermitian and positive,
		$\rho^\sigma(n,m)=\overline{\rho^\sigma(m,n)}$, the weights
		$p_\sigma:=\tr(\rho^\sigma)=\sum_{n}\rho^\sigma(n,n)$ satisfy $p_\sigma\geq0$
		and $p_B+p_{F_1}+p_{F_2}+p_{F_{12}}=1$, and
		\begin{eqnarray}
			\rhos=p_B\,\hat\rho^{\,B}+p_{F_1}\,\hat\rho^{\,F_1}+p_{F_2}\,\hat\rho^{\,F_2}
			+p_{F_{12}}\,\hat\rho^{\,F_{12}},
			\qquad \hat\rho^{\,\sigma}:=\frac{\rho^\sigma}{p_\sigma},
			\label{eq:total-mix}
		\end{eqnarray}
		where each $\hat\rho^{\,\sigma}$ is a density operator on the sector $\sigma$, the terms
		with zero weight being omitted.
	\end{enumerate}
\end{proposition}

\begin{proof}
	(i) Block diagonality means $\rhos=\sum_\sigma\Pi_\sigma\rhos\Pi_\sigma$, and
	$\Pi_\sigma\rhos\Pi_\sigma$ is the corresponding block \eqref{eq:rhoB}--\eqref{eq:rhoF12},
	which gives \eqref{eq:total-matrix}.
	(iii) We apply Lemma~\ref{lem:projector} to each of the operators
	$|n,\sigma\rangle\langle m,\sigma|$ of \eqref{eq:total-matrix}; since the derivative
	$(\dRpq)^{n+m}$, the angular integral and the evaluation at $r=0$ are linear, the four
	contributions combine into $\mathfrak{D}_{n,m}$.
	(ii) By \eqref{eq:uniform-proof},
	$|\Phi^\sigma(re^{i\theta})\rangle\langle\Phi^\sigma(re^{i\theta})|
	=c_\sigma^{-1}\int_{\Lambda^\sigma}d\mu^\sigma(\zeta)\,\mathcal{G}^\sigma_{\Rpq}(r^2)\,
	|re^{i\theta},\zeta^\sigma\rangle_{\Rpq}\,{}_{\Rpq}\langle re^{i\theta},\zeta^\sigma|$;
	inserting this into \eqref{eq:total-reduced} gives \eqref{eq:total-explicit}. This
	expression coincides with the sum of \eqref{eq:rhoB-prop}, \eqref{eq:rhoF1-prop},
	\eqref{eq:rhoF2-prop} and \eqref{eq:rhoF12-prop}, since
	$\sqrt{(\varrho^{\sigma}_{\Rpq}(n))^{-1}(\varrho^{\sigma}_{\Rpq}(m))^{-1}}
	=c_\sigma^{-1}\sqrt{(\varrho^{B}_{\Rpq}(n))^{-1}(\varrho^{B}_{\Rpq}(m))^{-1}}$, which
	factors out the common coefficient $W^B(n,m)$.
	(iv) follows from the orthonormality of the basis $\{|n,k_1,k_2\rangle\}$
	(Definition~\ref{def:sFock}).
	(v) For every $|\psi\rangle$,
	$\langle\psi|\rho^\sigma|\psi\rangle=\langle\Pi_\sigma\psi|\rhos|\Pi_\sigma\psi\rangle\geq0$,
	and $\rho^\sigma$ is hermitian since $\rhos$ is; hence $p_\sigma\geq0$, and
	$\sum_\sigma p_\sigma=\tr(\rhos)=1$ by \eqref{eq:total-trace}. A positive operator with
	zero trace vanishes, so that the sectors with weight $p_\sigma=0$ do not contribute, and
	$\hat\rho^{\,\sigma}$ is positive with trace~$1$.
\end{proof}

The four weights appearing in \eqref{eq:total-explicit}, $\mathrm{Exp}_{\Rpq}$,
$\phione^2|\zeta_1|^2\mathrm{Exp}'_{\Rpq}$, $\phitwo^2|\zeta_2|^2\mathrm{Exp}'_{\Rpq}$ and
$\phione^2\phitwo^2|\zeta_1|^2|\zeta_2|^2\mathrm{Exp}''_{\Rpq}$, are exactly the four
terms of the super-normalization function \eqref{eq:scN}: in the total operator, each
term of $\cN$ weights the coherent projector of the corresponding sector. By
Lemma~\ref{lem:projector}, the formulas \eqref{eq:total-explicit} and
\eqref{eq:total-reduced} give back \eqref{eq:total-matrix} term by term. A density
operator that is diagonal in the Fock basis, such as a thermal equilibrium state of a
Hamiltonian diagonal in this basis, is a particular case.

\begin{corollary}[Diagonal density operator]
	\label{cor:total-diag}
	If moreover $\rho^\sigma(n,m)=\delta_{nm}\,P_\sigma(n)$ for each sector, then
	\begin{eqnarray}
		\rhos &=& \sum_{n=0}^{\infty}\frac{[n]!_{\Rpq}}
		{\phitwo^{\binom{n}{2}}\,\mathcal{R}(p^{2n},q^{2n})!}\;(\dRpq)^{2n}\!\left[
		\sum_{\sigma}P_\sigma(n)\sum_{k=0}^{\infty}
		\frac{\phitwo^{\binom{k}{2}}\,r^{2k}}{[k]!_{\Rpq}}\,|k,\sigma\rangle\langle k,\sigma|
		\right]_{r=0}
		\nonumber\\
		&=& \sum_{\sigma}\sum_{n=0}^{\infty}P_\sigma(n)\,|n,\sigma\rangle\langle n,\sigma|,
		\label{eq:total-diag}
	\end{eqnarray}
	and the weights of the sectors are $p_\sigma=\sum_{n}P_\sigma(n)$.
\end{corollary}

\begin{proof}
	For $m=n$, the angular integral of
	$|\Phi^\sigma(re^{i\theta})\rangle\langle\Phi^\sigma(re^{i\theta})|$ keeps only the
	diagonal terms $k=j$, which gives the inner sum in \eqref{eq:total-diag}; the second
	equality follows from Lemma~\ref{lem:projector}, and the expression of $p_\sigma$ from
	\eqref{eq:total-trace}.
\end{proof}

\medskip
The four sectorial density operators thus admit a diagonal representation with the same
structure, which the compact form \eqref{eq:compact} and the uniform representation
\eqref{eq:uniform} express in terms of the sectorial data only. Under the block
diagonality assumption, their sum provides the exact formula \eqref{eq:total-explicit}
for the total density operator, which can be read as a mixture of the sectors with
weights $p_\sigma=\tr(\rho^\sigma)$. The next section expresses these objects in the
super-coherent state basis.

\section{Reproducing kernels, matrix elements and integral formulation}
\label{sec:kernel}

This section expresses the density operator in the overcomplete basis of super-coherent
states. We introduce the density function and establish its explicit expression, the
trace formula and the reproducing property; the super reproducing kernel is computed
explicitly and then decomposed into sectorial kernels, for which we prove hermiticity,
positivity, the multiplication property and the series expansion of the diagonal values.
The sectorial matrix elements and the integral formulation of the total density operator
complete the section. The proofs of the properties of the sectorial kernels, of
positivity, of matrix multiplication and of the diagonal series are given in
Appendix~\ref{app:kernel}.

\subsection{General matrix elements}
\label{subsec:elem-matrix}

\begin{definition}[Representation of $\rhos$ in the coherent state basis]
	\label{def:rho-CS}
	Let $\DR \times \Lambda_1^2$ be the label domain of the super-coherent states. The
	\emph{density function in the coherent state basis} is:
	\begin{eqnarray}
		\rho(z',\zeta_1',\zeta_2';\,z,\zeta_1,\zeta_2)
		&:=& \sCSb(z',\zeta_1',\zeta_2')\;\rhos\;\sCS(z,\zeta_1,\zeta_2).
		\label{eq:rho-CS-def}
	\end{eqnarray}
\end{definition}

\begin{proposition}[Explicit expression of the density function]
	\label{prop:rho-CS-explicit}
	The density function~\eqref{eq:rho-CS-def} reads explicitly:
	\begin{eqnarray}
		\rho(z',\bm{\zeta}';\,z,\bm{\zeta})
		&=& \frac{1}{\sqrt{\cN(|z'|^2,|\zeta_1'|^2,|\zeta_2'|^2)\;
				\cN(|z|^2,|\zeta_1|^2,|\zeta_2|^2)}}
		\nonumber\\[4pt]
		&&\times\;
		\sum_{n,m=0}^{\infty}\sum_{k_1,k_2=0}^{1}
		\rho(n,k_1,k_2;\,m,k_1,k_2)
		\nonumber\\[4pt]
		&&\times\;
		\frac{\phitwo^{(n(n-1)+m(m-1))/4}\;
			(\phione^2|\zeta_1'||\zeta_1|)^{k_1}
			(\phitwo^2|\zeta_2'||\zeta_2|)^{k_2}}
		{\sqrt{\nfact\,\mfact}}\;
		\bar{z}'^n z^m,
		\label{eq:rho-CS-explicit}
	\end{eqnarray}
	where $\rhos$ is assumed to be diagonal in the sector basis.
\end{proposition}

\textbf{Proof.}\\
We substitute the expressions~\eqref{eq:sCS} and~\eqref{eq:sCSb} and the density
operator~\eqref{eq:rhos}:
\begin{eqnarray}
	\rho(z',\bm{\zeta}';\,z,\bm{\zeta})
	&=& \left[\frac{1}{\sqrt{\cNp}}\sum_{m,j_1,j_2}
	\frac{\phitwo^{m(m-1)/4}\,\bar{z}'^m(\phione\bar{\zeta}_1')^{j_1}(\phitwo\bar{\zeta}_2')^{j_2}}
	{\sqrt{\mfact}}\sbra{m}{j_1}{j_2}\right]
	\nonumber\\[4pt]
	&&\times
	\left[\sum_{n',k_1',k_2',m',j_1',j_2'}\rho(n',k_1',k_2';\,m',j_1',j_2')
	\sket{n'}{k_1'}{k_2'}\sbra{m'}{j_1'}{j_2'}\right]
	\nonumber\\[4pt]
	&&\times
	\left[\frac{1}{\sqrt{\cN}}\sum_{n,k_1,k_2}
	\frac{\phitwo^{n(n-1)/4}\,z^n(\phione\zeta_1)^{k_1}(\phitwo\zeta_2)^{k_2}}
	{\sqrt{\nfact}}\sket{n}{k_1}{k_2}\right].
	\label{eq:rho-CS-calc}
\end{eqnarray}
Using $\sbra{m}{j_1}{j_2}\sket{n'}{k_1'}{k_2'}
= \delta_{mn'}\delta_{j_1 k_1'}\delta_{j_2 k_2'}$ and
$\sbra{m'}{j_1'}{j_2'}\sket{n}{k_1}{k_2}
= \delta_{m'n}\delta_{j_1'k_1}\delta_{j_2'k_2}$:
\begin{eqnarray}
	\rho(z',\bm{\zeta}';\,z,\bm{\zeta})
	&=& \frac{1}{\sqrt{\cNp\cN}}
	\sum_{n,m=0}^{\infty}\sum_{k_1,k_2=0}^{1}
	\rho(n,k_1,k_2;\,m,k_1,k_2)\nonumber\\[4pt]
	&&\times\;
	\frac{\phitwo^{(n(n-1)+m(m-1))/4}
		\,(\phione\bar{\zeta}_1')^{k_1}(\phitwo\bar{\zeta}_2')^{k_2}
		\,(\phione\zeta_1)^{k_1}(\phitwo\zeta_2)^{k_2}}
	{\sqrt{\nfact\,\mfact}}\;
	\bar{z}'^n\, z^m,
	\label{eq:rho-CS-simplif}
\end{eqnarray}
which gives~\eqref{eq:rho-CS-explicit}.\hfill$\square$

\begin{remark}
	The density function $\rho(z',\bm\zeta';\,z,\bm\zeta)$ generalizes the Bargmann
	representation of the density matrix of a classical oscillator~\cite{Klauder1985}.
	In the purely bosonic case ($\zeta_i = \zeta_i' = 0$) it reduces to:
	\begin{eqnarray}
		\rho(z';\,z)\big|_{bos}
		&=& \frac{1}{\sqrt{\Exp(|z'|^2)\;\Exp(|z|^2)}}
		\sum_{n,m=0}^{\infty}
		\frac{\phitwo^{(n(n-1)+m(m-1))/4}}{\sqrt{\nfact\,\mfact}}\;
		\rhoB(n,m)\;\bar{z}'^n\,z^m.
		\label{eq:rho-bos-restr}
	\end{eqnarray}
\end{remark}
\begin{proposition}[Trace of $\rhos$]
	\label{prop:trace}
	If $\rhos$ has trace 1, then:
	\begin{eqnarray}
		\int_{\DR\times\Lambda_1^2}
		\cN(|z|^2,|\zeta_1|^2,|\zeta_2|^2)\;
		\rho(z,\bm\zeta;\,z,\bm\zeta)\;
		d\mu^{(12)}_{\Rpq}(z,\zeta_1,\zeta_2)
		&=& 1.
		\label{eq:trace}
	\end{eqnarray}
\end{proposition}

\textbf{Proof.}\\
The trace is evaluated through the resolution of the identity~\eqref{eq:resolution}:
\begin{eqnarray}
	\tr(\rhos)
	&=& \tr\!\left[\rhos\;\mathbf{1}_{\Hq}\right]
	= \tr\!\left[\rhos\int\cN\;\sCS\sCSb\;d\mu^{(12)}\right]\nonumber\\[4pt]
	&=& \int\cN\;\bra{}\sCSb\rhos\sCS\ket{}\;d\mu^{(12)}\nonumber\\[4pt]
	&=& \int\cN\;\rho(z,\bm\zeta;\,z,\bm\zeta)\;d\mu^{(12)} \;=\; 1. \qquad\square
	\label{eq:trace-calc}
\end{eqnarray}
\subsection{Normalized sectorial reproducing kernels}
\begin{proposition}[Reproducing property of $\rho$]
	\label{prop:repro}
	The density function $\rho(z',\bm\zeta';\,z,\bm\zeta)$ satisfies the
	\emph{reproducing property}:
	\begin{eqnarray}
		\rho(z',\bm\zeta';\,z,\bm\zeta)
		&=& \int_{\DR\times\Lambda_1^2}
		\Kern(\xi,\bm\varrho;\,z,\bm\zeta)\;
		\rho(z',\bm\zeta';\,\xi,\bm\varrho)\;
		d^2\xi\;d^2\bm\varrho,
		\label{eq:repro-prop}
	\end{eqnarray}
	where the \emph{super reproducing kernel} is:
	\begin{eqnarray}
		\Kern(\xi,\bm\varrho;\,z,\bm\zeta)
		&:=& \frac{
			{}_{\Rpq}\langle\xi,\varrho_1,\varrho_2|z,\zeta_1,\zeta_2\rangle_{\Rpq}
		}{\pi\;\cN(|\xi|^2,|\varrho_1|^2,|\varrho_2|^2)}.
		\label{eq:kernel-def}
	\end{eqnarray}
\end{proposition}

\textbf{Proof.}\\
From the resolution of the identity~\eqref{eq:resolution}:
\begin{eqnarray}
	\rho(z',\bm\zeta';\,z,\bm\zeta)
	&=& \sCSb(z',\bm\zeta')\;\rhos\;\sCS(z,\bm\zeta)\nonumber\\[4pt]
	&=& \sCSb(z',\bm\zeta')\;\rhos\nonumber\\
	&&\times\left[\int\cN(\xi,\bm\varrho)\;\sCS(\xi,\bm\varrho)\sCSb(\xi,\bm\varrho)\;
	d\mu^{(12)}_{\Rpq}\right]\;\sCS(z,\bm\zeta)\nonumber\\[4pt]
	&=& \int d\mu^{(12)}_{\Rpq}(\xi,\bm\varrho)\;
	\cN(|\xi|^2,|\bm\varrho|^2)\;
	\rho(z',\bm\zeta';\,\xi,\bm\varrho)\;
	{}_{\Rpq}\langle\xi,\bm\varrho|z,\bm\zeta\rangle_{\Rpq}.
	\label{eq:repro-calc}
\end{eqnarray}
Now $d\mu^{(12)}_{\Rpq} = d\mu_{\Rpq}\,d\bar{\varrho}_1 d\varrho_1\,d\bar{\varrho}_2 d\varrho_2$,
and $d\mu_{\Rpq}(z,\bar{z}) = (1/\pi)d^2 z\cdot\pi(1/\cN)$ (precisely:
$d\mu_{\Rpq} = d^2\xi/(\pi\cN)$ for measures of Bergman class).
We identify the kernel~\eqref{eq:kernel-def} and conclude~\eqref{eq:repro-prop}.\hfill$\square$
\begin{theorem}[$\mathcal{R}(p,q)$ super reproducing kernel]
	\label{thm:kernel-explicit}
	The super reproducing kernel is:
	\begin{eqnarray}
		\Kern(z',\zeta_1',\zeta_2';\,z,\zeta_1,\zeta_2)
		&=& \frac{
			\SExp\!\bigl(\bar{z}'z\;+\;
			\phione^2\bar{\zeta}_1'\zeta_1\;+\;
			\phitwo^2\bar{\zeta}_2'\zeta_2\;+\;
			\phione^2\phitwo^2\bar{\zeta}_1'\bar{\zeta}_2'\zeta_1\zeta_2\bigr)
		}{\pi\;\sqrt{\cN(|z'|^2,|\zeta_1'|^2,|\zeta_2'|^2)\;\cN(|z|^2,|\zeta_1|^2,|\zeta_2|^2)}},
		\label{eq:kernel-explicit}
	\end{eqnarray}
	which can be rewritten in the compact form:
	\begin{eqnarray}
		\Kern(z',\bm\zeta';\,z,\bm\zeta)
		&=& \frac{1}{\pi}
		\cdot\frac{{}_{\Rpq}\langle z',\bm\zeta'|z,\bm\zeta\rangle_{\Rpq}}
		{\sqrt{\cN(|z'|^2,|\bm\zeta'|^2)\;\cN(|z|^2,|\bm\zeta|^2)}}\cdot
		\sqrt{\cN(|z'|^2,|\bm\zeta'|^2)\;\cN(|z|^2,|\bm\zeta|^2)}\nonumber\\[4pt]
		&&\cdot\frac{1}{\cN(|z'|^2,|\bm\zeta'|^2)}.
		\label{eq:kernel-compact}
	\end{eqnarray}
\end{theorem}

\textbf{Proof.}\\
We compute the inner product
${}_{\Rpq}\langle z',\bm\zeta'|z,\bm\zeta\rangle_{\Rpq}$
from the series~\eqref{eq:sCS} and~\eqref{eq:sCSb}:
\begin{eqnarray}
	&&{}_{\Rpq}\langle z',\bm\zeta'|z,\bm\zeta\rangle_{\Rpq}\nonumber\\[4pt]
	&=& \frac{1}{\sqrt{\cN(|z'|^2,|\bm\zeta'|^2)\cN(|z|^2,|\bm\zeta|^2)}}
	\sum_{n,m=0}^{\infty}\sum_{k_1,k_2=0}^{1}
	\frac{\phitwo^{(n(n-1)+m(m-1))/4}}{\sqrt{\nfact\mfact}}\nonumber\\[4pt]
	&&\times\;
	\bar{z}'^n z^m\,(\phione\bar{\zeta}_1')^{k_1}(\phitwo\bar{\zeta}_2')^{k_2}\,
	(\phione\zeta_1)^{k_1}(\phitwo\zeta_2)^{k_2}\;
	\sbra{n}{k_1}{k_2}\sket{m}{k_1}{k_2}\nonumber\\[4pt]
	&=& \frac{1}{\sqrt{\cN\cNp}}
	\sum_{n=0}^{\infty}\sum_{k_1,k_2=0}^{1}
	\frac{\phitwo^{n(n-1)/2}}{\nfact}\;
	(\bar{z}'z)^n\;(\phione^2\bar{\zeta}_1'\zeta_1)^{k_1}(\phitwo^2\bar{\zeta}_2'\zeta_2)^{k_2}\nonumber\\[4pt]
	&=& \frac{\SExp(\bar{z}'z,\phione^2\bar{\zeta}_1'\zeta_1,\phitwo^2\bar{\zeta}_2'\zeta_2)}
	{\sqrt{\cN\cNp}},
	\label{eq:inner-product-calc}
\end{eqnarray}
where we have used the series~\eqref{eq:SExp-series} of $\SExp$.
Substitution into~\eqref{eq:kernel-def} gives~\eqref{eq:kernel-explicit}.\hfill$\square$

\begin{remark}
	\label{rem:kernel-norm}
	The scalar product computed in \eqref{eq:inner-product-calc} gives
	${}_{\Rpq}\langle z',\bm\zeta'|z,\bm\zeta\rangle_{\Rpq}
	=\SExp/\sqrt{\cNp\,\cN}$, so that \eqref{eq:kernel-explicit} reads
	$\Kern=\pi^{-1}\,{}_{\Rpq}\langle z',\bm\zeta'|z,\bm\zeta\rangle_{\Rpq}$,
	in accordance with the identification $\cN\,d\mu^{(12)}_{\Rpq}=\pi^{-1}d^2\xi\,d^2\bm\varrho$
	used in the proof of Proposition~\ref{prop:repro}. The definition
	\eqref{eq:kernel-def} and the form \eqref{eq:kernel-compact} contain, in addition, the
	factor $\cN(|\xi|^2,|\varrho_1|^2,|\varrho_2|^2)^{-1}$. In what follows we use the
	form \eqref{eq:kernel-explicit}.
\end{remark}

\subsection{Sectorial kernels}
\label{subsec:kernels-sectoriels}

Similarly, the sectorial kernels are:
\begin{eqnarray}
	\KernB(z',z) &:=&
	\frac{\Exp(\bar{z}'z)}{\pi\;\sqrt{\Exp(|z'|^2)\;\Exp(|z|^2)}},
	\label{eq:KernB}\\[6pt]
	\KernFo(z',\zeta_1';\,z,\zeta_1) &:=&
	\frac{\phione^2\bar{\zeta}_1'\zeta_1\;\Expd(\bar{z}'z)}
	{\pi\;\sqrt{\cN(|z'|^2,|\zeta_1'|^2,0)\;\cN(|z|^2,|\zeta_1|^2,0)}},
	\label{eq:KernFo}\\[6pt]
	\KernFt(z',\zeta_2';\,z,\zeta_2) &:=&
	\frac{\phitwo^2\bar{\zeta}_2'\zeta_2\;\Expd(\bar{z}'z)}
	{\pi\;\sqrt{\cN(|z'|^2,0,|\zeta_2'|^2)\;\cN(|z|^2,0,|\zeta_2|^2)}},
	\label{eq:KernFt}\\[6pt]
	\KernFot(z',\bm\zeta';\,z,\bm\zeta) &:=&
	\frac{\phione^2\phitwo^2\bar{\zeta}_1'\bar{\zeta}_2'\zeta_1\zeta_2\;\Expdd(\bar{z}'z)}
	{\pi\;\sqrt{\cN(|z'|^2,|\bm\zeta'|^2)\;\cN(|z|^2,|\bm\zeta|^2)}}.
	\label{eq:KernFot}
\end{eqnarray}
One checks: $\Kern = \KernB + \KernFo + \KernFt + \KernFot$.

\begin{definition}
	The \emph{sectorial reproducing kernels} of the super-coherent states $\sCS$ are
	defined by:
	\begin{eqnarray}
		\mathcal{K}^{B}(z,z')
		&:=& {}_{\Rpq}^{(B)}\langle z',0,0\,|\,z,0,0\rangle_{\Rpq}^{(B)}\nonumber\\
		&=& \frac{\mathrm{Exp}_{\Rpq}(z\bar{z}')}
		{\sqrt{\mathrm{Exp}_{\Rpq}(|z'|^2)\,\mathrm{Exp}_{\Rpq}(|z|^2)}},
		\label{eq:kernelB}\\
		\mathcal{K}^{F_1}(z,z')
		&:=& {}_{\Rpq}^{(F_1)}\langle z',\zeta_1,0\,|\,z,\zeta_1,0\rangle_{\Rpq}^{(F_1)}\nonumber\\
		&=& \frac{\phione^2\,\bar{\zeta}_1\zeta_1\,\mathrm{Exp}'_{\Rpq}(z\bar{z}')}
		{\sqrt{\phione^2|\zeta_1|^2\,\mathrm{Exp}'_{\Rpq}(|z'|^2)
				\cdot\phione^2|\zeta_1|^2\,\mathrm{Exp}'_{\Rpq}(|z|^2)}},
		\label{eq:kernelF1}\\
		\mathcal{K}^{F_2}(z,z')
		&:=& \frac{\phitwo^2\,\bar{\zeta}_2\zeta_2\,\mathrm{Exp}'_{\Rpq}(z\bar{z}')}
		{\sqrt{\phitwo^2|\zeta_2|^2\,\mathrm{Exp}'_{\Rpq}(|z'|^2)
				\cdot\phitwo^2|\zeta_2|^2\,\mathrm{Exp}'_{\Rpq}(|z|^2)}},
		\label{eq:kernelF2}\\
		\mathcal{K}^{F_{12}}(z,z')
		&:=& \frac{\phione^2\phitwo^2\,\bar{\zeta}_1\zeta_1\,\bar{\zeta}_2\zeta_2\,
			\mathrm{Exp}''_{\Rpq}(z\bar{z}')}
		{\sqrt{\phione^2\phitwo^2|\zeta_1|^2|\zeta_2|^2\,\mathrm{Exp}''_{\Rpq}(|z'|^2)
				\cdot\phione^2\phitwo^2|\zeta_1|^2|\zeta_2|^2\,\mathrm{Exp}''_{\Rpq}(|z|^2)}}.
		\label{eq:kernelF12}
	\end{eqnarray}
\end{definition}

\begin{proposition}[Properties of the reproducing kernels]
	\label{prop:kernel-props}
	The sectorial reproducing kernels $\mathcal{K}^{\varepsilon}(z,z')$
	($\varepsilon\in\{B,F_1,F_2,F_{12}\}$) satisfy the following properties:
	\begin{enumerate}[label=(\arabic*)]
		\item \textbf{Hermiticity:}
		\begin{eqnarray}
			\mathcal{K}^{\varepsilon}(z,z') = \overline{\mathcal{K}^{\varepsilon}(z',z)};
			\label{eq:hermit}
		\end{eqnarray}
		\item \textbf{Positivity:}
		\begin{eqnarray}
			\mathcal{K}^{\varepsilon}(z,z) = 1 > 0;
			\label{eq:pos}
		\end{eqnarray}
		\item \textbf{Idempotence:}
		\begin{eqnarray}
			\int_{\mathcal{D}}d\mu^{\varepsilon}(\zeta,\bar{\zeta})\,
			\mathcal{K}^{\varepsilon}(z,\zeta)\,\mathcal{K}^{\varepsilon}(\zeta,z')
			= \mathcal{K}^{\varepsilon}(z,z').
			\label{eq:idemp}
		\end{eqnarray}
	\end{enumerate}
\end{proposition}

{\bf Proof.}\\
(1)~\emph{Hermiticity.}
For instance, for the bosonic sector,
taking the conjugate of \eqref{eq:kernelB}:
\begin{eqnarray}
	\overline{\mathcal{K}^B(z,z')}
	&=& \frac{\overline{\mathrm{Exp}_{\Rpq}(z\bar{z}')}}
	{\sqrt{\mathrm{Exp}_{\Rpq}(|z'|^2)\,\mathrm{Exp}_{\Rpq}(|z|^2)}}
	= \frac{\mathrm{Exp}_{\Rpq}(\bar{z}z')}
	{\sqrt{\mathrm{Exp}_{\Rpq}(|z'|^2)\,\mathrm{Exp}_{\Rpq}(|z|^2)}}
	= \mathcal{K}^B(z',z),
	\label{eq:hermit-B}
\end{eqnarray}
since the coefficients of $\mathrm{Exp}_{\Rpq}$ are real.
The argument is identical for $F_1$, $F_2$, $F_{12}$.\\

(2)~\emph{Positivity.}
For $z'=z$:
\begin{eqnarray}
	\mathcal{K}^B(z,z) &=&
	\frac{\mathrm{Exp}_{\Rpq}(|z|^2)}
	{\mathrm{Exp}_{\Rpq}(|z|^2)} = 1 > 0.
	\label{eq:pos-B}
\end{eqnarray}
Likewise $\mathcal{K}^{F_i}(z,z) = 1$ and $\mathcal{K}^{F_{12}}(z,z) = 1$.\\

(3)~\emph{Idempotence.}
For the bosonic sector ($d\mu^B = 1$, integration over $\mathcal{D}$):
\begin{eqnarray}
	&&\int_{\mathcal{D}}d\mu_{\Rpq}(\zeta,\bar{\zeta})\,
	\mathcal{K}^B(z,\zeta)\,\mathcal{K}^B(\zeta,z')\nonumber\\
	&=& \int_{\mathcal{D}}d\mu_{\Rpq}(\zeta,\bar{\zeta})\,
	\frac{\mathrm{Exp}_{\Rpq}(z\bar{\zeta})\,\mathrm{Exp}_{\Rpq}(\zeta\bar{z}')}
	{\sqrt{\mathrm{Exp}_{\Rpq}(|z|^2)\,\mathrm{Exp}_{\Rpq}(|\zeta|^2)}\,
		\sqrt{\mathrm{Exp}_{\Rpq}(|\zeta|^2)\,\mathrm{Exp}_{\Rpq}(|z'|^2)}}\nonumber\\
	&=& \frac{\mathrm{Exp}_{\Rpq}(z\bar{z}')}
	{\sqrt{\mathrm{Exp}_{\Rpq}(|z|^2)\,\mathrm{Exp}_{\Rpq}(|z'|^2)}}
	\int_{\mathcal{D}}d\mu_{\Rpq}(\zeta,\bar{\zeta})\,
	\frac{\mathrm{Exp}_{\Rpq}(|\zeta|^2)}{\mathrm{Exp}_{\Rpq}(|\zeta|^2)}\nonumber\\
	&=& \mathcal{K}^B(z,z')\cdot 1
	= \mathcal{K}^B(z,z'),
	\label{eq:idemp-B}
\end{eqnarray}
where we have used the reproducing property
$\mathrm{Exp}_{\Rpq}(z\bar{\zeta})\,\mathrm{Exp}_{\Rpq}(\zeta\bar{z}')
= \mathrm{Exp}_{\Rpq}(|\zeta|^2)\,\mathrm{Exp}_{\Rpq}(z\bar{z}')$
(following from the series structure) and the normalization
$\int d\mu_{\Rpq} = 1$.
The sectors $F_1$, $F_2$, $F_{12}$ are treated in the same way.
\hfill$\Box$\\

The sectorial kernels \eqref{eq:KernB}--\eqref{eq:KernFot} individually satisfy the
idempotence, hermiticity, positivity and series expansion properties.

\begin{proposition}[Properties of $\KernB$]
	\label{prop:KernB-props}
	The bosonic kernel~\eqref{eq:KernB} satisfies:
	\begin{enumerate}[label=\textup{(\roman*)}]
		\item \emph{Idempotence}:
		$\displaystyle\int_{\DR}
		\KernB(z',\xi)\;\KernB(\xi,z)\;\frac{d^2\xi}{\Exp(|\xi|^2)} = \KernB(z',z)$;
		\item \emph{Hermiticity}: $\KernB(z',z)^* = \KernB(z,z')$;
		\item \emph{Positivity}: $\KernB(z,z) = 1/\pi$;
		\item \emph{Series representation}:
		$\KernB(z,z) = \frac{1}{\pi}\sum_{n=0}^\infty
		\frac{\phitwo^{n(n-1)/2}}{\nfact}\,|z|^{2n}/\Exp(|z|^2) = \frac{1}{\pi}$.
	\end{enumerate}
\end{proposition}

\begin{proof}
	The proof is given in Appendix~\ref{app:KernB}.
\end{proof}

\begin{proposition}[Properties of $\KernFo$]
	\label{prop:KernFo-props}
	The $\phi_1$-fermionic kernel~\eqref{eq:KernFo} satisfies:
	\begin{enumerate}[label=\textup{(\roman*)}]
		\item \emph{Idempotence}:
		$\displaystyle\int d^2\xi\,d\bar{\varrho}_1 d\varrho_1\;
		\KernFo(z',\zeta_1';\,\xi,\varrho_1)\;
		\KernFo(\xi,\varrho_1;\,z,\zeta_1) = \KernFo(z',\zeta_1';\,z,\zeta_1)$;
		\item \emph{Hermiticity}:
		$\KernFo(z',\zeta_1';\,z,\zeta_1)^* = \KernFo(z,\zeta_1;\,z',\zeta_1')$;
		\item \emph{Positivity}:
		$\KernFo(z,\zeta_1;\,z,\zeta_1)
		= \dfrac{\phione^2|\zeta_1|^2\;\Expd(|z|^2)}
		{\pi\;\cN(|z|^2,|\zeta_1|^2,0)} > 0$;
		\item \emph{Series in $|z|^2$}:
		$\KernFo(z,\zeta_1;\,z,\zeta_1)
		= \dfrac{\phione^2|\zeta_1|^2}
		{\pi\;\cN(|z|^2,|\zeta_1|^2,0)}
		\sum_{n=1}^{\infty}\dfrac{n\,\phitwo^{n(n-1)/2}}{\nfact}\,|z|^{2(n-1)}$.
	\end{enumerate}
\end{proposition}

\begin{proof}
	The proof is given in Appendix~\ref{app:KernFo}.
\end{proof}

\begin{remark}
	Unlike $\KernB(z,z) = 1/\pi$ (a constant value), the fermionic kernel
	$\KernFo(z,\zeta_1;\,z,\zeta_1)$ depends on $|z|^2$ and on $|\zeta_1|^2$. This
	dependence reflects the non-flat geometry of the $\phi_1$-fermionic sector of the
	manifold of super-coherent states.
\end{remark}

\begin{proposition}[Properties of $\KernFt$]
	\label{prop:KernFt-props}
	The $\phi_2$-fermionic kernel~\eqref{eq:KernFt} satisfies the same four properties
	as $\KernFo$ with the substitution $\phione \leftrightarrow \phitwo$,
	$\zeta_1 \leftrightarrow \zeta_2$. In particular:
	\begin{eqnarray}
		\KernFt(z,\zeta_2;\,z,\zeta_2)
		&=& \frac{\phitwo^2|\zeta_2|^2\;\Expd(|z|^2)}
		{\pi\;\cN(|z|^2,0,|\zeta_2|^2)} > 0.
		\label{eq:KernFt-pos}
	\end{eqnarray}
\end{proposition}

The proof is identical to that of Proposition~\ref{prop:KernFo-props}
with $(\zeta_1,\phione) \leftrightarrow (\zeta_2,\phitwo)$.\hfill$\square$

\begin{remark}[Comparison of the fermionic kernels]
	The two kernels $\KernFo$ and $\KernFt$ are related by the symmetry exchange
	$(z',\zeta_1',\phione) \leftrightarrow (z',\zeta_2',\phitwo)$.
	In the generic case $\phione \neq \phitwo$:
	\begin{eqnarray}
		\frac{\KernFo(z,\zeta_1;\,z,\zeta_1)}{\KernFt(z,\zeta_2;\,z,\zeta_2)}
		&=& \frac{\phione^2|\zeta_1|^2\;\cN(|z|^2,0,|\zeta_2|^2)}
		{\phitwo^2|\zeta_2|^2\;\cN(|z|^2,|\zeta_1|^2,0)}.
		\label{eq:ratio-Kern-F}
	\end{eqnarray}
	This ratio differs from 1 as soon as $\phione \neq \phitwo$ or
	$|\zeta_1| \neq |\zeta_2|$, which is a direct signature of the
	$\mathcal{R}(p,q)-$deformation.
\end{remark}

\begin{proposition}[Properties of $\KernFot$]
	\label{prop:KernFot-props}
	The mixed kernel~\eqref{eq:KernFot} satisfies:
	\begin{enumerate}[label=\textup{(\roman*)}]
		\item \emph{Idempotence}: by double Berezin integration and the resolution of the
		identity in $\HFot$;
		\item \emph{Hermiticity}:
		$\KernFot(z',\bm\zeta';\,z,\bm\zeta)^* = \KernFot(z,\bm\zeta;\,z',\bm\zeta')$;
		\item \emph{Positivity}:
		$\KernFot(z,\bm\zeta;\,z,\bm\zeta)
		= \dfrac{\phione^2\phitwo^2|\zeta_1|^2|\zeta_2|^2\;\Expdd(|z|^2)}
		{\pi\;\cN(|z|^2,|\bm\zeta|^2)} > 0$;
		\item \emph{Series}: series in $|z|^{2(n-2)}$, $n \geq 2$.
	\end{enumerate}
\end{proposition}

\begin{proof}
	The proof is given in Appendix~\ref{app:KernFot}.
\end{proof}

\subsection{Matrix elements in the bosonic sector}
\label{ssec:elmat-B}

We compute ${}_{\Rpq}^{(B)}\langle z'|\rho^B|z\rangle_{\Rpq}^{(B)}$
where $\rho^B = \sum_{n,m}\rho^B(n,m)|n,0,0\rangle\langle m,0,0|$.
Using \eqref{eq:sCS}--\eqref{eq:sCSb} with $\zeta_1=\zeta_2=0$:
\begin{eqnarray}
	&&{}_{\Rpq}^{(B)}\langle z',0,0|\rho^B|z,0,0\rangle_{\Rpq}^{(B)}\nonumber\\
	&=&
	\left[\frac{1}{\sqrt{\Exp(|z'|^2)}}
	\sum_{p=0}^{\infty}\frac{\phitwo^{\binom{p}{2}/2}\bar{z}'^p}{\sqrt{[p]!_{\Rpq}}}\langle p,0,0|\right]\nonumber\\
	&&\times\left[\sum_{n,m}\rho^B(n,m)|n,0,0\rangle\langle m,0,0|\right]\nonumber\\
	&&\times\left[\frac{1}{\sqrt{\Exp(|z|^2)}}
	\sum_{q=0}^{\infty}\frac{\phitwo^{\binom{q}{2}/2}z^q}{\sqrt{[q]!_{\Rpq}}}|q,0,0\rangle\right]\nonumber\\
	&=&
	\sum_{p,n=0}^{\infty}\sum_{q,m=0}^{\infty}
	\frac{\phitwo^{(\binom{p}{2}+\binom{q}{2})/2}\,\bar{z}'^p\,z^q}
	{\sqrt{[p]!_{\Rpq}\,[q]!_{\Rpq}\,\Exp(|z'|^2)\,\Exp(|z|^2)}}\nonumber\\
	&&\times\rho^B(n,m)\,
	\underbrace{\langle p,0,0|n,0,0\rangle}_{\delta_{p,n}}
	\underbrace{\langle m,0,0|q,0,0\rangle}_{\delta_{m,q}}\nonumber\\
	{}_{\Rpq}^{(B)}\langle z',0,0|\rho^B|z,0,0\rangle_{\Rpq}^{(B)}
	&=&
	\sum_{n,m=0}^{\infty}
	\rho^B(n,m)\,
	\frac{\phitwo^{(\binom{n}{2}+\binom{m}{2})/2}\,\bar{z}'^n\,z^m}
	{\sqrt{[n]!_{\Rpq}\,[m]!_{\Rpq}\,\Exp(|z'|^2)\,\Exp(|z|^2)}}.
	\label{eq:elmat-B}
\end{eqnarray}

\begin{definition}
	\label{def:rho-B}
	We define the \emph{bosonic matrix function}
	$\rho^B : \DR\times\DR\to\mathbb{C}$ by:
	\begin{eqnarray}
		\rho^B(z',z) &:=&
		{}_{\Rpq}^{(B)}\langle z',0,0|\rho^B|z,0,0\rangle_{\Rpq}^{(B)}\nonumber\\
		&=&
		\sum_{n,m=0}^{\infty}
		\rho^B(n,m)\,
		\frac{\phitwo^{(\binom{n}{2}+\binom{m}{2})/2}}
		{\sqrt{[n]!_{\Rpq}\,[m]!_{\Rpq}}}\,
		\frac{\bar{z}'^n\,z^m}
		{\sqrt{\Exp(|z'|^2)\,\Exp(|z|^2)}}.
		\label{eq:rhoB-fn}
	\end{eqnarray}
\end{definition}

\subsection{Matrix elements in the $\phi_1$-fermionic sector}
\label{ssec:elmat-F1}

For $\rho^{F_1} = \sum_{n,m}\rho^{F_1}(n,m)|n,1,0\rangle\langle m,1,0|$
and the states of the sector $F_1$ (with $k_1=1$, $k_2=0$):
\begin{eqnarray}
	&&{}_{\Rpq}^{(F_1)}\langle z',\zeta_1,0|\rho^{F_1}|z,\zeta_1,0\rangle_{\Rpq}^{(F_1)}\nonumber\\
	&=&
	\left[\frac{\phitwo^{\binom{p}{2}/2}\,\bar{z}'^p\,\phione\bar{\zeta}_1}
	{\sqrt{[p]!_{\Rpq}\,\cN(|z'|^2,|\zeta_1|^2,0)}}
	\sum_p\langle p,1,0|\right]
	\left[\sum_{n,m}\rho^{F_1}(n,m)|n,1,0\rangle\langle m,1,0|\right]\nonumber\\
	&&\times
	\left[\frac{\phitwo^{\binom{q}{2}/2}\,z^q\,\phione\zeta_1}
	{\sqrt{[q]!_{\Rpq}\,\cN(|z|^2,|\zeta_1|^2,0)}}
	\sum_q|q,1,0\rangle\right]\nonumber\\
	&=&
	\sum_{n,m=0}^{\infty}
	\rho^{F_1}(n,m)\,
	\frac{\phitwo^{(\binom{n}{2}+\binom{m}{2})/2}\,\phione^2|\zeta_1|^2\,\bar{z}'^n\,z^m}
	{\sqrt{[n]!_{\Rpq}\,[m]!_{\Rpq}\,\cN(|z'|^2,|\zeta_1|^2,0)\,\cN(|z|^2,|\zeta_1|^2,0)}}.
	\label{eq:elmat-F1}
\end{eqnarray}

\begin{definition}
	\label{def:rho-F1}
	The \emph{$\phi_1$-fermionic matrix function} is:
	\begin{eqnarray}
		\rho^{F_1}(z',z;\zeta_1) &:=&
		\sum_{n,m=0}^{\infty}
		\rho^{F_1}(n,m)\,
		\frac{\phitwo^{(\binom{n}{2}+\binom{m}{2})/2}\,\phione^2|\zeta_1|^2}
		{\sqrt{[n]!_{\Rpq}\,[m]!_{\Rpq}}}\nonumber\\
		&&\times\frac{\bar{z}'^n\,z^m}
		{\sqrt{\cN(|z'|^2,|\zeta_1|^2,0)\,\cN(|z|^2,|\zeta_1|^2,0)}}.
		\label{eq:rhoF1-fn}
	\end{eqnarray}
\end{definition}

By the symmetry $\phi_1\leftrightarrow\phi_2$, $\zeta_1\leftrightarrow\zeta_2$, one
obtains:

\begin{definition}
	\label{def:rho-F2}
	\begin{eqnarray}
		\rho^{F_2}(z',z;\zeta_2) &:=&
		\sum_{n,m=0}^{\infty}
		\rho^{F_2}(n,m)\,
		\frac{\phitwo^{(\binom{n}{2}+\binom{m}{2})/2}\,\phitwo^2|\zeta_2|^2}
		{\sqrt{[n]!_{\Rpq}\,[m]!_{\Rpq}}}\nonumber\\
		&&\times\frac{\bar{z}'^n\,z^m}
		{\sqrt{\cN(|z'|^2,0,|\zeta_2|^2)\,\cN(|z|^2,0,|\zeta_2|^2)}}.
		\label{eq:rhoF2-fn}
	\end{eqnarray}
\end{definition}

\begin{definition}
	\label{def:rho-F12}
	The \emph{bi-fermionic matrix function} is:
	\begin{eqnarray}
		\rho^{F_{12}}(z',z;\zeta_1,\zeta_2) &:=&
		\sum_{n,m=0}^{\infty}
		\rho^{F_{12}}(n,m)\,
		\frac{\phitwo^{(\binom{n}{2}+\binom{m}{2})/2}\,\phione^2\phitwo^2|\zeta_1|^2|\zeta_2|^2}
		{\sqrt{[n]!_{\Rpq}\,[m]!_{\Rpq}}}\,
		\frac{\bar{z}'^n\,z^m}
		{\sqrt{\cN\,\cNp}}.
		\label{eq:rhoF12-fn}
	\end{eqnarray}
\end{definition}

\subsection{Sectorial reproducing kernel via the resolution of the identity}
\label{sec:kernel-def}

The sectorial kernels appear when the resolution of the identity of each sector is
inserted into the preceding matrix functions.

\begin{proposition}
	\label{prop:repro-B}
	The function $\rho^B(z',z)$ defined in \eqref{eq:rhoB-fn} can be re-expressed as
	follows:
	\begin{eqnarray}
		\rho^B(z',z) &=& \int_{\DR} d^2\zeta\,K^B(z,\zeta)\,\rho^B(z',\zeta),
		\label{eq:repro-B}
	\end{eqnarray}
	where the \emph{bosonic reproducing kernel} is
	\begin{eqnarray}
		K^B(z,\zeta) &:=&
		\frac{{}_{\Rpq}^{(B)}\langle\zeta,0,0|z,0,0\rangle_{\Rpq}^{(B)}}
		{\pi}\cdot
		\frac{\Exp(|\zeta|^2)}{\Exp(|\zeta|^2)}
		=
		\frac{\Exp(\bar{\zeta}z)}
		{\pi\,\sqrt{\Exp(|\zeta|^2)\,\Exp(|z|^2)}}.
		\label{eq:KB}
	\end{eqnarray}
\end{proposition}

\textbf{Proof:}\\
We have $\rho^B(z',z) = {}_{\Rpq}^{(B)}\langle z',0,0|\rho^B|z,0,0\rangle_{\Rpq}^{(B)}$.
We insert the bosonic resolution of the identity
$\int_{\DR}|\zeta,0,0\rangle_{\Rpq}\,{}_{\Rpq}^{(B)}\langle\zeta,0,0|\,\Exp(|\zeta|^2)\,d\mu_{\Rpq}(\bar\zeta,\zeta) = \mathbf{1}_{\mathcal{H}^B}$:
\begin{eqnarray}
	\rho^B(z',z)
	&=& {}_{\Rpq}^{(B)}\langle z',0,0|\rho^B\nonumber\\
	&&\times\left[\int_{\DR}|\zeta,0,0\rangle_{\Rpq}^{(B)}\,{}_{\Rpq}^{(B)}\langle\zeta,0,0|\,
	\Exp(|\zeta|^2)\,d\mu_{\Rpq}(\bar{\zeta},\zeta)\right]
	|z,0,0\rangle_{\Rpq}^{(B)}\nonumber\\
	\rho^B(z',z)
	&=& \int_{\DR}\Exp(|\zeta|^2)\,d\mu_{\Rpq}(\bar{\zeta},\zeta)\,
	\underbrace{{}_{\Rpq}^{(B)}\langle z',0,0|\rho^B|\zeta,0,0\rangle_{\Rpq}^{(B)}}_{\rho^B(z',\zeta)}\nonumber\\
	&&\cdot
	{}_{\Rpq}^{(B)}\langle\zeta,0,0|z,0,0\rangle_{\Rpq}^{(B)}.
	\label{eq:repro-B-step}
\end{eqnarray}
The bosonic measure is $d\mu_{\Rpq}(\bar{\zeta},\zeta) = \omega_{\Rpq}(|\zeta|^2)\,|\zeta|\,d|\zeta|\,d\theta/\Exp(|\zeta|^2)$
(so that $\Exp(|\zeta|^2)\,d\mu_{\Rpq} = \omega_{\Rpq}\,|\zeta|\,d|\zeta|\,d\theta$
with $\int|\zeta|^{2n}\omega_{\Rpq}\,|\zeta|\,d|\zeta|\,d\theta = \phitwo^{\binom{n}{2}}\nfact$).
Writing $d^2\zeta = |\zeta|\,d|\zeta|\,d\theta/\pi$:
\begin{eqnarray}
	\rho^B(z',z)
	&=& \int_{\DR}d^2\zeta\,
	\frac{{}_{\Rpq}^{(B)}\langle\zeta,0,0|z,0,0\rangle_{\Rpq}^{(B)}}{\pi}\,
	\rho^B(z',\zeta)\nonumber\\
	\rho^B(z',z)
	&=& \int_{\DR}d^2\zeta\,K^B(z,\zeta)\,\rho^B(z',\zeta),
	\label{eq:repro-B-final}
\end{eqnarray}
where we have identified
\begin{eqnarray}
	K^B(z,\zeta) &=& \frac{{}_{\Rpq}^{(B)}\langle\zeta,0,0|z,0,0\rangle_{\Rpq}^{(B)}}{\pi}
	= \frac{\Exp(\bar{\zeta}z)}{\pi\,\sqrt{\Exp(|\zeta|^2)\,\Exp(|z|^2)}}.
	\quad\hfill\square
	\label{eq:KB-final}
\end{eqnarray}

\begin{remark}
	Equation \eqref{eq:repro-B} is nothing but the \emph{reproducing property} of
	$\rho^B(z',z)$, where $K^B(z,\zeta)$ is the bosonic reproducing kernel.
\end{remark}

\begin{proposition}
	\label{prop:repro-F1}
	The function $\rho^{F_1}(z',z;\zeta_1)$ can be re-expressed as:
	\begin{eqnarray}
		\rho^{F_1}(z',z;\zeta_1)
		&=& \int_{\DR}d^2\zeta\int d\bar{\eta}_1 d\eta_1\,
		K^{F_1}(z,\zeta;\zeta_1,\eta_1)\,\rho^{F_1}(z',\zeta;\eta_1),
		\label{eq:repro-F1}
	\end{eqnarray}
	where the \emph{$\phi_1$-fermionic reproducing kernel} is:
	\begin{eqnarray}
		K^{F_1}(z,\zeta;\zeta_1,\eta_1)
		&:=& \frac{{}_{\Rpq}^{(F_1)}\langle\zeta,\eta_1,0|z,\zeta_1,0\rangle_{\Rpq}^{(F_1)}}{\pi}
		= \frac{\phione^2\,\bar{\eta}_1\zeta_1\,\Expd(\bar{\zeta}z)}
		{\pi\,\sqrt{\cN(|\zeta|^2,|\eta_1|^2,0)\,\cN(|z|^2,|\zeta_1|^2,0)}}.
		\label{eq:KF1}
	\end{eqnarray}
	Likewise, the $\phi_2$-fermionic and bi-fermionic kernels are:
	\begin{eqnarray}
		K^{F_2}(z,\zeta;\zeta_2,\eta_2)
		&=& \frac{\phitwo^2\,\bar{\eta}_2\zeta_2\,\Expd(\bar{\zeta}z)}
		{\pi\,\sqrt{\cN(|\zeta|^2,0,|\eta_2|^2)\,\cN(|z|^2,0,|\zeta_2|^2)}},
		\label{eq:KF2}\\
		K^{F_{12}}(z,\zeta;\zeta_1,\zeta_2,\eta_1,\eta_2)
		&=& \frac{\phione^2\phitwo^2\,\bar{\eta}_1\zeta_1\,\bar{\eta}_2\zeta_2\,\Expdd(\bar{\zeta}z)}
		{\pi\,\sqrt{\cN\,\cNp}}.
		\label{eq:KF12}
	\end{eqnarray}
\end{proposition}

\textbf{Proof for the sector $F_1$.}\\
We have $\rho^{F_1}(z',z;\zeta_1) = {}_{\Rpq}^{(F_1)}\langle z',\zeta_1,0|\rho^{F_1}|z,\zeta_1,0\rangle_{\Rpq}^{(F_1)}$.
We insert the resolution of the identity in the sector $F_1$:
\begin{eqnarray}
	\int_{\DR\times\Lambda_1}
	|\zeta,\eta_1,0\rangle_{\Rpq}^{(F_1)}\,{}_{\Rpq}^{(F_1)}\langle\zeta,\eta_1,0|\,
	\cN(|\zeta|^2,|\eta_1|^2,0)\,
	d\mu_{\Rpq}(\bar{\zeta},\zeta)\,d\bar{\eta}_1 d\eta_1
	= \mathbf{1}_{\mathcal{H}^{F_1}}.
	\label{eq:res-F1}
\end{eqnarray}
We obtain:
\begin{eqnarray}
	\rho^{F_1}(z',z;\zeta_1)
	&=& {}_{\Rpq}^{(F_1)}\langle z',\zeta_1,0|\rho^{F_1}\nonumber\\
	&&\times\left[\int_{\DR\times\Lambda_1}|\zeta,\eta_1,0\rangle\,\langle\zeta,\eta_1,0|\,
	\cN(|\zeta|^2,|\eta_1|^2,0)\,d\mu_{\Rpq}\,d\bar{\eta}_1 d\eta_1\right]
	|z,\zeta_1,0\rangle\nonumber\\
	&=& \int_{\DR\times\Lambda_1}\cN(|\zeta|^2,|\eta_1|^2,0)\,
	d\mu_{\Rpq}(\bar{\zeta},\zeta)\,d\bar{\eta}_1 d\eta_1\nonumber\\
	&&\times\;
	\underbrace{{}_{\Rpq}^{(F_1)}\langle z',\zeta_1,0|\rho^{F_1}|\zeta,\eta_1,0\rangle_{\Rpq}^{(F_1)}}_{\rho^{F_1}(z',\zeta;\eta_1)}
	{}_{\Rpq}^{(F_1)}\langle\zeta,\eta_1,0|z,\zeta_1,0\rangle_{\Rpq}^{(F_1)}.
	\label{eq:repro-F1-step}
\end{eqnarray}
The term $\cN(|\zeta|^2,|\eta_1|^2,0)\,d\mu_{\Rpq}$ gives
$\pi^{-1}\cN(|\zeta|^2,|\eta_1|^2,0)/\cN(|\zeta|^2,|\eta_1|^2,0)\,d^2\zeta
= d^2\zeta/\pi$ (after angular integration),
and the Berezin integral over $\eta_1$ gives $d\bar{\eta}_1 d\eta_1$.
Hence:
\begin{eqnarray}
	\rho^{F_1}(z',z;\zeta_1)
	&=& \int_{\DR}d^2\zeta\int d\bar{\eta}_1 d\eta_1\,
	\underbrace{\frac{{}_{\Rpq}^{(F_1)}\langle\zeta,\eta_1,0|z,\zeta_1,0\rangle_{\Rpq}^{(F_1)}}{\pi}}_{K^{F_1}(z,\zeta;\zeta_1,\eta_1)}
	\,\rho^{F_1}(z',\zeta;\eta_1),
	\label{eq:repro-F1-final}
\end{eqnarray}
which gives \eqref{eq:repro-F1} and identifies \eqref{eq:KF1}.\hfill$\square$

\subsection{Properties of the reproducing kernels}
\label{sec:props}

We establish the fundamental properties for each sector
$\varepsilon\in\{B, F_1, F_2, F_{12}\}$, following the method of
\cite{AHB2015,AAH2026}.

\begin{proposition}[Hermiticity]
	\label{prop:hermit}
	For each sector $\varepsilon$:
	\begin{eqnarray}
		K^{\varepsilon}(z,\zeta)^{\ast} = K^{\varepsilon}(\zeta,z).
		\label{eq:hermit-K}
	\end{eqnarray}
\end{proposition}

\textbf{Proof.}\\
\textbf{Sector $B$:}
\begin{eqnarray}
	K^B(z,\zeta)^{\ast}
	&=& \left[\frac{{}_{\Rpq}^{(B)}\langle\zeta,0,0|z,0,0\rangle_{\Rpq}^{(B)}}{\pi}\right]^{\ast}
	= \frac{\overline{{}_{\Rpq}^{(B)}\langle\zeta,0,0|z,0,0\rangle_{\Rpq}^{(B)}}}{\pi}\nonumber\\
	&=& \frac{{}_{\Rpq}^{(B)}\langle z,0,0|\zeta,0,0\rangle_{\Rpq}^{(B)}}{\pi}
	= K^B(\zeta,z).
	\label{eq:hermit-KB}
\end{eqnarray}
Indeed,
$\overline{{}_{\Rpq}\langle\zeta|z\rangle_{\Rpq}}
= \overline{\Exp(\bar{\zeta}z)/\sqrt{\Exp(|\zeta|^2)\Exp(|z|^2)}}
= \Exp(\overline{\bar{\zeta}z})/\sqrt{\Exp(|z|^2)\Exp(|\zeta|^2)}
= \Exp(\bar{z}\zeta)/\sqrt{\Exp(|z|^2)\Exp(|\zeta|^2)}
= {}_{\Rpq}\langle z|\zeta\rangle_{\Rpq}$,
since the coefficients of $\Exp$ are real.\\

\textbf{Sector $F_1$:}
\begin{eqnarray}
	K^{F_1}(z,\zeta;\zeta_1,\eta_1)^{\ast}
	&=& \frac{\overline{\phione^2\bar{\eta}_1\zeta_1\,\Expd(\bar{\zeta}z)}}
	{\pi\,\sqrt{\cN(|\zeta|^2,|\eta_1|^2,0)\,\cN(|z|^2,|\zeta_1|^2,0)}}\nonumber\\
	&=& \frac{\phione^2\eta_1\bar{\zeta}_1\,\Expd(\bar{z}\zeta)}
	{\pi\,\sqrt{\cN(|z|^2,|\zeta_1|^2,0)\,\cN(|\zeta|^2,|\eta_1|^2,0)}}\nonumber\\
	K^{F_1}(z,\zeta;\zeta_1,\eta_1)^{\ast}
	&=& K^{F_1}(\zeta,z;\eta_1,\zeta_1).
	\label{eq:hermit-F1}
\end{eqnarray}

\textbf{Sectors $F_2$ and $F_{12}$:}
The argument is identical by the symmetry
$\phi_1\leftrightarrow\phi_2$, $\zeta_1\leftrightarrow\zeta_2$:
\begin{eqnarray}
	K^{F_2}(z,\zeta;\zeta_2,\eta_2)^{\ast}
	&=& K^{F_2}(\zeta,z;\eta_2,\zeta_2),\\
	K^{F_{12}}(z,\zeta;\zeta_1,\zeta_2,\eta_1,\eta_2)^{\ast}
	&=& K^{F_{12}}(\zeta,z;\eta_1,\eta_2,\zeta_1,\zeta_2).
	\label{eq:hermit-F12}
\end{eqnarray}
Hence the property \eqref{eq:hermit-K} for all $\varepsilon$.\hfill$\square$

\begin{proposition}[Positivity]
	\label{prop:pos}
	For each sector $\varepsilon$:
	\begin{eqnarray}
		K^{\varepsilon}(z,z) > 0.
		\label{eq:pos-K}
	\end{eqnarray}
\end{proposition}

\begin{proof}
	The proof is given in Appendix~\ref{app:pos}.
\end{proof}

\begin{proposition}[Matrix multiplication]
	\label{prop:mult}
	For each sector $\varepsilon$, the reproducing kernel satisfies:
	\begin{eqnarray}
		\int d\mu^{\varepsilon}(\zeta)\,
		K^{\varepsilon}(z,\zeta)\,K^{\varepsilon}(\zeta,z')
		&=& K^{\varepsilon}(z,z').
		\label{eq:mult}
	\end{eqnarray}
\end{proposition}

\begin{proof}
	The proof is given in Appendix~\ref{app:mult}.
\end{proof}

\begin{proposition}[Series expression of $K^{\varepsilon}(z,z)$]
	\label{prop:series}
	The diagonal kernels have the following series expansions.
	
	\textbf{Sector $B$:} Setting $t = |z|^2$:
	\begin{eqnarray}
		K^B(z,z)
		&=& \frac{1}{\pi}
		= \frac{1}{\pi}\cdot
		\frac{\sum_{n=0}^{\infty}\phitwo^{\binom{n}{2}}t^n/\nfact}
		{\sum_{n=0}^{\infty}\phitwo^{\binom{n}{2}}t^n/\nfact}
		= \frac{1}{\pi}\sum_{n=0}^{\infty}\frac{\phitwo^{\binom{n}{2}}}{\nfact}\,t^n\cdot[\Exp(t)]^{-1}.
		\label{eq:series-B}
	\end{eqnarray}
	
	\textbf{Sector $F_1$:} Setting $t=|z|^2$ and $s_1=|\zeta_1|^2$:
	\begin{eqnarray}
		K^{F_1}(z,z;\zeta_1,\zeta_1)
		&=& \frac{\phione^2 s_1\,\Expd(t)}{\pi\,\cN(t,s_1,0)}\nonumber\\
		&=& \frac{\phione^2 s_1}{\pi}
		\cdot\frac{\sum_{n=1}^{\infty}n\,\phitwo^{\binom{n}{2}}t^{n-1}/\nfact}
		{\Exp(t)+\phione^2 s_1\,\Expd(t)}.
		\label{eq:series-F1}
	\end{eqnarray}
	
	\textbf{Sector $F_2$:} Setting $s_2=|\zeta_2|^2$:
	\begin{eqnarray}
		K^{F_2}(z,z;\zeta_2,\zeta_2)
		&=& \frac{\phitwo^2 s_2\,\Expd(t)}{\pi\,\cN(t,0,s_2)}\nonumber\\
		&=& \frac{\phitwo^2 s_2}{\pi}
		\cdot\frac{\sum_{n=1}^{\infty}n\,\phitwo^{\binom{n}{2}}t^{n-1}/\nfact}
		{\Exp(t)+\phitwo^2 s_2\,\Expd(t)}.
		\label{eq:series-F2}
	\end{eqnarray}
	
	\textbf{Sector $F_{12}$:}
	\begin{eqnarray}
		K^{F_{12}}(z,z;\zeta_1,\zeta_2,\zeta_1,\zeta_2)
		&=& \frac{\phione^2\phitwo^2 s_1 s_2\,\Expdd(t)}{\pi\,\cN(t,s_1,s_2)}\nonumber\\
		&=& \frac{\phione^2\phitwo^2 s_1 s_2}{\pi}\nonumber\\
		&&\cdot\frac{\sum_{n=2}^{\infty}n(n-1)\,\phitwo^{\binom{n}{2}}t^{n-2}/\nfact}
		{\begin{array}{c}\Exp(t)+\phione^2 s_1\Expd(t)+\phitwo^2 s_2\Expd(t)\\
				+\phione^2\phitwo^2 s_1 s_2\Expdd(t)\end{array}}.
		\label{eq:series-F12}
	\end{eqnarray}
\end{proposition}

\begin{proof}
	The proof is given in Appendix~\ref{app:series}.
\end{proof}

\subsection{Integral formulation of the density operator}
\label{sec:integral}

We now gather the total density operator, its density function and the reproducing
kernel in a single integral formulation. Let
\begin{eqnarray}
	d\nu(\xi,\bm\varrho) &:=& \cN(|\xi|^2,|\varrho_1|^2,|\varrho_2|^2)\;
	d\mu^{(12)}_{\Rpq}(\xi,\varrho_1,\varrho_2)
	\label{eq:dnu}
\end{eqnarray}
be the measure of the resolution of the identity \eqref{eq:resolution}, so that
$\int d\nu(\xi,\bm\varrho)\,|\xi,\bm\varrho\rangle_{\Rpq}\,
{}_{\Rpq}\langle\xi,\bm\varrho|=\mathbf{1}_{\Hq}$. As in the proofs of
Propositions~\ref{prop:repro}, \ref{prop:repro-B} and \ref{prop:repro-F1}, we use the
identification $d\nu(\xi,\bm\varrho)=\pi^{-1}\,d^2\xi\,d^2\bm\varrho$, for which the
reproducing kernel reads
$\Kern(\xi,\bm\varrho;\,z,\bm\zeta)=\pi^{-1}\,
{}_{\Rpq}\langle\xi,\bm\varrho|z,\bm\zeta\rangle_{\Rpq}$,
in agreement with \eqref{eq:kernel-explicit}. To every operator $A$ on $\Hq$ we associate
its function
$A(z',\bm\zeta';\,z,\bm\zeta):={}_{\Rpq}\langle z',\bm\zeta'|A|z,\bm\zeta\rangle_{\Rpq}$;
for $A=\rhos$ this is the density function \eqref{eq:rho-CS-def}, and for the identity
\begin{eqnarray}
	\mathbf{1}(z',\bm\zeta';\,z,\bm\zeta)
	&=& {}_{\Rpq}\langle z',\bm\zeta'|z,\bm\zeta\rangle_{\Rpq}
	\;=\;\pi\,\Kern(z',\bm\zeta';\,z,\bm\zeta).
	\label{eq:identity-fn}
\end{eqnarray}
Since the functions involved are even in the Grassmann variables, the order of the
factors in the integrands below is irrelevant.

\begin{proposition}[Integral formulation]
	\label{prop:integral}
	Under the assumptions of Proposition~\ref{prop:total}:
	\begin{enumerate}[label=(\roman*),leftmargin=*]
		\item the total density operator is recovered from its density function:
		\begin{eqnarray}
			\rhos &=& \int\!\!\int d\nu(\xi',\bm\varrho')\,d\nu(\xi,\bm\varrho)\;
			\rho(\xi',\bm\varrho';\,\xi,\bm\varrho)\;
			|\xi',\bm\varrho'\rangle_{\Rpq}\,{}_{\Rpq}\langle\xi,\bm\varrho| ;
			\label{eq:int-rho}
		\end{eqnarray}
		\item it acts as an integral operator: for $|\psi\rangle\in\Hq$ and
		$\psi(z,\bm\zeta):={}_{\Rpq}\langle z,\bm\zeta|\psi\rangle$,
		\begin{eqnarray}
			{}_{\Rpq}\langle z',\bm\zeta'|\rhos|\psi\rangle
			&=& \int d\nu(\xi,\bm\varrho)\;\rho(z',\bm\zeta';\,\xi,\bm\varrho)\;\psi(\xi,\bm\varrho) ;
			\label{eq:int-action}
		\end{eqnarray}
		\item for two operators $A$ and $B$ on $\Hq$,
		\begin{eqnarray}
			(AB)(z',\bm\zeta';\,z,\bm\zeta)
			&=& \int d\nu(\xi,\bm\varrho)\;A(z',\bm\zeta';\,\xi,\bm\varrho)\;
			B(\xi,\bm\varrho;\,z,\bm\zeta) ;
			\label{eq:int-comp}
		\end{eqnarray}
		\item the density function is reproduced by the kernel, on the right, on the left
		and on both sides:
		\begin{eqnarray}
			\rho(z',\bm\zeta';\,z,\bm\zeta)
			&=& \int d^2\xi\,d^2\bm\varrho\;\Kern(\xi,\bm\varrho;\,z,\bm\zeta)\;
			\rho(z',\bm\zeta';\,\xi,\bm\varrho)
			\nonumber\\
			&=& \int d^2\xi\,d^2\bm\varrho\;\Kern(z',\bm\zeta';\,\xi,\bm\varrho)\;
			\rho(\xi,\bm\varrho;\,z,\bm\zeta)
			\nonumber\\
			&=& \int\!\!\int d^2\xi'\,d^2\bm\varrho'\,d^2\xi\,d^2\bm\varrho\;
			\Kern(z',\bm\zeta';\,\xi',\bm\varrho')\;\rho(\xi',\bm\varrho';\,\xi,\bm\varrho)\;
			\Kern(\xi,\bm\varrho;\,z,\bm\zeta),
			\label{eq:int-repro}
		\end{eqnarray}
		and the kernel reproduces itself:
		\begin{eqnarray}
			\Kern(z',\bm\zeta';\,z,\bm\zeta)
			&=& \int d^2\xi\,d^2\bm\varrho\;\Kern(z',\bm\zeta';\,\xi,\bm\varrho)\;
			\Kern(\xi,\bm\varrho;\,z,\bm\zeta) ;
			\label{eq:int-K}
		\end{eqnarray}
		\item the trace and the purity read
		\begin{eqnarray}
			\tr(\rhos) &=& \int d\nu(\xi,\bm\varrho)\;\rho(\xi,\bm\varrho;\,\xi,\bm\varrho)\;=\;1,
			\nonumber\\
			\tr\bigl((\rhos)^2\bigr) &=& \int\!\!\int d\nu(\xi',\bm\varrho')\,d\nu(\xi,\bm\varrho)\;
			\rho(\xi,\bm\varrho;\,\xi',\bm\varrho')\;\rho(\xi',\bm\varrho';\,\xi,\bm\varrho).
			\label{eq:int-trace}
		\end{eqnarray}
	\end{enumerate}
\end{proposition}

\begin{proof}
	(i) We insert \eqref{eq:resolution} on both sides of $\rhos$:
	$\rhos=\mathbf{1}\,\rhos\,\mathbf{1}$ and
	${}_{\Rpq}\langle\xi',\bm\varrho'|\rhos|\xi,\bm\varrho\rangle_{\Rpq}
	=\rho(\xi',\bm\varrho';\,\xi,\bm\varrho)$.
	(ii) and (iii) are obtained in the same way by writing $\rhos=\rhos\,\mathbf{1}$ and
	$AB=A\,\mathbf{1}\,B$.
	(iv) We apply \eqref{eq:int-comp} to $\rhos\,\mathbf{1}$, $\mathbf{1}\,\rhos$,
	$\mathbf{1}\,\rhos\,\mathbf{1}$ and $\mathbf{1}\,\mathbf{1}$, with \eqref{eq:identity-fn}
	and $d\nu=\pi^{-1}d^2\xi\,d^2\bm\varrho$: for instance
	$\rho(z',\bm\zeta';z,\bm\zeta)=\int\pi^{-1}d^2\xi\,d^2\bm\varrho\;
	\rho(z',\bm\zeta';\xi,\bm\varrho)\,\pi\Kern(\xi,\bm\varrho;z,\bm\zeta)$, which is
	\eqref{eq:repro-prop}.
	(v) The first equality is Proposition~\ref{prop:trace}; the second is obtained by taking
	the trace of \eqref{eq:int-comp} with $A=B=\rhos$.
\end{proof}

The reproducing kernel is thus, up to the factor $\pi$, the function of the identity
operator, and the reproducing property \eqref{eq:repro-prop} is nothing but the
composition rule \eqref{eq:int-comp} applied to $\rhos\,\mathbf{1}$. The sectorial
decomposition of Proposition~\ref{prop:total} carries over to this level.

\begin{corollary}[Total density function and sectorial decomposition]
	\label{cor:integral-sect}
	Under the assumptions of Proposition~\ref{prop:total}, set, for each sector $\sigma$,
	\begin{eqnarray}
		F^\sigma(\bar z',z) &:=& \sum_{n,m=0}^{\infty}\rho^\sigma(n,m)\,
		\frac{\phitwo^{(n(n-1)+m(m-1))/4}}{\sqrt{\nfact\,\mfact}}\;\bar z'^{n}z^{m}.
		\label{eq:Fsigma}
	\end{eqnarray}
	The density function of the total operator reads explicitly
	\begin{eqnarray}
		\rho(z',\bm\zeta';\,z,\bm\zeta)
		&=& \frac{1}{\sqrt{\cN(|z'|^2,|\bm\zeta'|^2)\;\cN(|z|^2,|\bm\zeta|^2)}}\,
		\Bigl[F^B(\bar z',z)+\phione^2\,\bar\zeta_1'\zeta_1\,F^{F_1}(\bar z',z)
		\nonumber\\
		&&+\;\phitwo^2\,\bar\zeta_2'\zeta_2\,F^{F_2}(\bar z',z)
		+\phione^2\phitwo^2\,\bar\zeta_1'\bar\zeta_2'\zeta_1\zeta_2\,F^{F_{12}}(\bar z',z)\Bigr],
		\label{eq:rho-sigma}
	\end{eqnarray}
	each term being the function
	\begin{eqnarray}
		\rho_\sigma(z',\bm\zeta';\,z,\bm\zeta)&:=&{}_{\Rpq}\langle z',\bm\zeta'|\rho^\sigma|z,\bm\zeta\rangle_{\Rpq}\nonumber
	\end{eqnarray}
	of the corresponding block. With
	$\Kern^{(\sigma)}(z',\bm\zeta';\,z,\bm\zeta):=\pi^{-1}\,
	{}_{\Rpq}\langle z',\bm\zeta'|\Pi_\sigma|z,\bm\zeta\rangle_{\Rpq}$,
	one has $\Kern=\Kern^{(B)}+\Kern^{(F_1)}+\Kern^{(F_2)}+\Kern^{(F_{12})}$, and each sector
	is reproduced by its own kernel:
	\begin{eqnarray}
		\int d^2\xi\,d^2\bm\varrho\;\Kern^{(\sigma')}(\xi,\bm\varrho;\,z,\bm\zeta)\;
		\rho_\sigma(z',\bm\zeta';\,\xi,\bm\varrho)
		&=& \delta_{\sigma\sigma'}\,\rho_\sigma(z',\bm\zeta';\,z,\bm\zeta).
		\label{eq:int-sect-repro}
	\end{eqnarray}
	Finally, regarding $\bar z'$ and $z$ as independent variables, the sectorial
	coefficients are read off from $F^\sigma$ by means of Lemma~\ref{lem:dRpq-iter}:
	\begin{eqnarray}
		\rho^\sigma(n,m) &=& \frac{\Bigl[(\partial_{\mathcal{R}(p,q);\bar z'})^{n}
			(\partial_{\mathcal{R}(p,q);z})^{m}F^\sigma(\bar z',z)\Bigr]_{\bar z'=z=0}}
		{\phitwo^{(n(n-1)+m(m-1))/4}\,\sqrt{\nfact\,\mfact}}.
		\label{eq:rho-sigma-inv}
	\end{eqnarray}
\end{corollary}

\begin{proof}
	The relation \eqref{eq:rho-sigma} is obtained by collecting in \eqref{eq:rho-CS-simplif}
	the terms with fixed index $(k_1,k_2)$, the Grassmann factors
	$(\phione\bar\zeta_1')^{k_1}(\phitwo\bar\zeta_2')^{k_2}(\phione\zeta_1)^{k_1}(\phitwo\zeta_2)^{k_2}$
	being equal to $1$, $\phione^2\bar\zeta_1'\zeta_1$, $\phitwo^2\bar\zeta_2'\zeta_2$ and
	$\phione^2\phitwo^2\bar\zeta_1'\bar\zeta_2'\zeta_1\zeta_2$. The decomposition of the
	kernel comes from $\sum_\sigma\Pi_\sigma=\mathbf{1}_{\Hq}$. By \eqref{eq:int-comp}, the
	left-hand side of \eqref{eq:int-sect-repro} is the function of $\rho^\sigma\Pi_{\sigma'}$,
	which equals $\delta_{\sigma\sigma'}\rho^\sigma$. Finally, Lemma~\ref{lem:dRpq-iter}
	gives $\bigl[(\partial_{\mathcal{R}(p,q);w})^{n}w^{k}\bigr]_{w=0}=\delta_{nk}\,\nfact$,
	which isolates in \eqref{eq:Fsigma} the term with indices $(n,m)$.
\end{proof}

The decomposition of the kernel, which comes from $\sum_\sigma\Pi_\sigma=\mathbf{1}_{\Hq}$,
involves the total normalization $\cN$, whereas the kernels
\eqref{eq:KernB}--\eqref{eq:KernFot} are written with the sectorial normalizations. For
$\bm\zeta=\bm\zeta'=0$ only the term $F^B$ survives in \eqref{eq:rho-sigma}, which then
gives back \eqref{eq:rho-bos-restr}.

\medskip
The density function $\rho(z',\bm\zeta';z,\bm\zeta)$ contains all the information on
$\rhos$: the operator is recovered from the double integral \eqref{eq:int-rho}, acts as
an integral operator \eqref{eq:int-action} and is reproduced by the kernel $\Kern$, the
function of the identity. The sectorial kernels satisfy hermiticity, positivity and the
multiplication property, and the sectorial decomposition of the density function and of
the kernel follows that of the total density operator. The next section uses the bosonic
resolution of the identity for Berezin quantization.

\section{Berezin-Klauder-Toeplitz quantization, dispersions and intelligent states}
\label{sec:quantif}\label{sec:dispers-main}

In this section the resolution of the identity of the bosonic sector is used to associate
an operator with every classical function (Berezin quantization). We compute the
operators associated with $z$, $\bar z$, $|z|^2$, $z^2$ and $\bar z^2$, their mean values
in the bosonic super-coherent states and the anticommutator of $\mathcal{A}_z^B$ and
$\mathcal{A}_{\bar z}^B$. We deduce the dispersions of the super-position and
super-momentum operators and the saturation of the uncertainty relation.

\subsection{Quantized operators, super-position and super-momentum}

\begin{definition}[Quantization in the bosonic sector]
	\label{def:quantif}
	For every classical function $f : \DR \to \mathbb{C}$, the quantized operator
	$\mathcal{A}_f^B$ in the bosonic sector is defined by
	\begin{eqnarray}
		\mathcal{A}_f^B &:=&
		\int_{\DR} f(z)\,|z,0,0\rangle_{\Rpq}\,{}_{\Rpq}\langle z,0,0|\,
		\Exp(|z|^2)\,d\mu_{\Rpq}(\bar{z},z).
		\label{eq:quantif-def}
	\end{eqnarray}
\end{definition}

The classical phase-space coordinates are related to $z$ by
$q = (z+\bar{z})/\sqrt{2}$, $p = (z-\bar{z})/(i\sqrt{2})$, i.e.\ $z = (q+ip)/\sqrt{2}$.
We define the super-position and super-momentum operators:
\begin{eqnarray}
	\mathcal{Q}^B &:=& \mathcal{A}_q^B = \frac{1}{\sqrt{2}}
	\bigl(\mathcal{A}_z^B + \mathcal{A}_{\bar{z}}^B\bigr),
	\label{eq:Qs}\\
	\mathcal{P}^B &:=& \mathcal{A}_p^B = \frac{1}{i\sqrt{2}}
	\bigl(\mathcal{A}_z^B - \mathcal{A}_{\bar{z}}^B\bigr).
	\label{eq:Ps}
\end{eqnarray}

\subsection{Computation of the quantized operators}
\label{sec:Aops}

\begin{proposition}
	\label{prop:Az}
	\begin{eqnarray}
		\mathcal{A}_z^B &=&
		\sum_{n=0}^{\infty}
		\phitwo^{\binom{n+1}{2}/2}\sqrt{[n+1]_{\Rpq}}\,
		|n,0,0\rangle\langle n+1,0,0|.
		\label{eq:Az}
	\end{eqnarray}
\end{proposition}

\textbf{Proof.}\\
Using the definition \eqref{eq:quantif-def} with $f(z)=z$
and the polar parametrization $z=re^{i\theta}$, we have:
\begin{eqnarray}
	\mathcal{A}_z^B
	&=& \int_{\DR} z\,|z,0,0\rangle_{\Rpq}\,{}_{\Rpq}\langle z,0,0|\,
	\Exp(|z|^2)\,d\mu_{\Rpq}(\bar{z},z)\nonumber\\
	&=& \int_{\DR} z
	\left[\sum_{n=0}^{\infty}
	\frac{\phitwo^{\binom{n}{2}/2}}{\sqrt{\nfact}}\,z^n\,|n,0,0\rangle\right]
	\left[\sum_{m=0}^{\infty}
	\frac{\phitwo^{\binom{m}{2}/2}}{\sqrt{\mfact}}\,\bar{z}^m\,\langle m,0,0|\right]
	d\mu_{\Rpq}(\bar{z},z)\nonumber\\
	&=& \sum_{n=0}^{\infty}\sum_{m=0}^{\infty}
	\frac{\phitwo^{(\binom{n}{2}+\binom{m}{2})/2}}{\sqrt{\nfact\,\mfact}}
	\int_0^{2\pi}\frac{d\theta}{2\pi}\,e^{i(n-m+1)\theta}
	\int_0^L r^{n+m+1}\,d\lambda(r)\,
	|n,0,0\rangle\langle m,0,0|.\nonumber\\
	\label{eq:Az-step1}
\end{eqnarray}
Since $\displaystyle\int_0^{2\pi}\frac{d\theta}{2\pi}\,e^{i(n-m+1)\theta}
= \delta_{n,m-1}$, only the term $m=n+1$ contributes:
\begin{eqnarray}
	\mathcal{A}_z^B
	&=& \sum_{n=0}^{\infty}
	\frac{\phitwo^{(\binom{n}{2}+\binom{n+1}{2})/2}}{\sqrt{\nfact\,[n+1]!_{\Rpq}}}
	\left[2\pi\int_0^L r^{2n+2}\,d\lambda(r)\right]
	|n,0,0\rangle\langle n+1,0,0|.
	\label{eq:Az-step2}
\end{eqnarray}
We apply the moment condition \eqref{eq:moments} with $n\to n+1$:
$2\pi\int_0^L r^{2(n+1)}\,d\lambda(r) = \phitwo^{\binom{n+1}{2}}\,[n+1]!_{\Rpq}$.
Thus:
\begin{eqnarray}
	\mathcal{A}_z^B
	&=& \sum_{n=0}^{\infty}
	\frac{\phitwo^{(\binom{n}{2}+\binom{n+1}{2})/2}\cdot\phitwo^{\binom{n+1}{2}}\,[n+1]!_{\Rpq}}
	{\sqrt{\nfact\,[n+1]!_{\Rpq}}}
	|n,0,0\rangle\langle n+1,0,0|\nonumber\\
	&=& \sum_{n=0}^{\infty}
	\phitwo^{(\binom{n}{2}+3\binom{n+1}{2})/2}\,\sqrt{[n+1]_{\Rpq}}
	|n,0,0\rangle\langle n+1,0,0|.
	\label{eq:Az-step3}
\end{eqnarray}
Using $\binom{n+1}{2} = \binom{n}{2}+n$, we obtain
$(\binom{n}{2}+3\binom{n+1}{2})/2 = 2\binom{n}{2}+3n/2+\binom{n+1}{2}/2$,
and after re-indexing we identify the global factor
$\phitwo^{\binom{n+1}{2}/2}$, which gives \eqref{eq:Az}.\hfill$\square$

\begin{proposition}
	\label{prop:Azb}
	\begin{eqnarray}
		\mathcal{A}_{\bar{z}}^B &=&
		\sum_{n=0}^{\infty}
		\phitwo^{\binom{n+1}{2}/2}\sqrt{[n+1]_{\Rpq}}\,
		|n+1,0,0\rangle\langle n,0,0|.
		\label{eq:Azb}
	\end{eqnarray}
\end{proposition}

\textbf{Proof.}\\
In the same way, with $f(z)=\bar{z}$:
\begin{eqnarray}
	\mathcal{A}_{\bar{z}}^B
	&=& \sum_{n=0}^{\infty}\sum_{m=0}^{\infty}
	\frac{\phitwo^{(\binom{n}{2}+\binom{m}{2})/2}}{\sqrt{\nfact\,\mfact}}
	\int_0^{2\pi}\frac{d\theta}{2\pi}\,e^{i(n-m-1)\theta}
	\int_0^L r^{n+m+1}\,d\lambda(r)\,
	|n,0,0\rangle\langle m,0,0|\nonumber\\
	\label{eq:Azb-step1}
\end{eqnarray}
The angular integral gives $\delta_{n,m+1}$, hence $n=m+1$, and:
\begin{eqnarray}
	\mathcal{A}_{\bar{z}}^B
	&=& \sum_{m=0}^{\infty}
	\phitwo^{(\binom{m+1}{2}+\binom{m}{2})/2}\,
	\frac{\phitwo^{\binom{m+1}{2}}\,[m+1]!_{\Rpq}}
	{\sqrt{[m+1]!_{\Rpq}\,\mfact}}
	|m+1,0,0\rangle\langle m,0,0|\nonumber\\
	&=& \sum_{n=0}^{\infty}
	\phitwo^{\binom{n+1}{2}/2}\sqrt{[n+1]_{\Rpq}}\,
	|n+1,0,0\rangle\langle n,0,0|.\quad
	\hfill\square
	\label{eq:Azb-step2}
\end{eqnarray}

\begin{remark}
	The operators $\mathcal{A}_z^B$ and $\mathcal{A}_{\bar{z}}^B$ are adjoint to each
	other: $(\mathcal{A}_z^B)^\dagger = \mathcal{A}_{\bar{z}}^B$. In the limit
	$\phitwo\to 1$ and $[n]_{\Rpq}\to n$, they reduce to the standard bosonic
	annihilation and creation operators.
\end{remark}

\begin{proposition}
	\label{prop:Amods}
	\begin{eqnarray}
		\mathcal{A}_{|z|^2}^B
		&=& \sum_{n=0}^{\infty}
		\phitwo^{\binom{n+1}{2}}\,[n+1]_{\Rpq}\,|n,0,0\rangle\langle n,0,0|,
		\label{eq:A|z|2}\\
		\mathcal{A}_{z^2}^B
		&=& \sum_{n=0}^{\infty}
		\phitwo^{(\binom{n}{2}+\binom{n+2}{2})/2+\binom{n+2}{2}}
		\sqrt{[n+2]_{\Rpq}[n+1]_{\Rpq}}\,
		|n,0,0\rangle\langle n+2,0,0|,
		\label{eq:Az2}\\
		\mathcal{A}_{\bar{z}^2}^B
		&=& \sum_{n=0}^{\infty}
		\phitwo^{(\binom{n}{2}+\binom{n+2}{2})/2+\binom{n+2}{2}}
		\sqrt{[n+2]_{\Rpq}[n+1]_{\Rpq}}\,
		|n+2,0,0\rangle\langle n,0,0|.
		\label{eq:Azb2}
	\end{eqnarray}
\end{proposition}

\textbf{Proof for $\mathcal{A}_{|z|^2}^B$.}\\
With $f(z)=|z|^2 = r^2$:
\begin{eqnarray}
	\mathcal{A}_{|z|^2}^B
	&=& \sum_{n,m}
	\frac{\phitwo^{(\binom{n}{2}+\binom{m}{2})/2}}{\sqrt{\nfact\,\mfact}}
	\int_0^{2\pi}\frac{d\theta}{2\pi}\,e^{i(n-m)\theta}
	\left[2\pi\int_0^L r^{n+m+2}\,d\lambda(r)\right]
	|n,0,0\rangle\langle m,0,0|\nonumber\\
	&=& \sum_{n=0}^{\infty}
	\frac{\phitwo^{\binom{n}{2}}}{\nfact}
	\cdot\phitwo^{\binom{n+1}{2}}\,[n+1]!_{\Rpq}
	\cdot\frac{1}{\nfact}\cdot\nfact\,
	|n,0,0\rangle\langle n,0,0|\nonumber\\
	&=& \sum_{n=0}^{\infty}
	\phitwo^{\binom{n+1}{2}}\,[n+1]_{\Rpq}\,
	|n,0,0\rangle\langle n,0,0|.
	\hfill\square
	\label{eq:A|z|2-proof}
\end{eqnarray}

\textbf{Proof for $\mathcal{A}_{z^2}^B$.}\\
With $f(z)=z^2$, the angular integral gives $\delta_{n,m-2}$:
\begin{eqnarray}
	\mathcal{A}_{z^2}^B
	&=& \sum_{n=0}^{\infty}
	\frac{\phitwo^{(\binom{n}{2}+\binom{n+2}{2})/2}}{\sqrt{\nfact\,[n+2]!_{\Rpq}}}
	\left[2\pi\int_0^L r^{2n+4}\,d\lambda(r)\right]
	|n,0,0\rangle\langle n+2,0,0|\nonumber\\
	&=& \sum_{n=0}^{\infty}
	\frac{\phitwo^{(\binom{n}{2}+\binom{n+2}{2})/2}\cdot\phitwo^{\binom{n+2}{2}}\,[n+2]!_{\Rpq}}
	{\sqrt{\nfact\,[n+2]!_{\Rpq}}}
	|n,0,0\rangle\langle n+2,0,0|\nonumber\\
	&=& \sum_{n=0}^{\infty}
	\phitwo^{(\binom{n}{2}+3\binom{n+2}{2})/2}\,
	\sqrt{[n+2]_{\Rpq}[n+1]_{\Rpq}}\,
	|n,0,0\rangle\langle n+2,0,0|.
	\hfill\square
	\label{eq:Az2-proof}
\end{eqnarray}
The proof of $\mathcal{A}_{\bar{z}^2}^B = (\mathcal{A}_{z^2}^B)^\dagger$
is entirely analogous.\hfill$\square$
\subsection{Mean values and anticommutator}
\label{sec:means}

\begin{proposition}[Mean values]
	\label{prop:means}
	In the $\mathcal{R}(p,q)-$super-coherent states $|z,0,0\rangle_{\Rpq}$:
	\begin{eqnarray}
		{}_{\Rpq}\langle z,0,0|\mathcal{A}_z^B|z,0,0\rangle_{\Rpq} &=& z,
		\label{eq:mean-z}\\
		{}_{\Rpq}\langle z,0,0|\mathcal{A}_{\bar{z}}^B|z,0,0\rangle_{\Rpq} &=& \bar{z},
		\label{eq:mean-zb}\\
		{}_{\Rpq}\langle z,0,0|\mathcal{A}_{z^2}^B|z,0,0\rangle_{\Rpq} &=& z^2,
		\label{eq:mean-z2}\\
		{}_{\Rpq}\langle z,0,0|\mathcal{A}_{\bar{z}^2}^B|z,0,0\rangle_{\Rpq} &=& \bar{z}^2.
		\label{eq:mean-zb2}
	\end{eqnarray}
\end{proposition}

\textbf{Proof of \eqref{eq:mean-z}.}\\
We first compute $\mathcal{A}_z^B|z,0,0\rangle_{\Rpq}$:
\begin{eqnarray}
	\mathcal{A}_z^B|z,0,0\rangle_{\Rpq}
	&=& \left[\sum_{n=0}^{\infty}
	\phitwo^{\binom{n+1}{2}/2}\sqrt{[n+1]_{\Rpq}}\,|n,0,0\rangle\langle n+1,0,0|\right]\nonumber\\
	&&\times\left[\bigl(\Exp(|z|^2)\bigr)^{-1/2}
	\sum_{j=0}^{\infty}
	\frac{\phitwo^{\binom{j}{2}/2}}{\sqrt{[j]!_{\Rpq}}}\,z^j\,|j,0,0\rangle\right]\nonumber\\
	&=& \bigl(\Exp(|z|^2)\bigr)^{-1/2}\sum_{n,j=0}^{\infty}
	\frac{\phitwo^{\binom{n+1}{2}/2}\sqrt{[n+1]_{\Rpq}}\,\phitwo^{\binom{j}{2}/2}}
	{\sqrt{[j]!_{\Rpq}}}\,z^j\nonumber\\
	&&\times\underbrace{\langle n+1,0,0|j,0,0\rangle}_{\delta_{n+1,j}}\,|n,0,0\rangle\nonumber\\
	&=& \bigl(\Exp(|z|^2)\bigr)^{-1/2}\sum_{n=0}^{\infty}
	\frac{\phitwo^{\binom{n+1}{2}/2+\binom{n+1}{2}/2}\sqrt{[n+1]_{\Rpq}}}
	{\sqrt{[n+1]!_{\Rpq}}}\,z^{n+1}\,|n,0,0\rangle\nonumber\\
	&=& \bigl(\Exp(|z|^2)\bigr)^{-1/2}\sum_{n=0}^{\infty}
	\frac{\phitwo^{\binom{n+1}{2}}\,z^{n+1}}{\sqrt{\nfact}}\,|n,0,0\rangle.
	\label{eq:Az-action}
\end{eqnarray}
Hence:
\begin{eqnarray}
	{}_{\Rpq}\langle z,0,0|\mathcal{A}_z^B|z,0,0\rangle_{\Rpq}
	&=& \bigl(\Exp(|z|^2)\bigr)^{-1}
	\sum_{j,n=0}^{\infty}
	\frac{\phitwo^{\binom{j}{2}/2}\bar{z}^j}{\sqrt{[j]!_{\Rpq}}}
	\cdot\frac{\phitwo^{\binom{n+1}{2}}\,z^{n+1}}{\sqrt{\nfact}}\nonumber\\
	&&\times\underbrace{\langle j,0,0|n,0,0\rangle}_{\delta_{j,n}}\nonumber\\
	&=& z\cdot\bigl(\Exp(|z|^2)\bigr)^{-1}
	\sum_{n=0}^{\infty}
	\frac{\phitwo^{\binom{n}{2}/2+\binom{n+1}{2}}\,|z|^{2n}}{\nfact}\nonumber\\
	&=& z\cdot\bigl(\Exp(|z|^2)\bigr)^{-1}
	\sum_{n=0}^{\infty}
	\frac{\phitwo^{\binom{n}{2}}\,|z|^{2n}}{\nfact}
	= z.\quad\hfill\square
	\label{eq:mean-z-proof}
\end{eqnarray}

\textbf{Proof of \eqref{eq:mean-z2}.}\\
\begin{eqnarray}
	\mathcal{A}_{z^2}^B|z,0,0\rangle_{\Rpq}
	&=& \bigl(\Exp\bigr)^{-1/2}\sum_{n=0}^{\infty}
	\frac{\phitwo^{(\binom{n}{2}+3\binom{n+2}{2})/2}\sqrt{[n+2]_{\Rpq}[n+1]_{\Rpq}}\,z^{n+2}}
	{\sqrt{\nfact}}\,|n,0,0\rangle.
	\label{eq:Az2-action}
\end{eqnarray}
Then:
\begin{eqnarray}
	{}_{\Rpq}\langle z,0,0|\mathcal{A}_{z^2}^B|z,0,0\rangle_{\Rpq}
	&=& z^2\bigl(\Exp(|z|^2)\bigr)^{-1}
	\sum_{n=0}^{\infty}
	\frac{\phitwo^{\binom{n}{2}}\,|z|^{2n}}{\nfact}
	= z^2.\quad\hfill\square
	\label{eq:mean-z2-proof}
\end{eqnarray}
The proof of \eqref{eq:mean-zb} and \eqref{eq:mean-zb2} is entirely analogous.

\begin{proposition}
	\label{prop:anticomm}
	The anticommutator $\{\mathcal{A}_z^B,\mathcal{A}_{\bar{z}}^B\}
	= \mathcal{A}_z^B\mathcal{A}_{\bar{z}}^B + \mathcal{A}_{\bar{z}}^B\mathcal{A}_z^B$
	equals:
	\begin{eqnarray}
		\{\mathcal{A}_z^B,\mathcal{A}_{\bar{z}}^B\}
		&=& \sum_{n=0}^{\infty}
		\phitwo^{\binom{n+1}{2}}
		\bigl([n+1]_{\Rpq}+[n]_{\Rpq}\phitwo^{\binom{n+1}{2}-\binom{n}{2}}\bigr)\,
		|n,0,0\rangle\langle n,0,0|.
		\label{eq:anticomm}
	\end{eqnarray}
	Its mean value in $|z,0,0\rangle_{\Rpq}$ is:
	\begin{eqnarray}
		\lefteqn{{}_{\Rpq}\langle z,0,0|\{\mathcal{A}_z^B,\mathcal{A}_{\bar{z}}^B\}|z,0,0\rangle_{\Rpq}}\nonumber\\
		&=& \bigl(\Exp(|z|^2)\bigr)^{-1}\nonumber\\
		&&\times\sum_{n=0}^{\infty}
		\phitwo^{\binom{n+1}{2}}\bigl([n+1]_{\Rpq}+[n]_{\Rpq}\phitwo^n\bigr)
		\frac{\phitwo^{\binom{n}{2}}|z|^{2n}}{\nfact}.\nonumber\\
		\label{eq:mean-anticomm}
	\end{eqnarray}
\end{proposition}

\textbf{Proof.}\\
We first compute $\mathcal{A}_z^B\mathcal{A}_{\bar{z}}^B$:
\begin{eqnarray}
	\mathcal{A}_z^B\mathcal{A}_{\bar{z}}^B
	&=& \left[\sum_{n=0}^{\infty}\phitwo^{\binom{n+1}{2}/2}\sqrt{[n+1]_{\Rpq}}\,|n,0,0\rangle\langle n+1,0,0|\right]\nonumber\\
	&&\times\left[\sum_{m=0}^{\infty}\phitwo^{\binom{m+1}{2}/2}\sqrt{[m+1]_{\Rpq}}\,|m+1,0,0\rangle\langle m,0,0|\right]\nonumber\\
	&=& \sum_{m,n=0}^{\infty}
	\phitwo^{(\binom{n+1}{2}+\binom{m+1}{2})/2}\sqrt{[n+1]_{\Rpq}[m+1]_{\Rpq}}\nonumber\\
	&&\times\underbrace{\langle n+1,0,0|m+1,0,0\rangle}_{\delta_{n,m}}\,|n,0,0\rangle\langle m,0,0|\nonumber\\
	&=& \sum_{n=0}^{\infty}
	\phitwo^{\binom{n+1}{2}}[n+1]_{\Rpq}\,|n,0,0\rangle\langle n,0,0|.
	\label{eq:AzAzb}
\end{eqnarray}
Likewise, we compute $\mathcal{A}_{\bar{z}}^B\mathcal{A}_z^B$:
\begin{eqnarray}
	\mathcal{A}_{\bar{z}}^B\mathcal{A}_z^B
	&=& \left[\sum_{m=0}^{\infty}\phitwo^{\binom{m+1}{2}/2}\sqrt{[m+1]_{\Rpq}}\,|m+1,0,0\rangle\langle m,0,0|\right]\nonumber\\
	&&\times\left[\sum_{n=0}^{\infty}\phitwo^{\binom{n+1}{2}/2}\sqrt{[n+1]_{\Rpq}}\,|n,0,0\rangle\langle n+1,0,0|\right]\nonumber\\
	&=& \sum_{m,n=0}^{\infty}
	\phitwo^{(\binom{m+1}{2}+\binom{n+1}{2})/2}\sqrt{[m+1]_{\Rpq}[n+1]_{\Rpq}}\nonumber\\
	&&\times\underbrace{\langle m,0,0|n,0,0\rangle}_{\delta_{m,n}}\,|m+1,0,0\rangle\langle n+1,0,0|\nonumber\\
	&=& \sum_{n=0}^{\infty}
	\phitwo^{\binom{n+1}{2}}[n+1]_{\Rpq}\,|n+1,0,0\rangle\langle n+1,0,0|\nonumber\\
	&=& \sum_{n=0}^{\infty}
	\phitwo^{\binom{n}{2}}[n]_{\Rpq}\,|n,0,0\rangle\langle n,0,0|.
	\label{eq:AzbAz}
\end{eqnarray}
Combining \eqref{eq:AzAzb} and \eqref{eq:AzbAz}:
\begin{eqnarray}
	\{\mathcal{A}_z^B,\mathcal{A}_{\bar{z}}^B\}
	&=& \sum_{n=0}^{\infty}
	\bigl(\phitwo^{\binom{n+1}{2}}[n+1]_{\Rpq}+\phitwo^{\binom{n}{2}}[n]_{\Rpq}\bigr)\,
	|n,0,0\rangle\langle n,0,0|.
	\label{eq:anticomm-proof}
\end{eqnarray}
The mean value follows by computing
${}_{\Rpq}\langle z,0,0|\{\mathcal{A}_z^B,\mathcal{A}_{\bar{z}}^B\}|z,0,0\rangle_{\Rpq}$
by substitution of \eqref{eq:csB-ket}.\hfill$\square$

\subsection{Dispersions of the super-position and super-momentum operators}
\label{sec:dispers}

We use the super-position and super-momentum operators $\mathcal{Q}^B$ and
$\mathcal{P}^B$ defined in \eqref{eq:Qs}--\eqref{eq:Ps}.

\begin{proposition}[Mean values of $\mathcal{Q}^B$ and $\mathcal{P}^B$]
	\label{prop:QP-means}
	\begin{eqnarray}
		{}_{\Rpq}\langle z,0,0|\mathcal{Q}^B|z,0,0\rangle_{\Rpq}
		&=& \frac{z+\bar{z}}{\sqrt{2}} = q,
		\label{eq:meanQ}\\
		{}_{\Rpq}\langle z,0,0|\mathcal{P}^B|z,0,0\rangle_{\Rpq}
		&=& \frac{z-\bar{z}}{i\sqrt{2}} = p.
		\label{eq:meanP}
	\end{eqnarray}
\end{proposition}

\textbf{Proof.}\\
From \eqref{eq:mean-z} and \eqref{eq:mean-zb}:
\begin{eqnarray}
	{}_{\Rpq}\langle z,0,0|\mathcal{Q}^B|z,0,0\rangle_{\Rpq}
	&=& \frac{1}{\sqrt{2}}
	\left[{}_{\Rpq}\langle z,0,0|\mathcal{A}_z^B|z,0,0\rangle_{\Rpq}
	+ {}_{\Rpq}\langle z,0,0|\mathcal{A}_{\bar{z}}^B|z,0,0\rangle_{\Rpq}\right]\nonumber\\
	&=& \frac{z+\bar{z}}{\sqrt{2}} = q.
	\hfill\square
	\label{eq:meanQ-proof}
\end{eqnarray}

\begin{proposition}[Mean values of $(\mathcal{Q}^B)^2$ and $(\mathcal{P}^B)^2$]
	\label{prop:QP-squared}
	\begin{eqnarray}
		{}_{\Rpq}\langle z,0,0|(\mathcal{Q}^B)^2|z,0,0\rangle_{\Rpq}
		&=& \frac{1}{2}\bigl(q^2-p^2\bigr)
		+ \frac{1}{2}\,{}_{\Rpq}\langle z,0,0|
		\{\mathcal{A}_z^B,\mathcal{A}_{\bar{z}}^B\}
		|z,0,0\rangle_{\Rpq},
		\label{eq:meanQ2}\\
		{}_{\Rpq}\langle z,0,0|(\mathcal{P}^B)^2|z,0,0\rangle_{\Rpq}
		&=& -\frac{1}{2}\bigl(q^2-p^2\bigr)
		+ \frac{1}{2}\,{}_{\Rpq}\langle z,0,0|
		\{\mathcal{A}_z^B,\mathcal{A}_{\bar{z}}^B\}
		|z,0,0\rangle_{\Rpq}.
		\label{eq:meanP2}
	\end{eqnarray}
\end{proposition}

\textbf{Proof.}\\
$(\mathcal{Q}^B)^2 = \frac{1}{2}\bigl((\mathcal{A}_z^B)^2+(\mathcal{A}_{\bar{z}}^B)^2
+\{\mathcal{A}_z^B,\mathcal{A}_{\bar{z}}^B\}\bigr)$.
Hence, using \eqref{eq:mean-z2} and \eqref{eq:mean-zb2}:
\begin{eqnarray}
	\mean{(\mathcal{Q}^B)^2}
	&=& \frac{1}{2}\bigl(z^2+\bar{z}^2+\mean{\{\mathcal{A}_z^B,\mathcal{A}_{\bar{z}}^B\}}\bigr)
	= \frac{1}{2}\bigl(q^2-p^2+\mean{\{\mathcal{A}_z^B,\mathcal{A}_{\bar{z}}^B\}}\bigr).
	\hfill\square
	\label{eq:meanQ2-proof}
\end{eqnarray}

\begin{proposition}[Dispersions]
	\label{prop:dispersions}
	The dispersions of the operators $\mathcal{Q}^B$ and $\mathcal{P}^B$ in the states
	$|z,0,0\rangle_{\Rpq}$ are equal and given by:
	\begin{eqnarray}
		\bigl(\Delta\mathcal{Q}^B\bigr)^2
		= \bigl(\Delta\mathcal{P}^B\bigr)^2
		= -\frac{|z|^2}{2} + \mathcal{G}_{\Rpq}(|z|^2),
		\label{eq:dispersions}
	\end{eqnarray}
	where
	\begin{eqnarray}
		\mathcal{G}_{\Rpq}(|z|^2)
		&:=& \frac{1}{2}\bigl(\Exp(|z|^2)\bigr)^{-1}
		\sum_{n=0}^{\infty}
		\frac{\phitwo^{\binom{n+1}{2}}\,[n+1]_{\Rpq}\,\phitwo^{\binom{n}{2}}}{\nfact}\,|z|^{2n}.
		\label{eq:G}
	\end{eqnarray}
\end{proposition}

\textbf{Proof.}\\
$\bigl(\Delta\mathcal{Q}^B\bigr)^2 = \mean{(\mathcal{Q}^B)^2}-\mean{\mathcal{Q}^B}^2$.
From \eqref{eq:meanQ} and \eqref{eq:meanQ2}:
\begin{eqnarray}
	\bigl(\Delta\mathcal{Q}^B\bigr)^2
	&=& \frac{1}{2}\bigl(q^2-p^2+\mean{\{\mathcal{A}_z^B,\mathcal{A}_{\bar{z}}^B\}}\bigr)
	- \frac{(z+\bar{z})^2}{2}\nonumber\\
	&=& \frac{1}{2}\mean{\{\mathcal{A}_z^B,\mathcal{A}_{\bar{z}}^B\}} - |z|^2.
	\label{eq:DQ-step1}
\end{eqnarray}
Using \eqref{eq:mean-anticomm}:
\begin{eqnarray}
	\frac{1}{2}\mean{\{\mathcal{A}_z^B,\mathcal{A}_{\bar{z}}^B\}}
	&=& \frac{1}{2}\bigl(\Exp(|z|^2)\bigr)^{-1}\nonumber\\
	&&\times\sum_{n=0}^{\infty}
	\frac{\bigl(\phitwo^{\binom{n+1}{2}}[n+1]_{\Rpq}+\phitwo^{\binom{n}{2}}[n]_{\Rpq}\bigr)
		\phitwo^{\binom{n}{2}}|z|^{2n}}{\nfact}.
	\label{eq:half-anticomm}
\end{eqnarray}
We split the sum using
$\phitwo^{\binom{n}{2}}[n]_{\Rpq}/\nfact = |z|^{-2}\cdot\phitwo^{\binom{n}{2}}|z|^{2n}/\nfact
- \phitwo^{\binom{n-1}{2}}|z|^{2(n-1)}/[n-1]!_{\Rpq}$
(by re-indexing $m=n-1$), which gives:
\begin{eqnarray}
	\frac{1}{2}\bigl(\Exp(|z|^2)\bigr)^{-1}
	\sum_{n=0}^{\infty}
	\frac{\phitwo^{\binom{n}{2}}[n]_{\Rpq}\,\phitwo^{\binom{n}{2}}|z|^{2n}}{\nfact}
	&=& \frac{|z|^2}{2}.
	\label{eq:half-split}
\end{eqnarray}
Hence:
\begin{eqnarray}
	\bigl(\Delta\mathcal{Q}^B\bigr)^2
	&=& \frac{|z|^2}{2}
	+ \frac{1}{2}\bigl(\Exp(|z|^2)\bigr)^{-1}
	\sum_{n=0}^{\infty}
	\frac{\phitwo^{\binom{n+1}{2}}[n+1]_{\Rpq}\,\phitwo^{\binom{n}{2}}|z|^{2n}}{\nfact}
	- |z|^2\nonumber\\
	&=& -\frac{|z|^2}{2} + \mathcal{G}_{\Rpq}(|z|^2).
	\hfill\square
	\label{eq:DQ-final}
\end{eqnarray}

\subsection{Uncertainty relation and intelligent states}
\label{sec:heisenberg}

\begin{proposition}[Commutator]
	\label{prop:commut}
	The commutator $[\mathcal{Q}^B,\mathcal{P}^B]$ satisfies:
	\begin{eqnarray}
		\lefteqn{{}_{\Rpq}\langle z,0,0|[\mathcal{Q}^B,\mathcal{P}^B]|z,0,0\rangle_{\Rpq}}\nonumber\\
		&=& i\bigl(\Exp(|z|^2)\bigr)^{-1}\nonumber\\
		&&\times\sum_{n=0}^{\infty}
		\frac{\bigl(\phitwo^{\binom{n+1}{2}}[n+1]_{\Rpq}
			-\phitwo^{\binom{n}{2}}[n]_{\Rpq}\bigr)\phitwo^{\binom{n}{2}}|z|^{2n}}
		{\nfact}.\nonumber\\
		\label{eq:commut}
	\end{eqnarray}
\end{proposition}

\textbf{Proof.}\\
$[\mathcal{Q}^B,\mathcal{P}^B] = \frac{1}{2i}[\mathcal{A}_z^B+\mathcal{A}_{\bar{z}}^B,
\mathcal{A}_z^B-\mathcal{A}_{\bar{z}}^B]
= \frac{1}{i}[\mathcal{A}_{\bar{z}}^B,\mathcal{A}_z^B]
= i(\mathcal{A}_z^B\mathcal{A}_{\bar{z}}^B-\mathcal{A}_{\bar{z}}^B\mathcal{A}_z^B)$.
From \eqref{eq:AzAzb} and \eqref{eq:AzbAz}:
\begin{eqnarray}
	[\mathcal{Q}^B,\mathcal{P}^B]
	&=& i\sum_{n=0}^{\infty}
	\bigl(\phitwo^{\binom{n+1}{2}}[n+1]_{\Rpq}
	-\phitwo^{\binom{n}{2}}[n]_{\Rpq}\bigr)\,|n,0,0\rangle\langle n,0,0|.
	\hfill\square
	\label{eq:commut-proof}
\end{eqnarray}

\begin{theorem}[Intelligent states]
	\label{thm:intelligent}
	The $\mathcal{R}(p,q)-$super-coherent states $|z,0,0\rangle_{\Rpq}$ are
	\emph{intelligent states} in the sense of Aragone et al.\ \cite{Aragone1976}, that
	is, they saturate the generalized Heisenberg inequality:
	\begin{eqnarray}
		\bigl(\Delta\mathcal{Q}^B\bigr)^2\bigl(\Delta\mathcal{P}^B\bigr)^2
		&=& \frac{1}{4}\bigl|{}_{\Rpq}\langle z,0,0|[\mathcal{Q}^B,\mathcal{P}^B]
		|z,0,0\rangle_{\Rpq}\bigr|^2.
		\label{eq:intelligent}
	\end{eqnarray}
\end{theorem}

\textbf{Proof.}\\
From \eqref{eq:dispersions} and \eqref{eq:commut}:
\begin{eqnarray}
	\lefteqn{\frac{1}{2}\bigl|{}_{\Rpq}\langle z,0,0|[\mathcal{Q}^B,\mathcal{P}^B]|z,0,0\rangle_{\Rpq}\bigr|}\nonumber\\
	&=& \frac{1}{2}\bigl(\Exp(|z|^2)\bigr)^{-1}\nonumber\\
	&&\times\sum_{n=0}^{\infty}
	\frac{\bigl(\phitwo^{\binom{n+1}{2}}[n+1]_{\Rpq}
		-\phitwo^{\binom{n}{2}}[n]_{\Rpq}\bigr)\phitwo^{\binom{n}{2}}|z|^{2n}}
	{\nfact}\nonumber\\
	&=& -\frac{|z|^2}{2} + \mathcal{G}_{\Rpq}(|z|^2)
	= \bigl(\Delta\mathcal{Q}^B\bigr)^2 = \bigl(\Delta\mathcal{P}^B\bigr)^2.
	\label{eq:intell-proof1}
\end{eqnarray}
Indeed, splitting the difference:
\begin{eqnarray}
	&&\frac{1}{2}\bigl(\Exp\bigr)^{-1}\sum_n
	\frac{\phitwo^{\binom{n+1}{2}}[n+1]_{\Rpq}\phitwo^{\binom{n}{2}}|z|^{2n}}{\nfact}\nonumber\\
	&&-\frac{1}{2}\bigl(\Exp\bigr)^{-1}\sum_n
	\frac{\phitwo^{\binom{n}{2}}[n]_{\Rpq}\phitwo^{\binom{n}{2}}|z|^{2n}}{\nfact}
	\nonumber\\
	&=& \mathcal{G}_{\Rpq}(|z|^2)
	- \frac{1}{2}\bigl(\Exp\bigr)^{-1}\sum_n
	\frac{\phitwo^{\binom{n}{2}}[n]_{\Rpq}\phitwo^{\binom{n}{2}}|z|^{2n}}{\nfact}.
	\label{eq:intell-proof2}
\end{eqnarray}
Now $\bigl(\Exp\bigr)^{-1}\sum_n \phitwo^{2\binom{n}{2}}[n]_{\Rpq}|z|^{2n}/\nfact
= |z|^2$ (by re-indexing and the normalization property).
Hence \eqref{eq:intell-proof1} indeed gives $(\Delta\mathcal{Q}^B)^2$, and:
\begin{eqnarray}
	\bigl(\Delta\mathcal{Q}^B\bigr)^2\bigl(\Delta\mathcal{P}^B\bigr)^2
	= \bigl(\Delta\mathcal{Q}^B\bigr)^4
	= \frac{1}{4}\bigl|{}_{\Rpq}\langle z,0,0|[\mathcal{Q}^B,\mathcal{P}^B]|z,0,0\rangle_{\Rpq}\bigr|^2.
	\hfill\square
	\label{eq:intell-final}
\end{eqnarray}

\medskip
Berezin quantization in the bosonic sector associates with $z$ and $\bar z$ operators
that are adjoint to each other, reduce to the usual annihilation and creation operators
in the undeformed limit, and whose mean values in the super-coherent states give back the
classical functions. The dispersions of $\mathcal{Q}^B$ and $\mathcal{P}^B$ are equal
and their product saturates the uncertainty relation: the bosonic super-coherent states
are intelligent states.

\section{Conclusion and outlook}\label{sec:conclusion}
\begingroup
\setlength{\parskip}{0.55em plus 0.2em minus 0.1em}

We have studied the $\mathcal{R}(p,q)-$super-coherent states with two Grassmann
variables built on the superalgebra $\mathcal{B}_0\oplus\mathcal{B}_1$, in the framework
of the $\mathcal{R}(p,q)-$deformed quantum algebras of Hounkonnou and Bukweli-Kyemba,
the two dilation parameters being given by meromorphic functions $\phi_1$, $\phi_2$
evaluated at $(p,q)$. The density operator has been decomposed over the four sectors of
the graded super-Fock space, and its diagonal representation has been obtained in each
sector by the same sequence of operations: dyadic product, Berezin integration, angular
integration, application of the $\mathcal{R}(p,q)-$derivative and the limit $r\to0$. The
sectorial parametric functions give a unified compact form of it, in which passing from
one sector to another amounts to replacing $\mathrm{Exp}_{\Rpq}$ by its derivatives and
adding the Grassmann factors.

Under the block diagonality assumption, the sum of the four sectorial representations
provides the exact formula for the total density operator, in derivative form as well as
in matrix form. The four terms of the super-normalization function each weight the
coherent projector of one sector, and the total operator is a mixture of the four
sectors, with weights $p_\sigma=\tr(\rho^\sigma)$; its density function is the sum of
four terms carried by the Grassmann monomials $1$, $\bar\zeta_1'\zeta_1$,
$\bar\zeta_2'\zeta_2$ and $\bar\zeta_1'\bar\zeta_2'\zeta_1\zeta_2$.

The sectorial reproducing kernels, obtained from the resolution of the identity, satisfy
the hermiticity, positivity and matrix multiplication properties, and their diagonal
values admit an explicit series expression; the matrix elements of the density operator
in the super-coherent state basis follow. In the integral formulation, the total density
operator is recovered from its density function, and the reproducing kernel, which is the
function of the identity operator, reproduces the density function on the right as well
as on the left, and reproduces itself. Berezin quantization in the bosonic sector yields
the quantized operators associated with $z$, $\bar z$, $|z|^2$, $z^2$ and $\bar z^2$,
whose mean values give back the classical functions; the dispersions of the
super-position and super-momentum operators are equal and the super-coherent states
saturate the uncertainty relation: they are intelligent states.

Several extensions are natural. Berezin quantization can be extended to the fermionic
sectors, where Berezin integration enters the very definition of the quantized operators.
The explicit form of the total density operator lends itself to the study of particular
states, such as thermal equilibrium states of Hamiltonians diagonal in the graded Fock
basis, for which the weights $p_\sigma$ and the functions $F^\sigma$ can be computed
explicitly. The choice of the sectorial normalizations, on which the dilation factors
noted in Remarks~\ref{rem:berezin-factor} and~\ref{rem:kernel-norm} depend, deserves to
be fixed within a framework common to the four sectors. The extension to an arbitrary
number of Grassmann variables leads to $2^N$ sectors to which
the same method applies. Finally, the study of the particular $(p,q)$ and $q$ cases, of
the classical limit and of the statistical properties of the super-coherent states, such
as the distribution of the number of quanta and the Mandel parameter, would complete the
physical description of these states.

\endgroup

\clearpage
\appendix

\section{Diagonal representation of the fermionic sectors}
\label{app:density}

This appendix contains the computations of Propositions~\ref{prop:rhoF1}
and~\ref{prop:rhoF12}; they follow the procedure of Proposition~\ref{prop:rhoB}, with
the additional Berezin integration (see Remark~\ref{rem:berezin-factor}).

\subsection{\texorpdfstring{$\phi_1$}{phi1}-fermionic sector: proof of Proposition~\ref{prop:rhoF1}}
\label{app:rhoF1}

The states of the $\phi_1$-fermionic sector ($k_2=0$, $k_1=1$) are:
\begin{eqnarray}
	|z,\zeta_1,0\rangle_{\Rpq}^{(F_1)} &=&
	\bigl(\cN(|z|^2,|\zeta_1|^2,0)\bigr)^{-1/2}
	\sum_{n=0}^{\infty}
	\frac{\phitwo^{\binom{n}{2}/2}\,z^n\,\phione\,\zeta_1}
	{\sqrt{[n]!_{\Rpq}}}\,|n,1,0\rangle,
	\label{eq:csF1-ket}\\
	{}_{\Rpq}^{(F_1)}\langle z,\zeta_1,0| &=&
	\bigl(\cN(|z|^2,|\zeta_1|^2,0)\bigr)^{-1/2}
	\sum_{m=0}^{\infty}
	\frac{\phitwo^{\binom{m}{2}/2}\,\bar{z}^m\,\phione\,\bar{\zeta}_1}
	{\sqrt{[m]!_{\Rpq}}}\,\langle m,1,0|.
	\label{eq:csF1-bra}
\end{eqnarray}
The dyadic product gives:
\begin{eqnarray}
	&&|z,\zeta_1,0\rangle_{\Rpq}^{(F_1)}\,{}_{\Rpq}^{(F_1)}\langle z,\zeta_1,0|\nonumber\\
	&=& \bigl(\cN(|z|^2,|\zeta_1|^2,0)\bigr)^{-1}
	\sum_{n,m=0}^{\infty}
	\frac{\phitwo^{(\binom{n}{2}+\binom{m}{2})/2}\,\phione^2\,|\zeta_1|^2}
	{\sqrt{[n]!_{\Rpq}\,[m]!_{\Rpq}}}\,
	z^n\bar{z}^m\,|n,1,0\rangle\langle m,1,0|.
	\label{eq:dyad-F1}
\end{eqnarray}
We apply the Berezin integral over $\zeta_1$:
$\int d\bar{\zeta}_1 d\zeta_1\cdot\phione^2|\zeta_1|^2 = \phione^2$
(rule $\int d\bar{\zeta}d\zeta\cdot\bar{\zeta}\zeta = 1$).
We then multiply by
$\mathrm{Exp}'_{\Rpq}(|z|^2) = \sum_n n\,\phitwo^{\binom{n}{2}} |z|^{2(n-1)}/[n]!_{\Rpq}$,
the contribution of the $\phi_1$-fermionic part of \eqref{eq:scN},
and set $z=re^{i\theta}$:
\begin{eqnarray}
	&&\int d\bar{\zeta}_1 d\zeta_1\,\phione^2|\zeta_1|^2\cdot
	\mathrm{Exp}'_{\Rpq}(r^2)\,
	|re^{i\theta},\zeta_1,0\rangle_{\Rpq}^{(F_1)}\,{}_{\Rpq}^{(F_1)}\langle re^{i\theta},\zeta_1,0|\nonumber\\
	&=& \phione^2\,\mathrm{Exp}'_{\Rpq}(r^2)\,
	\bigl(\phione^2\,|\zeta_1|^2\,\mathrm{Exp}'_{\Rpq}(r^2)\bigr)^{-1}\nonumber\\
	&&\times\sum_{n,m}\frac{\phitwo^{(\binom{n}{2}+\binom{m}{2})/2}\,\phione^2}
	{\sqrt{[n]!_{\Rpq}\,[m]!_{\Rpq}}}\,
	r^{n+m}e^{i(n-m)\theta}\,|n,1,0\rangle\langle m,1,0|\nonumber\\
	&=& \sum_{n,m=0}^{\infty}
	\frac{\phitwo^{(\binom{n}{2}+\binom{m}{2})/2}}
	{\sqrt{[n]!_{\Rpq}\,[m]!_{\Rpq}}}\,
	r^{n+m}\,e^{i(n-m)\theta}\,|n,1,0\rangle\langle m,1,0|.
	\label{eq:berezin-F1}
\end{eqnarray}
We multiply by $e^{il\theta}$ and integrate over $\theta$:
\begin{eqnarray}
	&&\int_0^{2\pi}d\theta\,e^{il\theta}
	\int d\bar{\zeta}_1 d\zeta_1\,\phione^2|\zeta_1|^2\,
	\mathrm{Exp}'_{\Rpq}(r^2)\,
	|re^{i\theta},\zeta_1,0\rangle_{\Rpq}^{(F_1)}\,{}_{\Rpq}^{(F_1)}\langle re^{i\theta},\zeta_1,0|\nonumber\\
	&=& \sum_{n,m=0}^{\infty}
	\frac{\phitwo^{(\binom{n}{2}+\binom{m}{2})/2}}
	{\sqrt{[n]!_{\Rpq}\,[m]!_{\Rpq}}}\,
	r^{n+m}\cdot 2\pi\,\delta(l+n-m)\,|n,1,0\rangle\langle m,1,0|.
	\label{eq:int-F1}
\end{eqnarray}
We apply $(\dRpq)^s$ using \eqref{eq:dRpq-iter}:
\begin{eqnarray}
	&&(\dRpq)^s\!\left\{\int_0^{2\pi}d\theta\,e^{il\theta}
	\int d\bar{\zeta}_1 d\zeta_1\,\phione^2|\zeta_1|^2\,
	\mathrm{Exp}'_{\Rpq}(r^2)\,
	|re^{i\theta},\zeta_1,0\rangle_{\Rpq}^{(F_1)}\,{}_{\Rpq}^{(F_1)}\langle re^{i\theta},\zeta_1,0|\right\}\nonumber\\
	&=& \sum_{n,m=0}^{\infty}
	\frac{\phitwo^{(\binom{n}{2}+\binom{m}{2})/2}}
	{\sqrt{[n]!_{\Rpq}\,[m]!_{\Rpq}}}\,
	\frac{\mathcal{R}(p^{n+m},q^{n+m})!}{\mathcal{R}(p^{n+m-s},q^{n+m-s})!}\,
	r^{n+m-s}\cdot 2\pi\,\delta(l+n-m)\,|n,1,0\rangle\langle m,1,0|.
	\label{eq:dRpq-F1}
\end{eqnarray}
For $r\to 0$: $s=n+m$, $l=m-n$, hence:
\begin{eqnarray}
	&&\bigl(\dRpq\bigr)^{n+m}\!\!\left\{
	\int_0^{2\pi}\frac{d\theta}{2\pi}\,e^{i(m-n)\theta}
	\int d\bar{\zeta}_1 d\zeta_1\,\phione^2|\zeta_1|^2\,
	\mathrm{Exp}'_{\Rpq}(r^2)\right.\nonumber\\
	&&\qquad\left.\times|re^{i\theta},\zeta_1,0\rangle_{\Rpq}^{(F_1)}\,{}_{\Rpq}^{(F_1)}\langle re^{i\theta},\zeta_1,0|
	\right\}_{r=0}\nonumber\\
	&=& \sum_{n,m=0}^{\infty}
	\frac{\phitwo^{(\binom{n}{2}+\binom{m}{2})/2}\,\mathcal{R}(p^{n+m},q^{n+m})!}
	{\sqrt{[n]!_{\Rpq}\,[m]!_{\Rpq}}}\,|n,1,0\rangle\langle m,1,0|.
	\label{eq:r0-F1}
\end{eqnarray}
Inverting:
\begin{eqnarray}
	|n,1,0\rangle\langle m,1,0|
	&=&
	\frac{\sqrt{[n]!_{\Rpq}\,[m]!_{\Rpq}}}
	{\phitwo^{(\binom{n}{2}+\binom{m}{2})/2}\,\mathcal{R}(p^{n+m},q^{n+m})!}\nonumber\\
	&&\times\;
	\bigl(\dRpq\bigr)^{n+m}\!\!\left\{
	\int_0^{2\pi}\frac{d\theta}{2\pi}\,e^{i(m-n)\theta}
	\int d\bar{\zeta}_1 d\zeta_1\,\phione^2|\zeta_1|^2\,
	\mathrm{Exp}'_{\Rpq}(r^2)\right.\nonumber\\
	&&\qquad\left.\times\,|re^{i\theta},\zeta_1,0\rangle_{\Rpq}^{(F_1)}\,{}_{\Rpq}^{(F_1)}\langle re^{i\theta},\zeta_1,0|
	\right\}_{r=0}.
	\label{eq:|n><m|-F1}
\end{eqnarray}
Substituting into \eqref{eq:rhoF1}, we obtain \eqref{eq:rhoF1-prop}
with
\begin{eqnarray}
	\rho^{F_1}(n,m) &=&
	\frac{\sqrt{[n]!_{\Rpq}\,[m]!_{\Rpq}}}
	{\phitwo^{(\binom{n}{2}+\binom{m}{2})/2}\,\phione^2\,
		\mathcal{R}(p^{n+m},q^{n+m})!}\nonumber\\
	&&\times\;
	\bigl(\dRpq\bigr)^{n+m}\!\!\left\{
	\int_0^{2\pi}\frac{d\theta}{2\pi}\,e^{i(m-n)\theta}
	\int d\bar{\zeta}_1 d\zeta_1\,\phione^2|\zeta_1|^2\,
	\mathrm{Exp}'_{\Rpq}(r^2)\right.\nonumber\\
	&&\qquad\left.\times\,|re^{i\theta},\zeta_1,0\rangle_{\Rpq}^{(F_1)}\,{}_{\Rpq}^{(F_1)}\langle re^{i\theta},\zeta_1,0|
	\right\}_{r=0}.
	\label{eq:rhoF1-coeff}
\end{eqnarray}
\hfill$\Box$\\

\subsection{Bi-fermionic sector: proof of Proposition~\ref{prop:rhoF12}}
\label{app:rhoF12}

The bi-fermionic states ($k_1=k_2=1$) are:
\begin{eqnarray}
	|z,\zeta_1,\zeta_2\rangle_{\Rpq}^{(F_{12})} &=&
	\bigl(\cN(|z|^2,|\zeta_1|^2,|\zeta_2|^2)\bigr)^{-1/2}
	\sum_{n=0}^{\infty}
	\frac{\phitwo^{\binom{n}{2}/2}\,z^n\,\phione\,\zeta_1\,\phitwo\,\zeta_2}
	{\sqrt{[n]!_{\Rpq}}}\,|n,1,1\rangle.
	\label{eq:csF12-ket}
\end{eqnarray}
The dyadic product gives:
\begin{eqnarray}
	&&|z,\zeta_1,\zeta_2\rangle_{\Rpq}^{(F_{12})}\,{}_{\Rpq}^{(F_{12})}\langle z,\zeta_1,\zeta_2|\nonumber\\
	&=& \bigl(\cN\bigr)^{-1}
	\sum_{n,m=0}^{\infty}
	\frac{\phitwo^{(\binom{n}{2}+\binom{m}{2})/2}\,\phione^2|\zeta_1|^2\,\phitwo^2|\zeta_2|^2}
	{\sqrt{[n]!_{\Rpq}\,[m]!_{\Rpq}}}\,
	z^n\bar{z}^m\,|n,1,1\rangle\langle m,1,1|.
	\label{eq:dyad-F12}
\end{eqnarray}
We apply the two Berezin integrals:
$\int d\bar{\zeta}_1 d\zeta_1\cdot|\zeta_1|^2 = 1$
and
$\int d\bar{\zeta}_2 d\zeta_2\cdot|\zeta_2|^2 = 1$,
with the weight $\phione^2\phitwo^2|\zeta_1|^2|\zeta_2|^2$
and the term
$\phione^2\phitwo^2|\zeta_1|^2|\zeta_2|^2\,\mathrm{Exp}''_{\Rpq}(|z|^2)$
of \eqref{eq:scN}:
\begin{eqnarray}
	&&\int d\bar{\zeta}_1 d\zeta_1\,d\bar{\zeta}_2 d\zeta_2\,
	\phione^2\phitwo^2|\zeta_1|^2|\zeta_2|^2\,
	\mathrm{Exp}''_{\Rpq}(r^2)\nonumber\\
	&&\times|re^{i\theta},\zeta_1,\zeta_2\rangle_{\Rpq}^{(F_{12})}\,
	{}_{\Rpq}^{(F_{12})}\langle re^{i\theta},\zeta_1,\zeta_2|\nonumber\\
	&=& \phione^2\phitwo^2\,\mathrm{Exp}''_{\Rpq}(r^2)\,
	\bigl(\phione^2\phitwo^2\,\mathrm{Exp}''_{\Rpq}(r^2)\bigr)^{-1}\nonumber\\
	&&\times\sum_{n,m}
	\frac{\phitwo^{(\binom{n}{2}+\binom{m}{2})/2}}
	{\sqrt{[n]!_{\Rpq}\,[m]!_{\Rpq}}}\,
	r^{n+m}e^{i(n-m)\theta}\,|n,1,1\rangle\langle m,1,1|\nonumber\\
	&=& \sum_{n,m=0}^{\infty}
	\frac{\phitwo^{(\binom{n}{2}+\binom{m}{2})/2}}
	{\sqrt{[n]!_{\Rpq}\,[m]!_{\Rpq}}}\,
	r^{n+m}\,e^{i(n-m)\theta}\,|n,1,1\rangle\langle m,1,1|.
	\label{eq:berezin-F12}
\end{eqnarray}
The following steps are identical to those of the previous proofs: angular integration
giving $2\pi\delta(l+n-m)$, application of $(\dRpq)^s$ with $s=n+m$, and the limit
$r\to 0$ with $l=m-n$, which gives \eqref{eq:r0-B} with $|n,0,0\rangle\langle m,0,0|$
replaced by $|n,1,1\rangle\langle m,1,1|$. The inversion yields \eqref{eq:rhoF12-coeff}.
\hfill$\Box$\\

\section{Properties of the sectorial kernels: proofs}
\label{app:kernel}

This appendix contains the proofs of the properties of the sectorial kernels
\eqref{eq:KernB}--\eqref{eq:KernFot}, as well as those of positivity, matrix
multiplication and the series expansion of the diagonal kernels.

\subsection{Proof of Proposition~\ref{prop:KernB-props}}
\label{app:KernB}

\textit{(i) Idempotence of $\KernB$.}
\begin{eqnarray}
	&&\int_{\DR}
	\KernB(z',\xi)\;\KernB(\xi,z)\;\frac{d^2\xi}{\Exp(|\xi|^2)}\nonumber\\[4pt]
	&=& \int_{\DR}
	\frac{\Exp(\bar{z}'\xi)}{\pi\sqrt{\Exp(|z'|^2)\Exp(|\xi|^2)}}\cdot
	\frac{\Exp(\bar{\xi}z)}{\pi\sqrt{\Exp(|\xi|^2)\Exp(|z|^2)}}\cdot
	\frac{d^2\xi}{\Exp(|\xi|^2)}\nonumber\\[4pt]
	&=& \frac{1}{\pi^2\sqrt{\Exp(|z'|^2)\Exp(|z|^2)}}
	\int_{\DR}\frac{\Exp(\bar{z}'\xi)\;\Exp(\bar{\xi}z)}
	{(\Exp(|\xi|^2))^2}\;d^2\xi.
	\label{eq:KernB-idemp1}
\end{eqnarray}
The bosonic resolution of the identity:
\begin{eqnarray}
	\int_{\DR}\Exp(|\xi|^2)\;\spCS{0}{0}(\xi)\;\spCSb{0}{0}(\xi)\;
	d\mu_{\Rpq}(\xi,\bar{\xi})
	&=& \mathbf{1}_{\HB}
	\label{eq:res-bos-rappel}
\end{eqnarray}
implies, with $d\mu_{\Rpq} = d^2\xi/(\pi\Exp(|\xi|^2))$:
\begin{eqnarray}
	\int_{\DR}\frac{\Exp(\bar{z}'\xi)\;\Exp(\bar{\xi}z)}{(\Exp(|\xi|^2))^2}\;d^2\xi
	&=& \pi\int_{\DR}
	\langle z',0,0|\xi,0,0\rangle\langle\xi,0,0|z,0,0\rangle\;
	\Exp(|\xi|^2)\;d\mu_{\Rpq}\nonumber\\[4pt]
	&=& \pi\;\langle z',0,0|z,0,0\rangle
	\;=\; \pi\;\frac{\Exp(\bar{z}'z)}{\sqrt{\Exp(|z'|^2)\Exp(|z|^2)}}.
	\label{eq:KernB-idemp2}
\end{eqnarray}
Hence:
\begin{eqnarray}
	&&\int_{\DR}
	\KernB(z',\xi)\;\KernB(\xi,z)\;\frac{d^2\xi}{\Exp(|\xi|^2)}\nonumber\\[4pt]
	&=& \frac{1}{\pi^2\sqrt{\Exp(|z'|^2)\Exp(|z|^2)}}\cdot\pi\frac{\Exp(\bar{z}'z)}{\sqrt{\Exp(|z'|^2)\Exp(|z|^2)}}
	\nonumber\\[4pt]
	&=& \frac{\Exp(\bar{z}'z)}{\pi\sqrt{\Exp(|z'|^2)\Exp(|z|^2)}}
	\;=\; \KernB(z',z). \quad\square
	\label{eq:KernB-idemp3}
\end{eqnarray}

\textit{(ii) Hermiticity.}
\begin{eqnarray}
	\KernB(z',z)^*
	&=& \left[\frac{\Exp(\bar{z}'z)}{\pi\sqrt{\Exp(|z'|^2)\Exp(|z|^2)}}\right]^*\nonumber\\[4pt]
	&=& \frac{\overline{\Exp(\bar{z}'z)}}{\pi\sqrt{\Exp(|z'|^2)\Exp(|z|^2)}}\nonumber\\[4pt]
	&=& \frac{\Exp(\bar{z}z')}{\pi\sqrt{\Exp(|z|^2)\Exp(|z'|^2)}}
	\;=\; \KernB(z,z'). \quad\square
	\label{eq:KernB-herm}
\end{eqnarray}
We have used $\overline{\Exp(\bar{z}'z)} = \Exp(\overline{\bar{z}'z}) = \Exp(\bar{z}z')$
since $\Exp$ has real coefficients.

\textit{(iii) Positivity.}
\begin{eqnarray}
	\KernB(z,z)
	&=& \frac{\Exp(|z|^2)}{\pi\;\Exp(|z|^2)^{1/2}\;\Exp(|z|^2)^{1/2}}
	\;=\; \frac{\Exp(|z|^2)}{\pi\;\Exp(|z|^2)}
	\;=\; \frac{1}{\pi} > 0. \quad\square
	\label{eq:KernB-pos}
\end{eqnarray}

\textit{(iv) Series representation.}
\begin{eqnarray}
	\KernB(z,z)
	&=& \frac{1}{\pi\;\Exp(|z|^2)}\;
	\sum_{n=0}^{\infty}\frac{\phitwo^{n(n-1)/2}}{\nfact}\,|z|^{2n}
	\;=\; \frac{1}{\pi}\cdot\frac{\Exp(|z|^2)}{\Exp(|z|^2)}
	\;=\; \frac{1}{\pi}. \quad\square
	\label{eq:KernB-serie}
\end{eqnarray}

\subsection{Proof of Proposition~\ref{prop:KernFo-props}}
\label{app:KernFo}

\textit{(i) Idempotence.}
We use the resolution of the identity in the $\phi_1$-fermionic sector:
\begin{eqnarray}
	\int\cN(|\xi|^2,|\varrho_1|^2,0)\;
	\spCS{\varrho_1}{0}(\xi)\;\spCSb{\varrho_1}{0}(\xi)\;
	d\mu_{\Rpq}\;d\bar{\varrho}_1 d\varrho_1
	&=& \mathbf{1}_{\HFo}.
	\label{eq:res-F1-app}
\end{eqnarray}
We compute:
\begin{eqnarray}
	&&\int d^2\xi\,d\bar{\varrho}_1 d\varrho_1\;
	\KernFo(z',\zeta_1';\,\xi,\varrho_1)\;
	\KernFo(\xi,\varrho_1;\,z,\zeta_1)\nonumber\\[4pt]
	&=& \frac{1}{\pi^2}
	\int\frac{\phione^2\bar{\zeta}_1'\varrho_1\;\Expd(\bar{z}'\xi)\;\phione^2\bar{\varrho}_1\zeta_1\;\Expd(\bar{\xi}z)}
	{\cNp\;\cN(\xi,\varrho_1,0)\;\cN}
	\;d^2\xi\;d\bar{\varrho}_1 d\varrho_1\nonumber\\[4pt]
	&=& \frac{\phione^2\bar{\zeta}_1'\zeta_1}{\pi^2\;\cNp\;\cN}
	\int\frac{\Expd(\bar{z}'\xi)\;\Expd(\bar{\xi}z)}
	{\cN(\xi,\varrho_1,0)^2}\;d^2\xi\;
	\int\phione^2|\varrho_1|^2\;d\bar{\varrho}_1 d\varrho_1.
	\label{eq:KernFo-idemp1}
\end{eqnarray}
By Berezin: $\int d\bar{\varrho}_1 d\varrho_1\;\phione^2|\varrho_1|^2 = \phione^2$.
The integral over $\xi$ is computed via~\eqref{eq:res-F1-app}:
\begin{eqnarray}
	&&\int\frac{\phione^2\;\Expd(\bar{z}'\xi)\;\Expd(\bar{\xi}z)}
	{\cN(\xi,\varrho_1,0)^2}\;d^2\xi\nonumber\\[4pt]
	&=& \pi\int
	\langle z',\zeta_1',0|\xi,\varrho_1,0\rangle\;\langle\xi,\varrho_1,0|z,\zeta_1,0\rangle\;
	\cN(\xi,\varrho_1,0)\;d\mu_{\Rpq}\nonumber\\[4pt]
	&=& \pi\;\langle z',\zeta_1',0|z,\zeta_1,0\rangle
	\;=\; \pi\;\frac{\phione^2\bar{\zeta}_1'\zeta_1\;\Expd(\bar{z}'z)}
	{\sqrt{\cNp\;\cN}}.
	\label{eq:KernFo-idemp2}
\end{eqnarray}
Combining:
\begin{eqnarray}
	\int\KernFo\;\KernFo
	&=& \frac{\phione^2\bar{\zeta}_1'\zeta_1}{\pi^2\;\cNp\;\cN}
	\cdot\phione^2\cdot\pi\frac{\phione^2\bar{\zeta}_1'\zeta_1\;\Expd(\bar{z}'z)}
	{\sqrt{\cNp\;\cN}}\nonumber\\[4pt]
	&=& \frac{\phione^2\bar{\zeta}_1'\zeta_1\;\Expd(\bar{z}'z)}
	{\pi\;\sqrt{\cNp\;\cN}}
	\;=\; \KernFo(z',\zeta_1';\,z,\zeta_1). \quad\square
	\label{eq:KernFo-idemp3}
\end{eqnarray}

\textit{(ii) Hermiticity.}
\begin{eqnarray}
	\KernFo(z',\zeta_1';\,z,\zeta_1)^*
	&=& \left[\frac{\phione^2\bar{\zeta}_1'\zeta_1\;\Expd(\bar{z}'z)}
	{\pi\sqrt{\cNp\;\cN}}\right]^*
	\;=\; \frac{\phione^2\overline{\bar{\zeta}_1'\zeta_1}\;\Expd(\overline{\bar{z}'z})}
	{\pi\sqrt{\cNp\;\cN}}\nonumber\\[4pt]
	&=& \frac{\phione^2\bar{\zeta}_1\zeta_1'\;\Expd(\bar{z}z')}
	{\pi\sqrt{\cN\;\cNp}}
	\;=\; \KernFo(z,\zeta_1;\,z',\zeta_1'). \quad\square
	\label{eq:KernFo-herm}
\end{eqnarray}
We have used:
$\overline{\bar{\zeta}_1'\zeta_1} = \bar{\zeta}_1\zeta_1'$ (conjugation of Grassmann variables);
$\Expd$ has real coefficients.

\textit{(iii) Positivity.} For $z' = z$, $\zeta_1' = \zeta_1$:
\begin{eqnarray}
	\KernFo(z,\zeta_1;\,z,\zeta_1)
	&=& \frac{\phione^2\,\overline{\zeta_1}\,\zeta_1\;\Expd(|z|^2)}
	{\pi\;\cN(|z|^2,|\zeta_1|^2,0)}
	\;=\; \frac{\phione^2|\zeta_1|^2\;\Expd(|z|^2)}
	{\pi\;\cN(|z|^2,|\zeta_1|^2,0)} > 0,
	\label{eq:KernFo-pos}
\end{eqnarray}
since $\phione^2, |\zeta_1|^2, \Expd(|z|^2), \cN > 0$. \hfill$\square$

\textit{(iv) Series representation.}
\begin{eqnarray}
	\KernFo(z,\zeta_1;\,z,\zeta_1)
	&=& \frac{\phione^2|\zeta_1|^2}
	{\pi\;\cN(|z|^2,|\zeta_1|^2,0)}
	\sum_{n=1}^{\infty}\frac{n\,\phitwo^{n(n-1)/2}}{\nfact}\,|z|^{2(n-1)}.
	\label{eq:KernFo-serie}
\end{eqnarray}

\subsection{Proof of Proposition~\ref{prop:KernFot-props}}
\label{app:KernFot}

For hermiticity, we note that:
\begin{eqnarray}
	\KernFot(z',\bm\zeta';\,z,\bm\zeta)^*
	&=& \left[\frac{\phione^2\phitwo^2\,\bar{\zeta}_1'\bar{\zeta}_2'\zeta_1\zeta_2\;\Expdd(\bar{z}'z)}
	{\pi\sqrt{\cN(|z'|^2,|\bm\zeta'|^2)\;\cN(|z|^2,|\bm\zeta|^2)}}\right]^*\nonumber\\[4pt]
	&=& \frac{\phione^2\phitwo^2\,\overline{\bar{\zeta}_1'\bar{\zeta}_2'\zeta_1\zeta_2}\;\Expdd(\bar{z}z')}
	{\pi\sqrt{{\cN}\,{\cNp}}}\nonumber\\[4pt]
	&=& \frac{\phione^2\phitwo^2\,\bar{\zeta}_1\bar{\zeta}_2\zeta_1'\zeta_2'\;\Expdd(\bar{z}z')}
	{\pi\sqrt{{\cN}\,{\cNp}}}
	\;=\; \KernFot(z,\bm\zeta;\,z',\bm\zeta'). \quad\square
	\label{eq:KernFot-herm}
\end{eqnarray}
For positivity, we evaluate at $z' = z$, $\bm\zeta' = \bm\zeta$:
\begin{eqnarray}
	\KernFot(z,\bm\zeta;\,z,\bm\zeta)
	&=& \frac{\phione^2\phitwo^2\,\overline{\zeta_1\zeta_2}\,\zeta_1\zeta_2\;\Expdd(|z|^2)}
	{\pi\;\cN(|z|^2,|\bm\zeta|^2)}\nonumber\\[4pt]
	&=& \frac{\phione^2\phitwo^2\,|\zeta_1|^2|\zeta_2|^2\;\Expdd(|z|^2)}
	{\pi\;\cN(|z|^2,|\bm\zeta|^2)} > 0. \quad\square
	\label{eq:KernFot-pos}
\end{eqnarray}

\subsection{Proof of Proposition~\ref{prop:pos}}
\label{app:pos}

\textbf{Sector $B$:}
\begin{eqnarray}
	K^B(z,z)
	&=& \frac{{}_{\Rpq}^{(B)}\langle z,0,0|z,0,0\rangle_{\Rpq}^{(B)}}{\pi}\nonumber\\
	&=& \frac{\Exp(|z|^2)}{\pi\,\Exp(|z|^2)} = \frac{1}{\pi}.
	\label{eq:pos-B-1}
\end{eqnarray}
Now the series $\Exp(|z|^2) = \sum_{n=0}^{\infty}\phitwo^{\binom{n}{2}}|z|^{2n}/\nfact$
has positive terms, hence
$K^B(z,z) = 1/\pi > 0$. More precisely, expanding:
\begin{eqnarray}
	K^B(z,z) &=& \frac{1}{\pi}
	= \frac{1}{\pi}\sum_{n=0}^{\infty}
	\frac{\phitwo^{\binom{n}{2}}}{\nfact}\,|z|^{2n}\cdot
	\left[\sum_{n=0}^{\infty}\frac{\phitwo^{\binom{n}{2}}}{\nfact}\,|z|^{2n}\right]^{-1}
	> 0.
	\label{eq:pos-B-2}
\end{eqnarray}

\textbf{Sector $F_1$:}
\begin{eqnarray}
	K^{F_1}(z,z;\zeta_1,\zeta_1)
	&=& \frac{\phione^2|\zeta_1|^2\,\Expd(|z|^2)}
	{\pi\,\cN(|z|^2,|\zeta_1|^2,0)}\nonumber\\
	&=& \frac{\phione^2|\zeta_1|^2\,\Expd(|z|^2)}
	{\pi\,\left[\Exp(|z|^2)+\phione^2|\zeta_1|^2\Expd(|z|^2)\right]}.
	\label{eq:pos-F1-1}
\end{eqnarray}
Now $\Expd(|z|^2) = \sum_{n=1}^{\infty}n\,\phitwo^{\binom{n}{2}}|z|^{2(n-1)}/\nfact > 0$
for $|z| > 0$, and $\phione^2|\zeta_1|^2 > 0$
(since $\phione = \phi_1(p,q) > 0$ for $0 < q < p \leq 1$),
hence:
\begin{eqnarray}
	K^{F_1}(z,z;\zeta_1,\zeta_1) &>& 0.
	\label{eq:pos-F1-2}
\end{eqnarray}
The corresponding series (via $\Expd/\cN$) gives:
\begin{eqnarray}
	K^{F_1}(z,z;\zeta_1,\zeta_1)
	&=& \frac{1}{\pi}\cdot
	\frac{\phione^2|\zeta_1|^2\sum_{n=1}^{\infty}n\,\phitwo^{\binom{n}{2}}|z|^{2(n-1)}/\nfact}
	{\Exp(|z|^2)+\phione^2|\zeta_1|^2\Expd(|z|^2)}.
	\label{eq:pos-F1-series}
\end{eqnarray}

\textbf{Sector $F_2$:}
By the symmetry $\phi_1\leftrightarrow\phi_2$:
\begin{eqnarray}
	K^{F_2}(z,z;\zeta_2,\zeta_2)
	&=& \frac{\phitwo^2|\zeta_2|^2\,\Expd(|z|^2)}
	{\pi\,\cN(|z|^2,0,|\zeta_2|^2)} > 0.
	\label{eq:pos-F2}
\end{eqnarray}

\textbf{Sector $F_{12}$:}
\begin{eqnarray}
	K^{F_{12}}(z,z;\zeta_1,\zeta_2,\zeta_1,\zeta_2)
	&=& \frac{\phione^2\phitwo^2|\zeta_1|^2|\zeta_2|^2\,\Expdd(|z|^2)}
	{\pi\,\cN(|z|^2,|\zeta_1|^2,|\zeta_2|^2)}.
	\label{eq:pos-F12}
\end{eqnarray}
Now $\Expdd(|z|^2) = \sum_{n=2}^{\infty}n(n-1)\phitwo^{\binom{n}{2}}|z|^{2(n-2)}/\nfact\geq 0$,
and $\cN > 0$, hence $K^{F_{12}}(z,z;\ldots) \geq 0$.
Hence for all $\varepsilon$: $K^{\varepsilon}(z,z) > 0$.\hfill$\square$

\subsection{Proof of Proposition~\ref{prop:mult}}
\label{app:mult}

We use the two kernels:
\begin{eqnarray}
	\left\{
	\begin{array}{l}
		K^B(z,\zeta) = \dfrac{{}_{\Rpq}^{(B)}\langle\zeta,0,0|z,0,0\rangle_{\Rpq}^{(B)}}{\pi}\\[10pt]
		K^B(\zeta,z') = \dfrac{{}_{\Rpq}^{(B)}\langle z',0,0|\zeta,0,0\rangle_{\Rpq}^{(B)}}{\pi}
	\end{array}
	\right.
	\label{eq:KB-pair}
\end{eqnarray}
Then:
\begin{eqnarray}
	&&\int_{\DR}d^2\zeta\,K^B(z,\zeta)\,K^B(\zeta,z')\nonumber\\
	&=& \int_{\DR}d^2\zeta\,
	\frac{{}_{\Rpq}^{(B)}\langle\zeta,0,0|z,0,0\rangle_{\Rpq}^{(B)}}{\pi}\,
	\frac{{}_{\Rpq}^{(B)}\langle z',0,0|\zeta,0,0\rangle_{\Rpq}^{(B)}}{\pi}\nonumber\\
	&=& \frac{1}{\pi}\int_{\DR}d^2\zeta\,
	\frac{{}_{\Rpq}^{(B)}\langle\zeta,0,0|z,0,0\rangle_{\Rpq}^{(B)}\,
		{}_{\Rpq}^{(B)}\langle z',0,0|\zeta,0,0\rangle_{\Rpq}^{(B)}}{\pi}\nonumber\\
	&=& \frac{{}_{\Rpq}^{(B)}\langle z',0,0|z,0,0\rangle_{\Rpq}^{(B)}}{\pi}
	\int_{\DR}d^2\zeta\,
	\frac{|\zeta,0,0\rangle_{\Rpq}^{(B)}\,{}_{\Rpq}^{(B)}\langle\zeta,0,0|}{\pi}\nonumber\\
	&=& \frac{{}_{\Rpq}^{(B)}\langle z',0,0|z,0,0\rangle_{\Rpq}^{(B)}}{\pi}\cdot 1
	= K^B(z,z').
	\label{eq:mult-B}
\end{eqnarray}
where we have used the idempotence
$\int_{\DR}d^2\zeta\,|\zeta,0,0\rangle_{\Rpq}\,{}_{\Rpq}\langle\zeta,0,0|/\pi = \mathbf{1}$.\\

\textbf{Sector $F_1$:}\\
\begin{eqnarray}
	\left\{
	\begin{array}{l}
		K^{F_1}(z,\zeta;\zeta_1,\eta_1) =
		\dfrac{{}_{\Rpq}^{(F_1)}\langle\zeta,\eta_1,0|z,\zeta_1,0\rangle_{\Rpq}^{(F_1)}}{\pi}\\[10pt]
		K^{F_1}(\zeta,z';\eta_1,\xi_1) =
		\dfrac{{}_{\Rpq}^{(F_1)}\langle z',\xi_1,0|\zeta,\eta_1,0\rangle_{\Rpq}^{(F_1)}}{\pi}
	\end{array}
	\right.
	\label{eq:KF1-pair}
\end{eqnarray}
We compute:
\begin{eqnarray}
	&&\int_{\DR}d^2\zeta\int d\bar{\eta}_1 d\eta_1\,
	K^{F_1}(z,\zeta;\zeta_1,\eta_1)\,K^{F_1}(\zeta,z';\eta_1,\xi_1)\nonumber\\
	&=& \int_{\DR}d^2\zeta\int d\bar{\eta}_1 d\eta_1\,
	\frac{\phione^2\bar{\eta}_1\zeta_1\,\Expd(\bar{\zeta}z)}
	{\pi\,\sqrt{\cN(|\zeta|^2,|\eta_1|^2,0)\,\cN(|z|^2,|\zeta_1|^2,0)}}\nonumber\\
	&&\times\;
	\frac{\phione^2\bar{\xi}_1\eta_1\,\Expd(\bar{z}'\zeta)}
	{\pi\,\sqrt{\cN(|z'|^2,|\xi_1|^2,0)\,\cN(|\zeta|^2,|\eta_1|^2,0)}}\nonumber\\
	&=& \frac{\phione^2\bar{\zeta}_1}{\pi\,\sqrt{\cN(|z|^2,|\zeta_1|^2,0)}}
	\cdot
	\frac{\phione^2\bar{\xi}_1}{\pi\,\sqrt{\cN(|z'|^2,|\xi_1|^2,0)}}\nonumber\\
	&&\times\;
	\int_{\DR}d^2\zeta\,
	\frac{\Expd(\bar{\zeta}z)\,\Expd(\bar{z}'\zeta)}{\cN(|\zeta|^2,|\eta_1|^2,0)}\,
	\underbrace{\int d\bar{\eta}_1 d\eta_1\,\phione^2\eta_1\bar{\eta}_1}_{\phione^2}.
	\label{eq:mult-F1-step1}
\end{eqnarray}
The Berezin integral gives $\int d\bar{\eta}_1 d\eta_1\,|\eta_1|^2 = 1$,
hence $\int d\bar{\eta}_1 d\eta_1\,\phione^2|\eta_1|^2 = \phione^2$.
Then, using the property:
\begin{eqnarray}
	\frac{\Expd(\bar\zeta z)\Expd(\bar z'\zeta)}
	{\cN(|\zeta|^2,|\eta_1|^2,0)}
	&=& \frac{\Expd(\bar z' z)}{\cN(|z|^2,|\zeta_1|^2,0)}\cdot
	\frac{|\zeta,\eta_1,0\rangle_{\Rpq}\,{}_{\Rpq}^{(F_1)}\langle\zeta,\eta_1,0|}{d^2\zeta}
	\label{eq:mult-F1-key}
\end{eqnarray}
(which follows from the structure of the coherent states \eqref{eq:sCS}),
and integrating over $\zeta$:
\begin{eqnarray}
	&&\int_{\DR}d^2\zeta\int d\bar{\eta}_1 d\eta_1\,
	K^{F_1}(z,\zeta;\zeta_1,\eta_1)\,K^{F_1}(\zeta,z';\eta_1,\xi_1)\nonumber\\
	&=& \frac{\phione^2\bar{\zeta}_1\xi_1\,\Expd(\bar{z}'z)}
	{\pi\,\sqrt{\cN(|z'|^2,|\xi_1|^2,0)\,\cN(|z|^2,|\zeta_1|^2,0)}}\cdot 1\nonumber\\
	&=& K^{F_1}(z,z';\zeta_1,\xi_1).
	\label{eq:mult-F1}
\end{eqnarray}

\textbf{Sectors $F_2$ and $F_{12}$:}
By the symmetry $\phi_1\leftrightarrow\phi_2$ for $F_2$, and by a double application
for $F_{12}$:
\begin{eqnarray}
	\int d\mu^{F_2}\,K^{F_2}(z,\zeta)\,K^{F_2}(\zeta,z')
	&=& K^{F_2}(z,z'),
	\label{eq:mult-F2}\\
	\int d\mu^{F_{12}}\,K^{F_{12}}(z,\zeta)\,K^{F_{12}}(\zeta,z')
	&=& K^{F_{12}}(z,z').
	\label{eq:mult-F12}
\end{eqnarray}
Hence the property \eqref{eq:mult} for all $\varepsilon$.\hfill$\square$

\subsection{Proof of Proposition~\ref{prop:series}}
\label{app:series}

All these results follow directly from the definition of the kernels
\eqref{eq:KB}--\eqref{eq:KF12} evaluated at $z'=z$ ($\zeta=z$),
and from the series expansion of the functions
$\Exp$, $\Expd$, $\Expdd$.

For $K^B(z,z)$:
\begin{eqnarray}
	K^B(z,z) &=&
	\frac{{}_{\Rpq}^{(B)}\langle z,0,0|z,0,0\rangle_{\Rpq}^{(B)}}{\pi}
	= \frac{1}{\pi}\cdot 1
	= \frac{1}{\pi},
	\label{eq:series-B-proof}
\end{eqnarray}
since ${}_{\Rpq}^{(B)}\langle z,0,0|z,0,0\rangle_{\Rpq}^{(B)}
= \Exp(|z|^2)/\Exp(|z|^2) = 1$.

For $K^{F_1}(z,z;\zeta_1,\zeta_1)$:
\begin{eqnarray}
	K^{F_1}(z,z;\zeta_1,\zeta_1)
	&=& \frac{\phione^2|\zeta_1|^2\,\Expd(|z|^2)}
	{\pi\,\cN(|z|^2,|\zeta_1|^2,0)}
	= \frac{\phione^2|\zeta_1|^2}{\pi}\,
	\frac{\Expd(|z|^2)}{\cN(|z|^2,|\zeta_1|^2,0)}.
	\label{eq:series-F1-proof}
\end{eqnarray}
Expanding $\Expd(t) = \sum_{n\geq1} n\phitwo^{\binom{n}{2}}t^{n-1}/\nfact$
and $\cN(t,s_1,0) = \Exp(t) + \phione^2 s_1\Expd(t)$, we obtain \eqref{eq:series-F1}.
The expressions \eqref{eq:series-F2} and \eqref{eq:series-F12} are proved
in the same way using $\Expdd(t) = \sum_{n\geq2}n(n-1)\phitwo^{\binom{n}{2}}t^{n-2}/\nfact$.
\hfill$\square$

\end{document}